  \documentclass[5p,times,twocolumn]{elsarticle}
  \usepackage{setspace}
  \usepackage[switch]{lineno}

\usepackage{comment}
\usepackage{amsmath}
\usepackage{graphicx}
\usepackage{hyperref}
\usepackage{float}
\usepackage{subcaption}
\usepackage{multirow}
\usepackage{afterpage}
\usepackage{etoolbox}
\usepackage[english]{babel}
\usepackage{natbib}
\usepackage{hhline,colortbl}

\newcommand{\fat}[1]{\mathbf{#1}} 
\newcommand{\bldgr}[1]{\boldsymbol{#1}} 
\newcommand{\transp}{^T} 
\newcommand{\ud}{\,\mathrm{d}}

\definecolor{todoColor}{rgb}{0.0,0.0,1.0}
\newcommand{\todo}[1]{\color{todoColor}[TODO: {#1}]\normalcolor}

\definecolor{RevisionColor}{rgb}{0.0, 0.0, 0.0} 

\definecolor{RevisionColorV2}{rgb}{0.0, 0.0, 0.0}

\newcommand{\revision}[1]{\color{RevisionColor}{#1}\normalcolor}

\newcommand{\revisionTwo}[1]{\color{RevisionColorV2}{#1}\normalcolor}

\definecolor{noteToSelfColor}{rgb}{0.6,0.6,0.6}

\definecolor{todoDoneColor}{rgb}{0.0,0.6,0.0}

\definecolor{unresolvedColor}{rgb}{0.6,0.0,0.0}

\definecolor{questionColor}{rgb}{1.0,1.0,0.0}

\definecolor{answerColor}{rgb}{0.0,1.0,0.0}

\setcitestyle{authoryear,round,semicolon} 

\let\today\relax
\makeatletter
\def\ps@pprintTitle{%
    \let\@oddhead\@empty
    \let\@evenhead\@empty
    \def\@oddfoot{\footnotesize\itshape
         {Submitted preprint to ???} \hfill\today}%
    \let\@evenfoot\@oddfoot
    }
\renewcommand{\fnum@figure}{Fig. \thefigure}
\addto\captionsenglish{}
\makeatother

\newcommand{\numberOfRegions}{41 } 

\begin{document}

\hyphenation{SAMSEG}

\begin{frontmatter}

\title{A 
Contrast-Adaptive 
Method
for Simultaneous 
Whole-Brain and Lesion
Segmentation in Multiple Sclerosis}

\author[add1,add2]{Stefano Cerri\corref{cor1}}
    \ead{stce@dtu.dk}
\author[add2]{Oula Puonti}
\author[add3]{Dominik S. Meier}
\author[add3]{Jens Wuerfel}
\author[add4]{Mark M\"{u}hlau}
\author[add2,add5,add6]{Hartwig R. Siebner}
\author[add1,add7]{Koen Van Leemput}
\address[add1]{Department of Health Technology, Technical University of Denmark, Denmark}
\address[add2]{Danish Research Centre for Magnetic Resonance, Copenhagen University Hospital Hvidovre, Denmark}
\address[add3]{Medical Image Analysis Center (MIAC AG) and Department of Biomedical Engineering, University Basel, Switzerland}
\address[add4]{Department of Neurology and TUM-Neuroimaging Center, School of Medicine, Technical University of Munich, Germany}
\address[add5]{Department of Neurology, Copenhagen University Hospital Bispebjerg, Denmark }
\address[add6]{Institute for Clinical Medicine, Faculty of Medical and Health Sciences, University of Copenhagen, Denmark}
\address[add7]{Athinoula A. Martinos Center for Biomedical Imaging, Massachusetts General Hospital, Harvard Medical School, USA}

\cortext[cor1]{Corresponding author}

\begin{abstract}
Here
we present a method for the simultaneous segmentation of white matter lesions and normal-appearing neuroanatomical structures from multi-contrast brain MRI scans of multiple sclerosis patients.
The method 
integrates 
a 
novel 
model for white matter lesions
into 
a 
previously validated
generative 
model for whole-brain segmentation.
By using separate models 
for the shape of anatomical structures and their appearance in MRI,
the algorithm
can
adapt to data acquired with different scanners and imaging protocols
without
retraining.
We validate the method using
\revision{four}
disparate
datasets, 
showing 
\revision{
robust
}
performance in white matter lesion segmentation while simultaneously segmenting
dozens of
other brain structures.
We further demonstrate that 
the contrast-adaptive method
can
also be 
\revision{safely}
applied 
to MRI scans of healthy controls,
and replicate previously documented atrophy patterns in 
deep gray matter structures in MS.
The
algorithm is publicly available as part of the open-source neuroimaging package FreeSurfer.\\


\end{abstract}

\end{frontmatter}

\section{Introduction}
\label{sec:Intro}

%
%
Multiple sclerosis (MS) is 
the most frequent
chronic
inflammatory
autoimmune
disorder of the central nervous system,
causing progressive damage and disability.
%
The disease
affects nearly half a million 
Americans
and 2.5 million individuals
world-wide~\citep{rosati2001prevalence,goldenberg2012multiple}, generating more than \$10 billion in annual healthcare spending in the United States alone~\citep{adelman2013cost}.

%
%
The ability to diagnose MS and track its progression has been greatly enhanced by magnetic resonance
imaging (MRI), which can detect 
characteristic brain lesions in white and gray matter%
~\citep{bakshi2008mri,lovblad2010mr,blystad2015quantitative,Garcia-Lorenzo2013}.
Lesions visualized by 
MRI 
are
up to an order of magnitude more sensitive in detecting disease activity
compared to
clinical assessment~\citep{filippi2006efns}.
The
prevalence and dynamics of white matter lesions
are 
thus
used clinically
to 
diagnose MS~\citep{Thompson2018},
define
disease
stages
and to determine
the efficacy of
\revisionTwo{a}
therapeutic regimen~\citep{Sormani}.
MRI is also
an unparalleled tool for characterizing brain atrophy, which occurs at a faster rate in patients with MS 
compared to
healthy controls~\citep{Barkhof2009,Azevedo2018} and, especially in deep gray matter structures and the cerebral cortex, has been shown to
correlate with measures of disability~\citep{geurts2012measurement}.

Although manual labeling 
remains 
the most accurate way\footnote{\revision{Although selectively fusing several automatic methods has recently been shown to approach human performance~\citep{Carass2020}.}} 
of delineating white matter lesions in MS~\citep{Commowick2018},
this approach is very cumbersome and 
in itself prone
to considerable intra- and inter-rater disagreement~\citep{Zijdenbos_MICCAI}.
%
%
Furthermore, manually labeling various normal-appearing brain structures to assess atrophy is simply too time consuming to be practically feasible.
%
%
%
Therefore,
there is a clear need for automated tools that can 
reliably and efficiently 
characterize the morphometry of white matter lesions, various neuroanatomical structures, and their changes over time directly from in vivo MRI.
Such tools 
are
of great potential value for diagnosing disease, tracking
progression, and evaluating treatment. 
They 
can also
help in obtaining
a better
understanding of 
underlying disease mechanisms, and 
to 
facilitate 
more
efficient testing in clinical trials. 
Ultimately, automated software tools may
help 
clinicians to prospectively identify which patients are at highest risk of future disability accrual, leading
to better counseling of patients and better overall clinical outcomes.

%

%
%
%
%

Despite decades of methodological development (cf.~\citep{Garcia-Lorenzo2013} or~\citep{Danelakis2018}%
),
currently available computational tools for analyzing MRI scans of MS patients 
remain limited in a number of 
important ways:
\begin{description}

  \item[ - ] Poor generalizability: 
  Existing tools are often developed and tested on very specific imaging protocols%
  ,
  and may not 
  be able to work on
  data that is acquired differently.
  Especially with the strong surge of 
  supervised 
  learning in recent years, 
  where the 
  relationship between image appearance and segmentation labels 
  in 
  training 
  scans 
  is 
  directly and
  statically
  encoded,
  the segmentation performance of many
  state-of-the-art algorithms will degrade substantially when applied to data from different scanners and acquisition
  protocols%
  ~\citep{Garcia-Lorenzo2013,Valverde2019}, severely limiting their usefulness in practice. 

%
%
%
%
%

  \item[-] Dearth of available software:
  Despite the 
  very large
  number of proposed methods,
  most algorithms are only developed and tested in-house, 
  and 
  very few 
  tools
  are 
  made 
  publicly available~\citep{Shiee2010,Schmidt2012,Bianca,Valverde2017}. 
  In order 
  to secure that
  computational methods
  will
  make 
  a
  real
  practical impact,
  they must be accompanied by
  %
  software implementations 
  that 
  work robustly 
  across
  a wide array of 
  image acquisitions;
  that are 
  made
  publicly available;
  and that are
  open-sourced, rigorously tested and comprehensively documented.
  

%
%
%

  \item[-] 
  Limitations
  in assessing atrophy:
   There is a lack of dedicated tools for characterizing brain atrophy patterns 
   in MS: 
   many
   existing methods
   characterize
   only aggregate measures such as global brain or gray matter volume~\citep{smith2002accurate,smeets2016reliable}
   rather than 
   individual brain structures,
   or require 
   that lesions are pre-segmented so that their MRI intensities can be replaced with placeholder values to avoid biased atrophy measures%
~\citep{sdika2009nonrigid,chard2010reducing,battaglini2012evaluating,gelineau2012effect,ceccarelli2012impact,Azevedo2018}
   (so-called lesion filling).
  
\end{description}

\begin{figure}
  \begin{center}
    \includegraphics[width=0.48\textwidth]{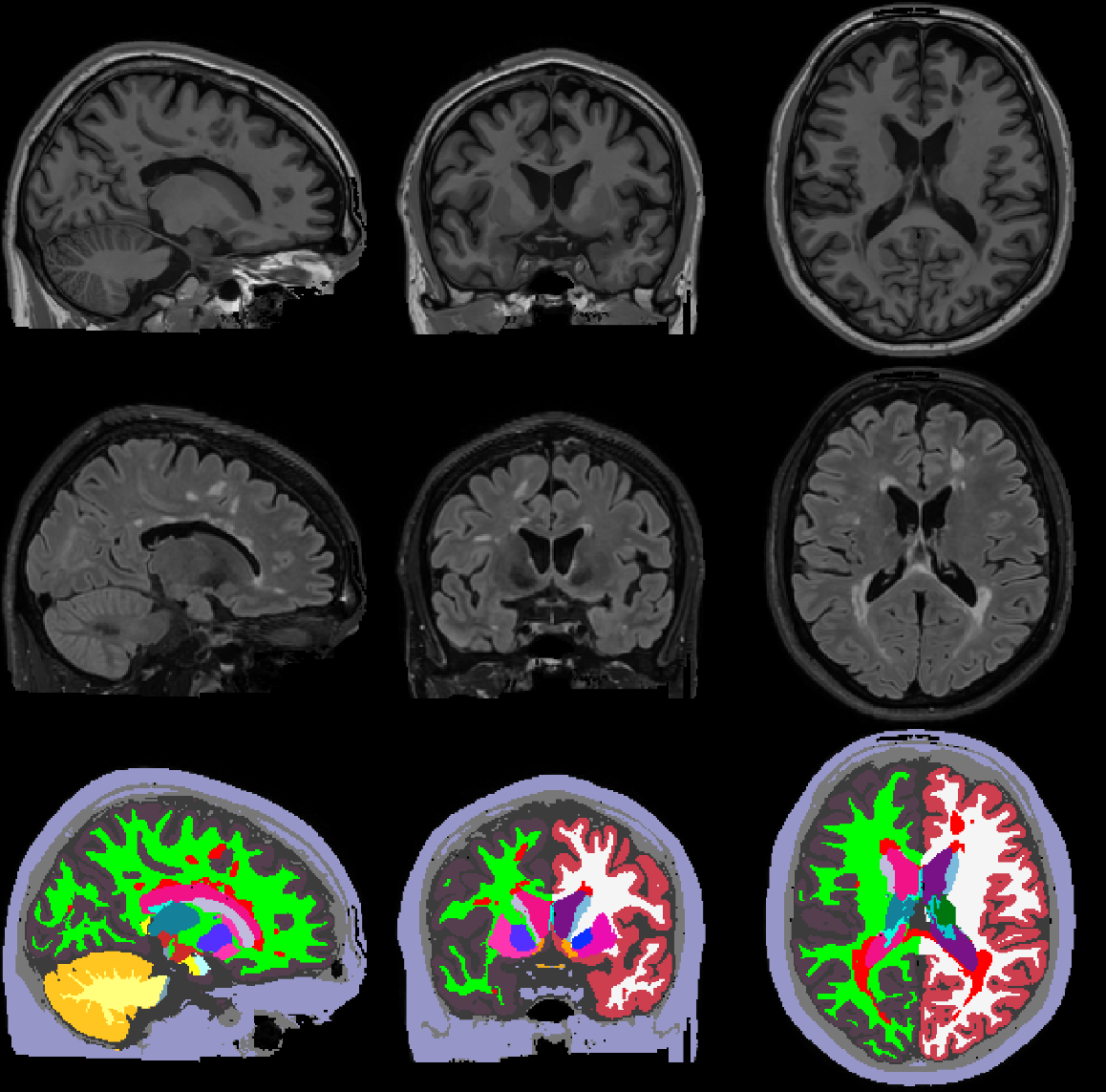}
  \end{center}
  \caption{Segmentation of white matter lesions and \numberOfRegions different brain structures from the proposed method on T1w-FLAIR input. 
  From left to right: sagittal, coronal, axial view. From top to bottom: T1w, FLAIR, automatic segmentation.
  } 
  \label{fig:example}
\end{figure}

In order to address these limitations, 
we describe 
a new open-source software tool 
for 
simultaneously 
segmenting 
white matter lesions and \numberOfRegions neuroanatomical structures
from MRI scans of 
MS patients.
An example segmentation 
produced by this tool
is shown in Fig.~\ref{fig:example}.
By performing lesion segmentation in the full context of whole-brain modeling,
the method 
obviates the need 
to segment lesions and assess atrophy in 
two separate processing phases,
as currently required in  lesion filling approaches.
%
%
%
%
The method 
works
robustly 
across a wide range of imaging hardware and protocols
by completely decoupling computational models of anatomy from models of the imaging process,
thereby side-stepping the 
intrinsic generalization difficulties of 
supervised methods such as convolutional neural networks.
Our software implementation is
freely available as part of the FreeSurfer neuroimaging analysis package~\citep{FreeSurfer}.

%
%


To the best of our knowledge, only two other methods have been 
developed
for joint whole-brain and white matter lesion segmentation in MS. 
\citep{Shiee2010} model lesions as an extra tissue class in an unsupervised whole-brain segmentation method~\citep{Bazin},
removing false positive detections of lesions using a combination of topological constraints and hand-crafted rules 
implementing
various intensity- and distance-based heuristics.
However, the method segments only a small set of neuroanatomical structures (10), and validation of this aspect 
was limited to a simulated MRI scan of a single subject. 	
\citep{Mckinley2019} use a cascade of two convolutional neural networks,
with the first one skull-stripping individual 
image modalities and the second one generating the 
actual segmentation.
However, the whole-brain segmentation performance 
of this method
was only evaluated on a 
few
structures (7).
Furthermore, 
as a supervised method 
its applicability 
on 
data that differs substantially from its training data 
will necessarily be 
limited.

A preliminary version of this work was presented in \cite{Puonti2016a}.
Compared to this earlier work, the current article employs more advanced models for the shape and appearance of white matter lesions, and includes a more thorough validation of the segmentation performance of the proposed method, including an evaluation of the whole-brain segmentation component and comparisons with human inter-rater variability.

\section{Contrast-adaptive whole-brain segmentation}
\label{sec:ModelNoLes}

We build upon 
a method for whole-brain segmentation called Sequence Adaptive
Multimodal SEGmentation (SAMSEG) that we previously developed~\citep{Puonti2016},
and that we propose to extend with the capability to handle white matter lesions.
SAMSEG robustly segments 
\numberOfRegions structures 
from head MRI scans
without any form of preprocessing
or prior assumptions on the scanning platform or the number and type of pulse sequences used.
%
Since we build heavily on this method for the remainder of the paper, 
we briefly 
outline its main 
characteristics
here.

%
%
%
%



SAMSEG is based on a generative approach, in which a forward probabilistic model is inverted to obtain automated segmentations.
Let $ \fat{D}=(\fat{d}_{1}, \dots , \fat{d}_{I})$ denote a matrix collecting the
intensities in a multi-contrast brain MR scan with $I$ voxels, where the vector $ \fat{d}_{i} = (d_{i}^{1}, \dots , d_{i}^{N})^{T} $ contains the intensities in voxel $i$ for each of the available $N$ contrasts. Furthermore, let $\fat{l} = (l_{1}, \dots , l_{I})^{T}$ be the corresponding labels, where $ l_{i} \in \{1, \dots K\}$ denotes one of the $K$ possible segmentation labels assigned to voxel $i$.
%
SAMSEG
estimates a segmentation $\fat{l}$ from MRI data $\fat{D}$ by using a generative model, illustrated 
in black
in~Fig.~\ref{fig:graphicalModel}.
According to this model, 
$\fat{l}$ is sampled from a  
segmentation prior $p(\fat{l} | \bldgr{\theta}_{\fat{l}})$,
after which $\fat{D}$ is obtained by sampling from a
likelihood function $p(\fat{D} | \fat{l}, \bldgr{\theta}_{\fat{d}})$,
where $\bldgr{\theta}_{\fat{l}}$ and $\bldgr{\theta}_{\fat{d}}$ are 
model parameters with priors $p( \bldgr{\theta}_{\fat{l}})$ and $p( \bldgr{\theta}_{\fat{d}})$.
Segmentation then consists of inferring 
the unknown
$\fat{l}$ from 
the observed
$\fat{D}$ under this model.
In the following,
we summarize the segmentation prior and the likelihood used in
SAMSEG,
as well as the way the resulting model is used to obtain automated segmentations.


\subsection{Segmentation prior}
\label{subsec:SegmentationPrior}

To model the spatial configuration of 
various
neuroanatomical
structures, we use a 
deformable
probabilistic atlas 
as detailed
in~\citep{Puonti2016}. 
In short,
the atlas
is based on a 
tetrahedral mesh, 
where the parameters $\bldgr{\theta}_{\fat{l}}$ are the spatial positions of the
mesh's
vertices,
and 
$p( \bldgr{\theta}_{\fat{l}} )$
is a topology-preserving deformation prior that prevents the 
mesh from tearing or folding~\citep{Ashburner2000}.
%
The model assumes conditional independence of the labels between voxels 
for a given deformation:
\begin{linenomath}
\begin{align*}
  p(\fat{l}| \bldgr{\theta}_{\fat{l}}) 
  = \prod_{i=1}^{I} 
    p( l_i | \bldgr{\theta}_{\fat{l}} )
  ,
\end{align*}
\end{linenomath}
and computes
the probability of observing label $k$ at voxel $i$
as
\begin{linenomath}
\begin{align}
  p( l_i = k | \bldgr{\theta}_{\fat{l}} )
  = \sum_{j=1}^J \alpha_j^k 
  \psi_j^i( \bldgr{\theta}_{\fat{l}} )
  \label{eq:atlasPrior}
  ,
\end{align}
\end{linenomath}
where
$\alpha_j^k$ 
are
label probabilities 
defined at 
the 
$J$ 
vertices of the mesh,
and
$\psi_j^i( \bldgr{\theta}_{\fat{l}} )$ denotes a spatially compact, piecewise-linear interpolation basis function attached to the $j^{th}$ 
vertex
and evaluated at the $i^{th}$ voxel%
~\citep{VanLeemput2009}.
%


The topology of the mesh, the mode of the deformation prior $p(\bldgr{\theta}_{\fat{l}})$, and the label probabilities $\alpha_j^k$ 
can be 
learned automatically 
from a set of 
segmentations provided as training data~\citep{VanLeemput2009}.
This
involves an iterative process that combines a mesh simplification 
operation
with 
a
group-wise 
nonrigid registration 
step
to warp 
the atlas to each of \revision{the} training subjects,
and 
an Expectation Maximization (EM) algorithm~\citep{Dempster1977JRSS} to estimate the label probabilities 
$\alpha_j^k$
in the mesh vertices%
.
The result is a sparse mesh that encodes high-dimensional atlas deformations through a compact set of vertex displacements.
As 
described
in~\citep{Puonti2016},
the atlas
used in SAMSEG was derived from 
manual whole-brain segmentations of 20 subjects,
representing a mix of healthy individuals and subjects with questionable or probable Alzheimer's disease.%

\subsection{Likelihood function}
\label{subsec:likelihood}

For the likelihood function we use a Gaussian model for each of the $K$ different structures.
We assume that the bias field artifact can be modelled as a multiplicative and spatially smooth effect \citep{Wells1996}.
For computational reasons, we use log-transformed image intensities in $\fat{D}$, and model the bias field as a linear combination of spatially smooth basis functions that is added to the local voxel intensities \citep{VanLeemput1999}. Letting $\bldgr{\theta}_{\fat{d}}$ collect all bias field parameters and Gaussian means and variances, the likelihood is defined as
\begin{linenomath}
\begin{align*}
  p(\fat{D} | \fat{l}, \bldgr{\theta}_{\fat{d}}) = \prod_{i=1}^{I} p(\fat{d}_{i} | l_{i}, \bldgr{\theta}_{\fat{d}}),
  \quad
\end{align*}
\end{linenomath}
\begin{linenomath}
\begin{align*}
  p(\fat{d}_{i} | l_i=k, \bldgr{\theta}_{\fat{d}} ) 
  = 
  \mathcal{N}( \fat{d}_{i} | \bldgr{\mu}_k + \fat{C}\bldgr{\phi}_{i}, \bldgr{\Sigma}_k ),
\end{align*}
\end{linenomath}
\begin{linenomath}
\begin{align*}
  \fat{C} = 
  \left(
    \begin{array}{c}
      \fat{c}_1^T \\
      \vdots \\
      \fat{c}_N^T
    \end{array}
  \right),
  \quad
  \fat{c}_n = 
  \left(
    \begin{array}{c}
      c_{n,1} \\
      \vdots \\
      c_{n,P}
    \end{array}
  \right),
  \quad
  \bldgr{\phi}_i = 
  \left(
    \begin{array}{c}
      \phi_1^i \\
      \vdots \\
      \phi_P^i
    \end{array}
  \right),
\end{align*}
\end{linenomath}
where $P$ denotes the number of bias field basis functions, $\phi_p^i$ is the basis function $p$ evaluated at voxel $i$, and $\fat{c}_n$ holds the bias field coefficients for MRI contrast $n$.
We use a flat prior for the parameters of the likelihood: $p(\bldgr{\theta}_{\fat{d}}) \propto 1$.

\subsection{Segmentation}
\label{subsec:Inference}

For a given MRI scan $\fat{D}$,
segmentation proceeds by 
computing
a point estimate of 
the unknown 
model
parameters $\bldgr{\theta} = \{ \bldgr{\theta}_{\bldgr{d}}, \bldgr{\theta}_{\bldgr{l}} \}$:
\begin{linenomath}
\begin{align*}
  \bldgr{\hat{\theta}}
  = \arg\max_{\bldgr{\theta}}
    p( \bldgr{\theta} | \fat{D} )
    ,
\end{align*}
\end{linenomath}
which effectively fits the model to the data.
Details of this 
procedure are given in~\ref{app:GEMS}.
%
Once 
$\bldgr{\hat{\theta}}$ 
is found, 
the
corresponding maximum a posteriori (MAP) segmentation
\begin{linenomath}
\begin{align*}
  \fat{\hat{l}} 
  = 
  \arg \max_{\fat{l}} p( \fat{l} | \fat{D}, \bldgr{\hat{\theta}} )
\end{align*}
\end{linenomath}
is obtained by
assigning each voxel to the label with the highest 
probability, i.e., 
$
\hat{l}_i = \arg \max_k 
\hat{w}_{i,k}
$,
where
$0 \leq \hat{w}_{i,k} \leq 1$
are 
probabilistic
label assignments
\begin{align}
  w_{i,k}
  =
  \frac{
    \mathcal{N}( \fat{d}_i | \bldgr{\mu}_{k} + \fat{C} \bldgr{\phi}_i, \bldgr{\Sigma}_{k} ) p( l_i = k | \bldgr{\theta}_{\fat{l}} ) 
  }{
    \sum_{k'=1}^K \mathcal{N}( \fat{d}_i | \bldgr{\mu}_{k'} + \fat{C} \bldgr{\phi}_i, \bldgr{\Sigma}_{k'} ) p( l_i = k' | \bldgr{\theta}_{\fat{l}} )
  }
  \label{eq:healthySeg}
\end{align}
evaluated at the estimated parameters $\bldgr{\hat{\theta}}$.
It is worth emphasizing that, 
since 
the class means and variances $\{\bldgr{\mu}_k, \bldgr{\Sigma}_k\}$
are estimated from each target scan individually, the model automatically adapts to each scan's specific intensity characteristics -- a property that we demonstrated experimentally on several data sets acquired with different 
imaging protocols,
scanners and field strengths in~\cite{Puonti2016}.

Our implementation of this method,
written in Python with the exception of C++ parts for the computationally demanding optimization of the atlas mesh deformation, 
is available as part of the open-source package FreeSurfer%
\footnote{\url{http://freesurfer.net/}}.
%
It segments MRI brain scans without any form of preprocessing such as skull stripping or bias field correction, 
taking around 10 minutes to process one subject on a state-of-the-art computer (measured on a machine with an Intel 12-core i7-8700K processor).
As explained in~\citep{Puonti2016}, 
in our implementation we make use of the fact that many neuroanatomical structures share the same intensity characteristics in MRI 
to reduce the number of free parameters in the model
(e.g., all white matter structures share the same Gaussian mean $\bldgr{\mu}_{k}$ and variance $\bldgr{\Sigma}_k$,
as do most gray matter structures).
Furthermore, for some structures (e.g., non-brain tissue) we use Gaussian mixture models instead of a single Gaussian. 
In addition to using full covariance matrices $\bldgr{\Sigma}_k$, our implementation also supports diagonal covariances, which
is currently selected as the default behavior.





\section{Modeling lesions}
\label{sec:modelingLesion}

\begin{figure}
	\begin{center}
		\includegraphics[width=0.48\textwidth]{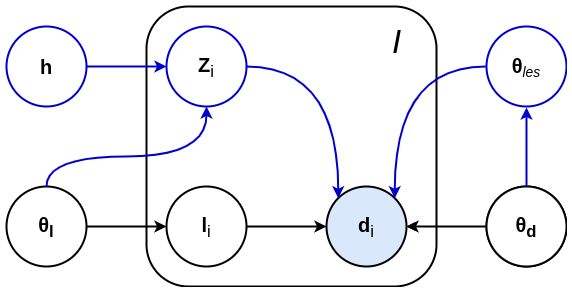}
	\end{center}
	\caption{Graphical model of the proposed method. In black the existing contrast-adaptive whole-brain segmentation method SAMSEG (without lesion modeling), in blue the proposed additional components to 
	\revisionTwo{also model}
	white matter lesions. Shading indicates observed variables. The plate indicates $I$ repetitions of the included variables%
	\revision{, where $I$ is the number of voxels.
	}
	} 
	\label{fig:graphicalModel}
\end{figure}

In order to make SAMSEG capable of 
additionally 
segmenting white matter lesions,
we augment its generative model 
by introducing a binary lesion map $\fat{z} = ( z_1, \ldots, z_I )\transp$, where $z_i \in \{0,1\}$ indicates the presence of a lesion in voxel $i$.
The augmented model is depicted in Fig.~\ref{fig:graphicalModel}, where the blue parts indicate the additional components compared to the original SAMSEG method.
The complete model consists of
a joint (i.e., over both $\fat{l}$ and $\fat{z}$ simultaneously) segmentation prior 
$p(\fat{l},\fat{z} | \fat{h}, \bldgr{\theta}_{\fat{l}})$,
where $\fat{h}$ is 
a 
new
latent variable that helps 
constrain the \emph{shape} of lesions,
as well as  
a joint likelihood 
$
p(\fat{D} | \fat{l},\fat{z}, \bldgr{\theta}_{\fat{d}}, \bldgr{\theta}_{les} )
$ 
,
where
$\bldgr{\theta}_{les}$ 
are 
new
parameters that govern their \emph{appearance}. 
In the following, we summarize the segmentation prior and the likelihood used in the augmented model, as well as the way the resulting model is used to obtain automated segmentations.

\subsection{Segmentation prior}

We use a joint segmentation prior of the form
\begin{linenomath}
\begin{align*}
  p(\fat{l},\fat{z} | \fat{h}, \bldgr{\theta}_{\fat{l}}) 
  =
  p(\fat{z} | \fat{h}, \bldgr{\theta}_{\fat{l}})
  p(\fat{l} | \bldgr{\theta}_{\fat{l}})
  ,
\end{align*}
\end{linenomath}
where 
$
p(\fat{l} | \bldgr{\theta}_{\fat{l}})
$
is the deformable atlas model 
defined in Section~\ref{subsec:SegmentationPrior},
and
\begin{linenomath}
\begin{align*}
p(\fat{z} | \fat{h}, \bldgr{\theta}_{\fat{l}}) 
= \prod_{i=1}^I p( z_i | \fat{h}, \bldgr{\theta}_{\fat{l}})
\end{align*}
\end{linenomath}
is a factorized model where
the probability that a voxel is part of a lesion
is given by:
\begin{linenomath}
\begin{align*}
p( z_i = 1 | | \fat{h}, \bldgr{\theta}_{\fat{l}} )
=
f_i( \fat{h} )
\,
\rho_i( \bldgr{\theta}_{\fat{l}} )
.
\end{align*}
\end{linenomath}
Here 
$0 \leq f_i( \fat{h} )  \leq 1$
aims to enforce
\emph{shape} constraints 
on lesions,
whereas
$0 \leq \rho_i( \bldgr{\theta}_{\fat{l}} ) \leq 1$
takes into account 
a 
voxel's 
spatial \emph{location} 
within 
its 
neuroanatomical context.
Below we provide more details on both these components of the model.

\subsubsection{Modeling lesion shapes}


In order to model lesion shapes, we use a variational auto-encoder
~\citep{Kingma2013,Rezende2014}
according to which lesion segmentation maps $\fat{z}$ are generated in a two-step process:
An unobserved, low-dimensional code $\fat{h}$ is 
first
sampled 
from a spherical Gaussian distribution 
$
p( \fat{h} ) = \mathcal{N}( \fat{h} | \fat{0}, \fat{I} )
$%
,
and
subsequently ``decoded'' into $\fat{z}$ 
by sampling from a factorized Bernoulli model:
\begin{linenomath}
\begin{align*}
p_{\omega}( \fat{z} | \fat{h} ) 
= \prod_{i=1}^I f_i( \fat{h} )^{z_i} \left( 1 - f_i( \fat{h} ) \right)^{(1-z_i)}
.
\end{align*}
\end{linenomath}
Here
$f_i( \fat{h})$ are the outputs of a 
``decoder''
convolutional neural network (CNN)
with 
filter weights
$\bldgr{\omega}$,
which parameterize the model.

Given a training data set in the form of $N$ binary segmentation maps
$\mathcal{D} = \{ \fat{z}^{(n)} \}_{n=1}^N $,
suitable
network parameters $\bldgr{\omega}$ 
can 
in principle
be estimated by maximizing the log-probability assigned to the data by the model%
:
\begin{linenomath}
\begin{align*}
    \log p_{\omega}( \mathcal{D} ) = 
\sum_{\fat{z} \in \mathcal{D}} \log p_{\omega}( \fat{z} )
,
\,\,
\mathrm{where} 
\,\,
p_{\omega}( \fat{z} )
= 
\int_{\fat{h}} p_{\omega}( \fat{z}_n | \fat{h} ) p( \fat{h} ) \ud \fat{h}
.
\end{align*}
\end{linenomath}
However,
because the integral over the latent 
codes
makes
this
intractable, 
we
use amortized variational inference in the form of 
stochastic gradient variational Bayes~\citep{Kingma2013,Rezende2014}.
In particular, we introduce an approximate posterior 
\begin{linenomath}
\begin{align*}
q_{\upsilon}(\fat{h}|\fat{z})
= \mathcal{N}\left( 
  \,
  \fat{h}
  \,
  | 
  \,
  \bldgr{\mu}_{\upsilon}(\fat{z}), 
  \,
  \mathrm{diag}( \bldgr{\sigma}^2_{\upsilon}(\fat{z}) ) 
\right)
,
\end{align*}
\end{linenomath}
where
the functions
$\bldgr{\mu}_{\upsilon}(\fat{z})$ 
and
$\bldgr{\sigma}_{\upsilon}(\fat{z})$
are 
implemented as an ``encoder'' CNN
parameterized by $\bldgr{\upsilon}$%
.
The variational parameters
$\bldgr{\upsilon}$
are
then
learned jointly with the model parameters $\bldgr{\omega}$
by maximizing a variational lower bound 
$
\sum_{\fat{z} \in \mathcal{D}} \mathcal{L_{\omega,\upsilon}( \fat{z} )}
\leq \log p_{\omega}( \mathcal{D} )
$
using stochastic gradient descent,
where
\begin{eqnarray}
\mathcal{L_{\omega,\upsilon}( \fat{z} )}
=
-D_{KL}(q_{\upsilon}(\fat{h}|\fat{z} ) || p(\fat{h}) )
+
\mathbb{E}_{q_{\upsilon}(\fat{h}|\fat{z} )}\left[ \log p_{\omega}( \fat{z} | \fat{h} ) \right]
.
\label{eq:VAELowerBound}
\end{eqnarray}
The first term is the Kullback-Leibler divergence between the approximate posterior and the prior,
which can be evaluated analytically.
The expectation in the last term is approximated using Monte Carlo sampling, 
using a change of variables (known as the ``reparameterization trick'') 
to reduce the variance in the computation of the gradient with respect to $\bldgr{\upsilon}$~\citep{Kingma2013,Rezende2014}.

\begin{figure*}[ht]
    \centering
    \begin{subfigure}[t]{0.9\textwidth}
        \centering
        \includegraphics[width=\linewidth]{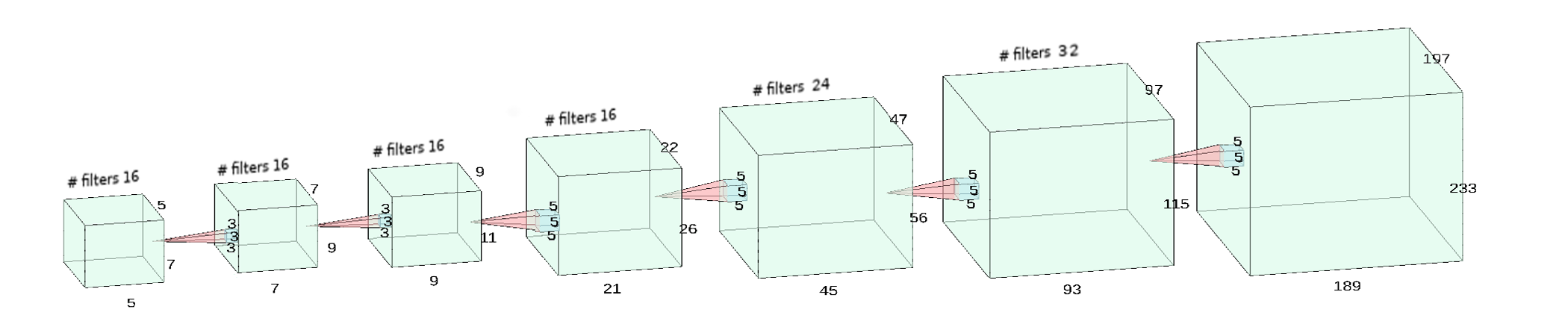} 
        \caption{Decoder network architecture.} 
        \label{fig:decoder}
    \end{subfigure}
    \begin{subfigure}[t]{0.9\textwidth}
        \centering
        \includegraphics[width=\linewidth]{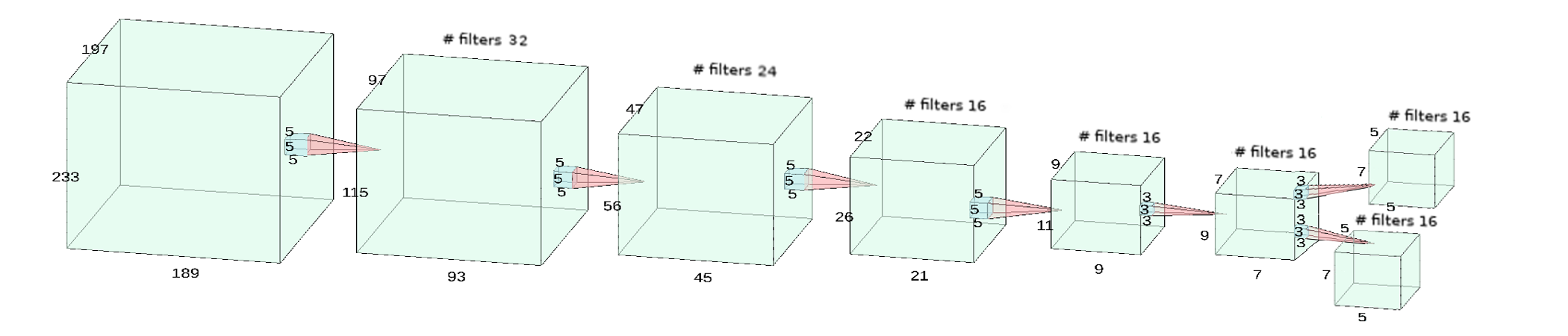} 
        \caption{Encoder network architecture.} 
        \label{fig:encoder}
    \end{subfigure}
	\caption{
	Lesion shape model architecture consisting of two symmetrical convolutional neural networks: (a) decoder network and (b) encoder network.
	The decoder network generates lesion segmentations from a low-dimensional code.
	Its architecture has ReLU activation functions ($f(x) = max(0, x)$) and batch normalization \citep{Ioffe2015} between each deconvolution layer, with the last layer having a sigmoid activation function, ensuring $0 \leq f_i(\fat{h}) \leq 1 $.
	The encoder network 
	encodes
	lesion segmentations into a
	latent
	code. 
	The main differences compared to the decoder network are the use of convolutional layers instead of deconvolutional layers and, 
	to encode the mean and variance parameters, the last layer has been 
	\revision{split}
	in two, with no activation function for the mean and a softplus activation function ($f(x) = \ln ( 1 + e^{x})$) for the variance.
	}
	\label{fig:VAEarchitecture}
\end{figure*}

Our training data set $\mathcal{D}$ was derived from manual lesion segmentations in 212 MS subjects, obtained from the University Hospital of Basel, Switzerland. The segmentations were all affinely registered and resampled to a 1mm isotropic grid of size 197 $\times$ 233 $\times$ 189. 
In order to reduce the risk of overfitting
to the training data,
we augmented 
each segmentation in the training data set by applying a rotation of 10 degrees around each axis, obtaining a total of 1484 
segmentations.
%
The architecture for our encoder and decoder networks is detailed in 
Fig.~\ref{fig:VAEarchitecture}. 
We trained the model for 1000 epochs 
with mini-batch size of 10 
using 
Adam optimizer~\citep{Kingma}
with a learning rate of 1e-4.
We approximated the expectation in the variational lower bound of Eq~\eqref{eq:VAELowerBound} by using a single Monte Carlo sample in each step.

\subsubsection{Modeling the spatial location of lesions}
\label{sec:VAE}

In order to encode the 
spatially 
varying 
frequency of
occurrence of lesions
across 
the brain,
we model the probability of finding a lesion in voxel $i$, based on its location alone, as
\begin{linenomath}
\begin{align*}
\rho_i( \bldgr{\theta}_{\mathbf{l}} )
  = 
  \sum_{j=1}^J 
  \beta_j
  \psi_j^i( \bldgr{\theta}_{\fat{l}} )
  ,
\end{align*}
\end{linenomath}
where lesion probabilities 
$0 \leq \beta_j \leq 1$ 
defined in the vertices of the SAMSEG 
atlas mesh are interpolated at the voxel location.
This effectively defines a 
lesion probability map that deforms in conjunction with the SAMSEG atlas to match the neuroanatomy in each image being segmented, allowing the model to impose contextual constraints on where lesions are expected to be found.

We estimated the parameters $\beta_j$ by 
running SAMSEG 
on
MRI scans (T1-weighted (T1w) and FLAIR) of 54 MS subjects in whom lesions had been manually annotated (data from the University Hospital of Basel, Switzerland), and recording the estimated atlas deformations. 
The parameters $\beta_j$ were then 
computed from the manual lesion segmentations
by applying the same 
technique 
we used
to estimate the $\alpha_j^k$ parameters in the SAMSEG atlas training phase (cf. Sec.~\ref{subsec:SegmentationPrior}).

\subsection{Likelihood function}

For the likelihood, which links joint segmentations $\{\fat{l},\fat{z}\}$ to intensities $\fat{D}$,
we use the same model as SAMSEG in voxels that do not contain lesion ($z_i=0$), but 
draw
intensities in 
lesions
($z_i=1$) 
from a separate Gaussian with parameters $\bldgr{\theta}_{les} = \{ \bldgr{\mu}_{les}, \bldgr{\Sigma}_{les} \}$:
\begin{linenomath}
\begin{align*}
  p(\fat{D} | \fat{l}, \fat{z}, \bldgr{\theta}_{\fat{d}}, \bldgr{\theta}_{les} ) 
  = \prod_{i=1}^{I} p(\fat{d}_{i} | l_{i}, z_i, \bldgr{\theta}_{\fat{d}}, \bldgr{\theta}_{les} ),
  \quad
\end{align*}
\end{linenomath}
where
\begin{linenomath}
\begin{align*}
  p(\fat{d}_{i} | l_i=k, z_i, \bldgr{\theta}_{\fat{d}}, \bldgr{\theta}_{les} ) 
  = 
  \begin{cases}
    \mathcal{N}( \fat{d}_{i} | \bldgr{\mu}_{les} + \fat{C}\bldgr{\phi}_{i}, \bldgr{\Sigma}_{les} ) & \text{if } z_i=1, \\
    \mathcal{N}( \fat{d}_{i} | \bldgr{\mu}_k + \fat{C}\bldgr{\phi}_{i}, \bldgr{\Sigma}_k ) & \text{otherwise}.
  \end{cases}
\end{align*}
\end{linenomath}
In order to constrain the 
values that the 
lesion intensity
parameters $\bldgr{\theta}_{les}$ can take, 
we make them conditional on the remaining intensity parameters
using a 
normal-inverse-Wishart 
distribution%
:
\begin{linenomath}
\begin{align}
   p( \bldgr{\theta}_{les} | \bldgr{\theta}_{\fat{d}} )
   = 
   \mathcal{N}( \bldgr{\mu}_{les} | \bldgr{\mu}_{WM}, \nu^{-1}\bldgr{\Sigma}_{les} ) \mathrm{IW}( \bldgr{\Sigma}_{les} | \kappa \nu \bldgr{\Sigma}_{WM}, \nu-N-2 )
   .
   \label{eq:NIW}
\end{align}
\end{linenomath}
Here 
the subscript ``WM'' denotes the white matter Gaussian 
and 
$\kappa > 1$ and $\nu \geq 0$ 
are hyperparameters in the model. 

This choice of model
is motivated by the fact that 
the normal-inverse-Wishart distribution is a conjugate prior for the parameters of a Gaussian distribution:
Eq.~\eqref{eq:NIW}
can be interpreted 
\revision{as}
providing 
$\nu$ ``pseudo-voxels'' with 
empirical
mean 
$\bldgr{\mu}_{WM}$ and variance 
$\kappa\bldgr{\Sigma}_{WM}$
in scenarios where
the lesion intensity parameters $\bldgr{\mu}_{les}$ and $\bldgr{\Sigma}_{les}$
need to be estimated from data.
In the absence of any such pseudo-voxels ($\nu = 0$), 
Eq.~\eqref{eq:NIW}
reduces to a flat prior on $\bldgr{\theta}_{les}$ and lesions are modeled as a completely 
independent class. 
Although such models have been 
used
in the literature~\citep{Guttmann1999,Kikinis1999,Shiee2010,Sudre2015}
their robustness may suffer
when applied to subjects with no or very few lesions, such as controls or patients with early disease, since there is essentially no data to estimate 
the lesion intensity parameters
from.
In the other extreme case%
, the number of pseudo-voxels can be set to such a high value ($\nu \rightarrow \infty$) that the 
intensity
parameters of the lesions are fully determined by those of WM.
This
effectively
replaces the Gaussian intensity model for WM 
in SAMSEG
by 
a distribution with longer tails,
in the form of 
a mixture of two Gaussians with identical means 
($\bldgr{\mu}_{les} \equiv \bldgr{\mu}_{WM}$)
but 
variances that differ by a constant 
factor
($\bldgr{\Sigma}_{les} \equiv \kappa\bldgr{\Sigma}_{WM}$ vs.~$\bldgr{\Sigma}_{WM}$).
In this scenario, 
MS lesions are detected as
model
outliers 
in a method using robust model parameter estimation\revision{~\citep{Huber1981}},
another technique that has also frequently been used in the literature~\citep{VanLeemput2001,Ait-Ali2005,Bricq2008,Rousseau2008,Prastawa2008,Liu2009,Garcia-Lorenzo2011}.

Based on pilot experiments on a variety of 
datasets
(distinct from the ones used in the results section), 
we found that 
good results are obtained by using an intermediate value of $\nu=500$ pseudo-voxels
for 1mm$^3$ isotropic scans, together with a scaling factor $\kappa=50$.
In order to adapt to different image resolutions, $\nu$ is scaled inversely proportionally with the voxel size in our implementation. 
We 
will visually demonstrate the 
role
of these hyperparameters 
in constraining
the lesion intensity parameters
in Sec.~\ref{sec:illustrationMethod}.

\subsection{Segmentation}
\label{sec:inferenceLesion}

As in the original SAMSEG method, segmentation proceeds 
by 
first
obtaining
point estimates $\bldgr{\hat{\theta}}$ that fit the model to the data,
and 
then
inferring
the corresponding segmentation posterior:
\begin{linenomath}
\begin{align*}
  p( \fat{l}, \fat{z} | \fat{D}, \bldgr{\hat{\theta}} )
  ,
\end{align*}
\end{linenomath}
which is now jointly over $\fat{l}$ and $\fat{z}$ simultaneously. 
Unlike in SAMSEG, however, both steps are 
made intractable
by the 
presence of
the new variables $\bldgr{\theta}_{les}$ and $\fat{h}$ in the model.
In order to side-step this difficulty, 
we obtain $\bldgr{\hat{\theta}}$ through 
a joint optimization over 
both
$\bldgr{\theta}$ and
$\bldgr{\theta}_{les}$:
\begin{linenomath}
\begin{align*}
\{ \bldgr{\hat{\theta}},  \bldgr{\hat{\theta}}_{les} \}
=
\arg \max_{
\{ \bldgr{\theta},  \bldgr{\theta}_{les} \}
}
p( \bldgr{\theta},  \bldgr{\theta}_{les} | \fat{D} )
\end{align*}
\end{linenomath}
in 
a simplified model in which the 
constraints on lesion \emph{shape} have been removed,
by clamping all decoder network outputs
$f_i(\fat{h})$ to value $1$.
This simplification is 
defensible since the aim here is merely to find appropriate
model parameters, rather than highly accurate lesion segmentations. 
By doing so, 
the latent code
$\fat{h}$ is effectively removed from the model and
the optimization
simplifies 
into 
the one
used in the original SAMSEG method, with only minor modifications due to the prior $p(\bldgr{\theta}_{les}|\bldgr{\theta}_{\fat{d}})$. 
Details are provided in~\ref{app:optimizationWithLesionPrior}.

Once parameter estimates $\bldgr{\hat{\theta}}$ are available,
we compute segmentations using the factorization 
\begin{linenomath}
\begin{align*}
p( \fat{l}, \fat{z} | \fat{D}, \bldgr{\hat{\theta}} )
=
p( \fat{z} | \fat{D}, \bldgr{\hat{\theta}} )
p( \fat{l} | , \fat{z}, \fat{D}, \bldgr{\hat{\theta}} )
,
\end{align*}
\end{linenomath}
first estimating $\fat{z}$ from 
$p( \fat{z} | \fat{D}, \bldgr{\hat{\theta}} )$
\revision{(Step 1 below)},
and 
then
plugging
this into $p( \fat{l} | , \fat{z}, \fat{D}, \bldgr{\hat{\theta}} )$ to 
estimate 
$\fat{l}$
\revision{(Step 2)}:
\begin{itemize}
  \item[\revision{Step 1:}] 
  Evaluating $p( \fat{z} | \fat{D}, \bldgr{\hat{\theta}} )$ involves marginalizing over both 
  $\fat{h}$ and $\bldgr{\theta}_{les}$, which we approximate by drawing $S$ 
  Monte Carlo 
  samples
  $\{\fat{h}^{(s)},\bldgr{\theta}_{les}^{(s)}\}_{s=1}^S$ from 
  $p( \fat{h}, \bldgr{\theta}_{les} | \fat{D}, \bldgr{\hat{\theta}})$:
  \begin{linenomath}
  \begin{align*}
    p( \fat{z} | \fat{D}, \bldgr{\hat{\theta}} )
    & =
    \int_{\fat{h}, \bldgr{\theta}_{les}}
    p( \fat{z} | \fat{D}, \bldgr{\hat{\theta}}, \fat{h}, \bldgr{\theta}_{les} )
    p( \fat{h}, \bldgr{\theta}_{les} | \fat{D}, \bldgr{\hat{\theta}})
    \ud \fat{h}, \bldgr{\theta}_{les} \nonumber
    \\
    & \simeq
    \frac{1}{S}
    \sum_{s=1}^S
    p( \fat{z} | \fat{D}, \bldgr{\hat{\theta}}, \fat{h}^{(s)}, \bldgr{\theta}_{les}^{(s)} )
    .
    \label{eq:approxLesionPosterior}
  \end{align*}
  \end{linenomath}
  This allows us to 
  estimate the probability of lesion occurrence in 
  each voxel,
  which we then 
  compare with a user-specified threshold value $\gamma$
  \begin{linenomath}
  \begin{align*}
  p( z_i = 1 | \fat{d}_i, \bldgr{\hat{\theta}} )
  \,\,
  \gtrless
  \,\,
  \gamma    
  \end{align*}
  \end{linenomath}
  to obtain the final lesion segmentation $\hat{z}_i$.
%
  Details on how we approximate $p( z_i = 1 | \fat{d}_i, \bldgr{\hat{\theta}} )$ using Monte Carlo sampling are provided in~\ref{app:MCMC}.
    
  \item[\revision{Step 2:}] 
  Voxels that are not assigned to lesion ($\hat{z}_i=0$) 
  in the previous step
  are 
  finally
  assigned to 
  the 
  neuroanatomical structure 
  with the highest 
  probability 
  $p( l_i = k | z_i=0, \fat{d}_i, \bldgr{\hat{\theta}} )$,
  which simply involves 
  computing
  $
  \hat{l}_i = \arg \max_k \hat{w}_{i,k}
  $
  with $\hat{w}_{i,k}$ defined in Eq.~\eqref{eq:healthySeg}.
  
\end{itemize}
%
In agreement with
other work%
~\citep{VanLeemput2001,Ait-Ali2005,Prastawa2008,Shiee2010,Garcia-Lorenzo2011,Jain2015}, 
we have found that using 
known
prior information regarding the 
expected
intensity profile of MS lesions in various MRI contrasts can help reduce the number of false positive
detections.
Therefore, we prevent some voxels from being assigned to lesion 
(i.e., forcing $\hat{z}_i = 0$)
based on 
their intensities 
in relation to the estimated intensity parameters $\{\bldgr{\hat{\mu}}_k,\bldgr{\hat{\Sigma}}_k\}_{k=1}^K$:
In our current implementation 
only voxels with an intensity higher than the mean of the gray matter Gaussian in FLAIR and/or T2
(if these modalities are present)
are
considered candidate lesions.

Since 
estimating $p( z_i = 1 | \fat{d}_i, \bldgr{\hat{\theta}} )$
involves repeatedly invoking the decoder and encoder networks of the 
lesion shape model,
as detailed in~\ref{app:MCMC},
we implemented the proposed method 
as an add-on to SAMSEG 
in Python using the Tensorflow library~\citep{Abadi2016}.
Estimating $\bldgr{\hat{\theta}}$ 
has the same computational complexity 
as running SAMSEG (i.e., taking approximately 10 minutes on a 
state-of-the-art
machine\revision{ with an Intel 12-core i7-8700K CPU}), 
while the Monte Carlo sampling takes an additional 5 minutes on a GeForce GTX 1060 graphics card,
bringing the total computation time to around 15 minutes per subject.

\section{
\revision{Evaluation datasets and benchmark methods}%
}
\label{sec:Experiments}
In this section, we describe 
\revision{four}
datasets that we will use for the experiments in this paper%
\revision{, including two taken from public challenges}%
.
We also outline two 
\revision{relevant}
methods for MS lesion segmentation
that the proposed method is 
\revision{compared to in detail}%
,
as well as
the metrics and measures used in our experiments.

\subsection{Datasets}
\label{subsec:datasets}
In order to test the proposed method and demonstrate its contrast-adaptiveness,
we conducted experiments on 
\revision{four}
datasets 
acquired 
with different
scanner platforms, field strengths, 
acquisition protocols and image resolution: 
\begin{description}
    \item[ - ] \textbf{MSSeg}: This dataset is the publicly available training set of the MS lesion segmentation challenge that was held in conjunction with the MICCAI 2016 conference \citep{Commowick2018}.
    It consists of 15 MS cases from three different scanners, all 
    acquired using
    a 
    harmonized
    imaging
    protocol \citep{Cotton2015}.
    For each patient a 3D T1w sequence%
    \revision{, a}
    contrast-enhanced (T1c) \revision{sequence},
    an axial dual PD-T2-weighted (T2w) sequence and a 3D fluid attenuation inversion recovery (FLAIR) sequence were acquired. 
    Each subject's lesions were delineated by seven different raters on the FLAIR scan
    and, if necessary, corrected using the T2w scan.
    These delineated images were then fused to create a consensus lesion segmentation for each subject.
    Both raw images and pre-processed images (pre-processing steps: denoising, rigid registration, brain extraction and bias field correction -- see \cite{Commowick2018} for details) were made available by the challenge organizers. 
    In our experiments we used the pre-processed data,
    which required only minor modifications in our software to remove non-brain tissues from the model.
    We note that the original challenge also included a separate set of 38 test subjects,
    but at the time of writing this data is no longer 
    available.

    \item[ - ] \textbf{Trio}: This dataset consists of 40 MS cases acquired on a Siemens Trio 3T scanner at the Danish Research Center of Magnetic Resonance (DRCMR). 
    For each patient, a 3D T1w sequence, a T2w sequence and a FLAIR sequence were acquired. Ground truth lesion segmentations were automatically delineated on the FLAIR images using Jim software\footnote{\url{http://www.xinapse.com/}}, and then checked and, if necessary, corrected by and expert rater at DRCMR using the T2w and MPRAGE images.

    \item[ - ] \textbf{Achieva}: This dataset consists of 50 MS cases 
    and 25 healthy controls 
    acquired on a Philips Achieva 3T scanner at DRCMR.
    After a visual inspection of the images, we decided to remove 2 healthy controls from the dataset as they present marked gray matter atrophy and white matter hyperintensities.
    For each patient, a 3D T1w sequence, a T2w sequence and a FLAIR sequence were acquired. Ground truth lesion segmentations were delineated using the same protocol as the one used for the Trio dataset.
    
    \revision{
    \item[ - ] \textbf{ISBI}: This dataset is the publicly available 
    test set of the MS lesion segmentation challenge that was held at the 2015 International Symposium on Biomedical Imaging~\citep{Carass2017}. It consists of 14 longitudinal MS cases, with 4 to 6 time points each, separated by approximately one year. Images were acquired on a Philips 3T scanner. For each patient, a 3D T1w sequence, a T2w sequence, a PDw sequence and a FLAIR sequence were acquired. Images were first preprocessed (inhomogeneity correction, skull stripping, dura stripping, again inhomogeneity correction -- see~\cite{Carass2017} for details), and then registered 
    to a 1mm MNI template. Each subject's lesions were delineated by two different raters on the FLAIR scan, and, if necessary, corrected using the other contrasts. 
    As part of the challenge, a training dataset of 5 additional longitudinal MS cases is also available, with the same scanner, imaging protocols and delineation procedure as the test dataset.
    }
    
\end{description}

A summary of the datasets, with scanner 
type,
image modalities and voxel resolution details, can be found in Table~\ref{tab:dataset}.
For each subject all the contrasts were co-registered and resampled to the FLAIR scan for MSSeg, and to the T1w scan for Trio,
Achieva \revision{and ISBI}. This is the only preprocessing step required by the proposed method.

\begin{table}[t]
\resizebox{\columnwidth}{!}{
\begin{tabular}[t]{l|llll}
Dataset & Scanner & Modality & Voxel resolution {[}mm{]} & Subjects \\ \hline
\multirow{15}{*}{MSSeg} & \multirow{5}{*}{Philips Ingenia 3T} & 3D FLAIR & 0.74 $\times$ 0.74 $\times$ 0.7 & \multirow{5}{*}{5} \\
 &  & 3D T1w & 0.74 $\times$ 0.74 $\times$ 0.85 &  \\
 &  & 3D T1c & 0.74 $\times$ 0.74 $\times$ 0.85 &  \\
 &  & 2D T2w & 0.45 $\times$ 0.45 $\times$ 3 &  \\
 &  & 2D PD & 0.45 $\times$ 0.45 $\times$ 3 &  \\ \cline{2-5}
 & \multirow{5}{*}{Siemens Aera 1.5T} & 3D FLAIR & 1.03 $\times$ 1.03 $\times$ 1.25 & \multirow{5}{*}{5} \\
 &  & 3D T1w & 1.08 $\times$ 1.08 $\times$ 0.9 &  \\
 &  & 3D T1c & 1.08 $\times$ 1.08 $\times$ 0.9 &  \\
 &  & 2D T2w & 0.72 $\times$ 0.72 $\times$ 4 (Gap: 1.2) &  \\ 
 &  & 2D PD & 0.72 $\times$ 0.72 $\times$ 4 (Gap: 1.2) &  \\ \cline{2-5}
 & \multirow{5}{*}{Siemens Verio 3T} & 3D FLAIR & 0.5 $\times$ 0.5 $\times$ 1.1 & \multirow{5}{*}{5} \\
 &  & 3D T1w & 1 $\times$ 1 $\times$ 1 &  \\
 &  & 3D T1c & 1 $\times$ 1 $\times$ 1 &  \\
 &  & 2D T2w & 0.69 $\times$ 0.69 $\times$ 3 &  \\
 &  & 2D PD & 0.69 $\times$ 0.69 $\times$ 3 &  \\ \hline
\multirow{3}{*}{Trio} & \multirow{3}{*}{Siemens Trio 3T} & 2D FLAIR & 0.7 $\times$ 0.7 $\times$ 4 & \multirow{3}{*}{40} \\
 &  & 3D T1w & 1 $\times$ 1 $\times$ 1 &  \\
 &  & 2D T2w & 0.7 $\times$ 0.7 $\times$ 4 &  \\ \hline
\multirow{3}{*}{Achieva} & \multirow{3}{*}{Philips Achieva 3T} & 3D FLAIR & 1 $\times$ 1 $\times$ 1 & \multirow{3}{*}{73} \\
 &  & 3D T1w & 0.85 $\times$ 0.85 $\times$ 0.85 &  \\
 &  & 3D T2w & 0.85 $\times$ 0.85 $\times$ 0.85 &  \\
\hline
\multirow{4}{*}{\revision{ISBI}}  & \multirow{4}{*}{\revision{Philips 3T}}
 & \revision{2D FLAIR} &  \revision{0.82 $\times$ 0.82 $\times$ 2.2} & \multirow{4}{*}{\revision{14}} \\
 &  & \revision{3D T1w} & \revision{0.82 $\times$ 0.82 $\times$ 1.17} &  \\
 &  &  \revision{2D T2w} & \revision{0.82 $\times$ 0.82 $\times$ 2.2} &  \\
 &  &  \revision{2D PDw} &  \revision{0.82 $\times$ 0.82 $\times$ 2.2}  &  \\ 
\end{tabular}
}
\caption{Summary of the datasets used in our experiments.}
\label{tab:dataset}
\end{table}

\subsection{%
Benchmark 
methods for lesion segmentation}

In order to 
evaluate the lesion segmentation component of the proposed method 
\revision{in detail},
we 
\revision{compared it to}
two publicly available and widely used algorithms for
MS lesion segmentation:
\begin{description}
	\item[ - ] \textbf{LST-lga}%
	\footnote{\url{https://www.applied-statistics.de/lst.html}} \citep{Schmidt2012}: 
	This lesion growth algorithm 
	starts by
	segmenting a
	T1w image into three main tissue classes (CSF, GM and WM) using SPM12\footnote{\url{https://www.fil.ion.ucl.ac.uk/spm/software/spm12/}},
	and combines the resulting segmentation with 
	co-registered FLAIR intensities to calculate a lesion belief map. A pre-chosen initial threshold $\kappa$ is then used to create an initial binary lesion map, which is subsequently grown along voxels that appear hyperintense in the FLAIR image. We set $\kappa$ to its recommended default value of 0.3, which was also used in previous studies \citep{Muhlau2013,Rissanen2014}.

	\item[ - ] \textbf{NicMsLesions}%
	\footnote{\url{https://github.com/sergivalverde/nicMsLesions}} 
	\citep{Valverde2017,Valverde2019}: 
	This deep learning method is based on a cascade of two 3D convolutional neural networks, where the first one reveals possible candidate lesion voxels, and the second one reduces the number of false positive outcomes. 
	Both networks were trained 
	\revision{by the authors of the method}
	on T1w and FLAIR scans coming from a publicly available training dataset of the MS lesion segmentation challenge held in conjunction with the MICCAI 2008 conference \citep{MICCAI2008} (20 cases) and the MSSeg dataset (15 cases). 
	This method was one of the top 
	performers
	on the test dataset of the MICCAI 2016 
	challenge \citep{Commowick2018}, and one of the few
	methods
	for which an implementation is publicly available. 
	
\end{description}
We note that both these benchmark methods 
%
are specifically targeting
T1w-FLAIR input, whereas the proposed method is not tuned to any 
particular
combination of input modalities.


\revisionTwo{
%
%
%
%
%

Although we only compared our method in detail to these 
two benchmarks, 
many more 
good
methods for MS lesion segmentation exist.
We refer the reader to 
the MSSeg paper~\citep{Commowick2018}, the ISBI challenge paper~\citep{Carass2017} and the ISBI challenge website\footnote{\url{https://smart-stats-tools.org/lesion-challenge}}
to compare the 
reported performance further with other ones.

%
}

\subsection{Metrics and measures}
\label{subsec:MetricsAndMeasures}

%
%
In order to evaluate the influence of varying the input modalities on the 
segmentation performance of
the proposed method, and 
to
assess 
segmentation 
accuracy
with respect to that of 
other
methods and human raters,
we 
used a combination of
segmentation volume estimates, 
Pearson correlation coefficients between such 
estimates and reference values,
and Dice scores.
Volumes were computed by counting the number of voxels assigned to a specific structure and 
converting into mm$^3$, whereas Dice coefficients were computed as
\begin{linenomath}
\begin{align*}
	\text{Dice}_{X,Y} = \frac{ 2 \cdot |X \cap Y|}{|X| + |Y|}
	,
\end{align*}
\end{linenomath}
where $X$ and $Y$ denote segmentation masks, and $| \cdot |$ counts the number of voxels in a mask.

The proposed method 
and
both
benchmark algorithms produce a probabilistic lesion map that needs to be thresholded to obtain a final lesion segmentation. 
This requires
an appropriate threshold value 
to be set
for this purpose
(variable $\gamma$ in the proposed method).
In order to ensure 
an objective
comparison between the methods, 
we 
used
a leave-one-out cross-validation strategy
in which the threshold 
for each 
test
image was set 
to the value that maximizes
the average Dice overlap with manual segmentations in all the other images of the same dataset.
For the reported performance of \revisionTwo{the methods} on the ISBI dataset, \revisionTwo{the thresholds were tuned} on the 5 training subjects that are part of the challenge instead.

\section{Results}
\label{sec:Results}

In this section, we first 
\revision{illustrate the effect of the various components of our model. We then}
evaluate
how the proposed model 
adapts
to
different 
input modalities
and 
acquisition platforms.
\revision{Subsequently we}
compare the lesion segmentation performance of our model against 
that of the 
two benchmark
methods,
relate it to human inter-rater variability,
\revision{and analyze its performance on the ISBI challenge data.}
Finally, we perform an indirect validation of the whole-brain segmentation component of the method.

Throughout 
the
section we use boxplots to show some of the results. 
In these plots,
the median is indicated by a horizontal line,
plotted inside
boxes that extend from the first to the third quartile values of the data.
The range of the data is indicated by whiskers extending from the boxes,
with outliers represented by circles. 

\subsection{Illustration of the method}
\label{sec:illustrationMethod}

\begin{figure*}
    \centering
    \includegraphics[width=\linewidth]{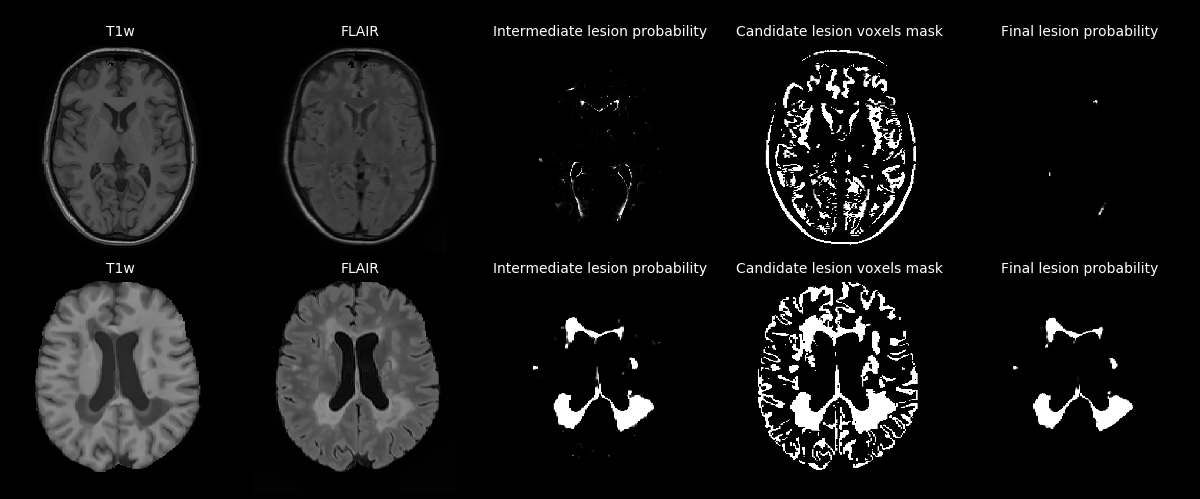}
    \caption{
    Illustration of how intensity constraints 
    and the lesion shape model help reduce false positive lesion detections in the method.
    Top row: a subject with a low lesion load; Bottom row: a subject with a high lesion load. 
    From left to right: 
    T1w and FLAIR input; 
    intermediate lesion probability obtained with the simplified model used to estimate $\bldgr{\hat{\theta}}$;
    mask of candidate voxels based on intensity alone
    (intensity higher than the mean gray matter intensity in FLAIR);
    and 
    final lesion probability estimate $p( z_i = 1 | \fat{d}_i, \bldgr{\hat{\theta}} )$
    produced by the method.
    }
    \label{fig:PrePost}
\end{figure*}

In order to illustrate the effect of the various components of the method, 
here we analyze 
its behaviour when segmenting T1w-FLAIR scans of two MS subjects --
one with a low and one with a high lesion load.
Fig.~\ref{fig:PrePost} shows, in addition to the 
input data and the final lesion probability estimate 
$p( z_i = 1 | \fat{d}_i, \bldgr{\hat{\theta}} )$, 
also an 
intermediate 
lesion probability 
obtained with the simplified model used to estimate $\bldgr{\hat{\theta}}$,
i.e., before the FLAIR-based intensity constraints and the lesion shape constraints are applied.
From these images we can see that the lesion shape model and the intensity constraints help remove false positive detections  and enforce more realistic shapes  of lesions,  especially for the case  with  low lesion load.

\begin{figure*}
    \centering
    \begin{subfigure}[t]{0.45\textwidth}
        \centering
        \includegraphics[width=\linewidth]{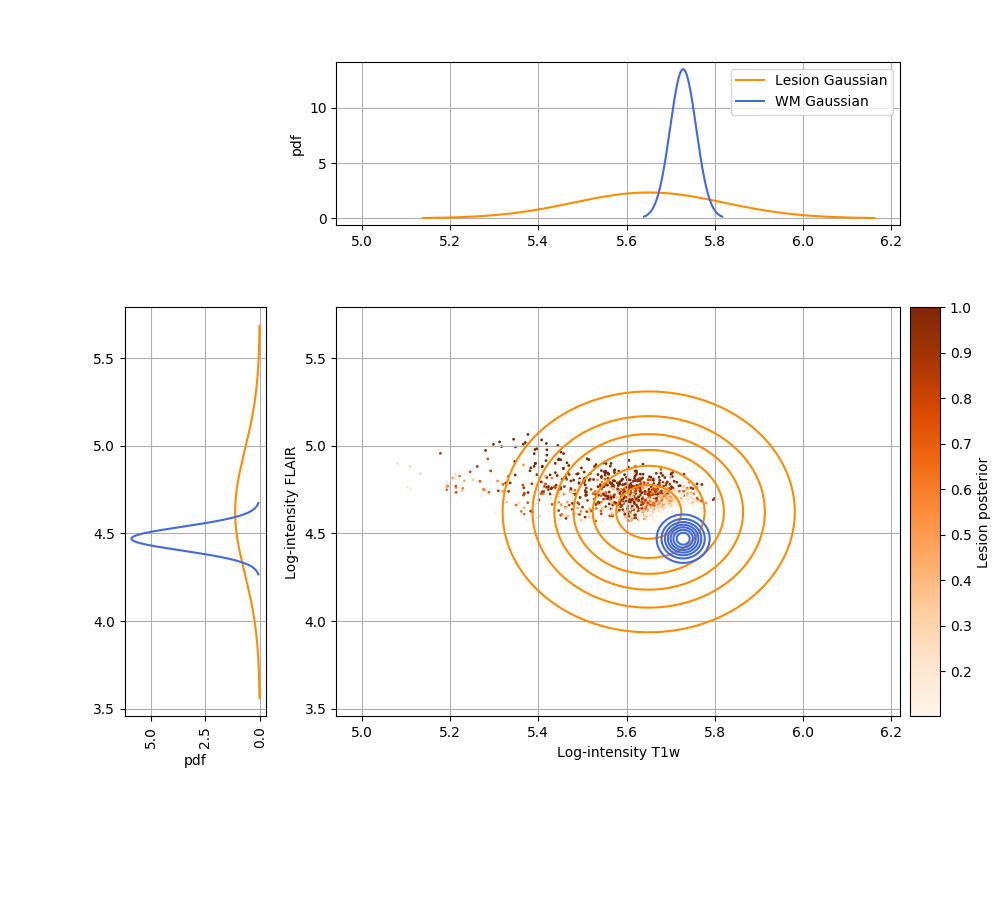} 
    \end{subfigure}
    \qquad
    \begin{subfigure}[t]{0.45\textwidth}
        \centering
        \includegraphics[width=\linewidth]{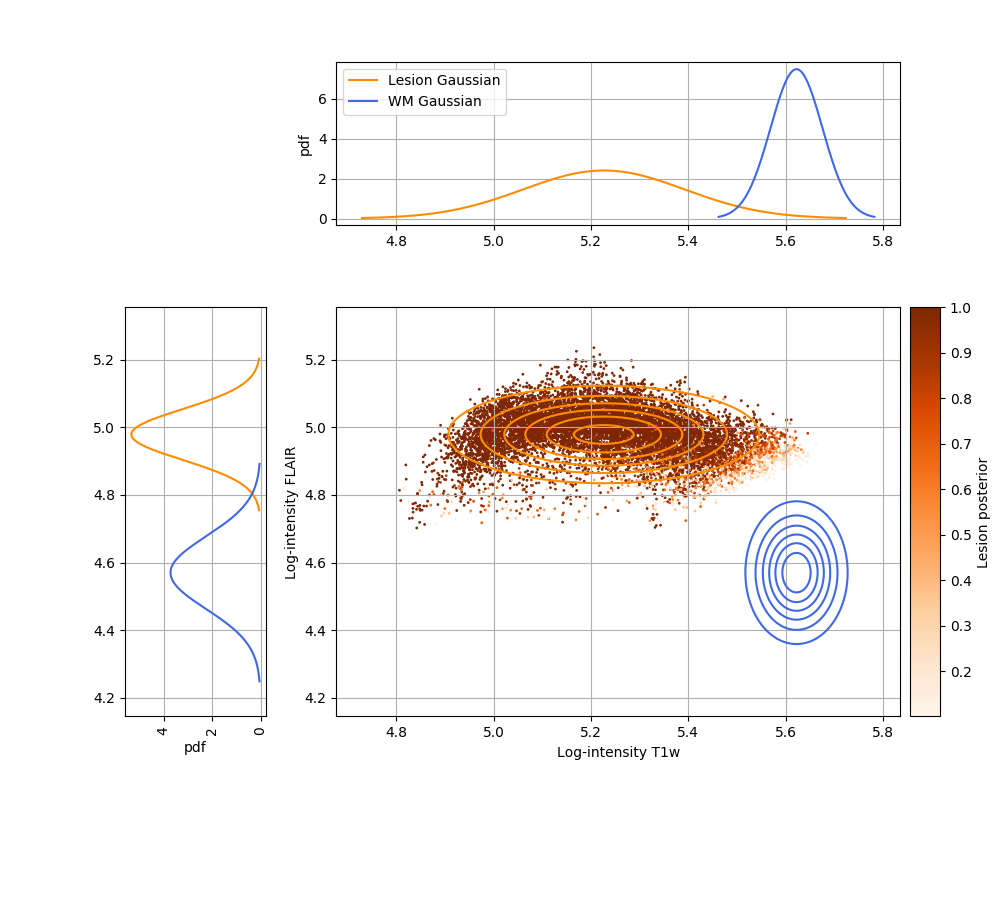}
    \end{subfigure}
    \caption{
    Illustration of the
    effect of the 
    prior 
    $p( \bldgr{\theta}_{les} | \bldgr{\theta}_{\fat{d}})$
    on the 
    lesion intensity parameters, 
    both in the case of 
    a lesion load that is
    low (left, corresponding to the subject in the top row of Fig.~\ref{fig:PrePost}) and high (right, corresponding to the subject in the bottom row of Fig.~\ref{fig:PrePost}).
    %
    %
    %
    The illustration is from the Monte Carlo sampling phase of the method:
    In each case, the value of the 
    parameters of the 
    lesion Gaussian is taken as the average over the Monte Carlo samples 
    $\{\bldgr{\theta}_{les}^{(s)}\}_{s=1}^S$,
    and the points represent the resulting lesion posterior estimate $p( z_i = 1 | \fat{d}_i, \bldgr{\hat{\theta}} )$ in each voxel.
    }
    \label{fig:TLLGauss}
\end{figure*}

Fig.~\ref{fig:TLLGauss} 
analyzes
the effect of the prior $p( \bldgr{\theta}_{les} | \bldgr{\theta}_{\fat{d}})$ on the lesion intensity parameters  
$\bldgr{\theta}_{les}$
for 
the two subjects shown in Fig.~\ref{fig:PrePost}.
When the lesion load is high, the prior does not have a strong influence, leaving the lesion Gaussian ``free'' to fit the data.
However, 
when the lesion load is low, the lesion Gaussian is constrained to retain a wide variance and 
a mean close to the mean of WM, effectively 
turning the model into an outlier detection method for WM lesions.
This behavior is important in cases when few 
lesions are present in the images, 
ensuring the method works robustly even when only limited 
data is available to estimate the lesion Gaussian parameters.

\revision{
In order to analyze the effect of the lesion shape prior, 
we 
compared the lesion
segmentation performance 
of the proposed method
with that obtained when
the shape prior was intentionally removed 
from the model
(i.e., all the decoder network outputs $f_i(\fat{h})$ 
clamped to value $1$).
%
For a fair comparison, 
the lesion threshold value $\gamma$ was 
re-tuned 
to maximize performance
for the method without shape prior,
in the way
described in Sec.~\ref{subsec:MetricsAndMeasures}.
%
Table~\ref{table:VAEvsNoVAE} summarizes the results across the MSSeg, Trio and Achieva datasets, for different ranges of lesion load.
In addition to Dice scores, the table also reports results for precision and recall, defined as
\begin{linenomath}
\begin{align*}
precision = \frac{TP}{TP + FP} \quad recall = \frac{TP}{TP + FN},
\end{align*}
\end{linenomath}
where $TP$, $FP$ and $FN$ count the true positive, false positive and false negative voxels compared to the manual segmentation.
%
The results 
indicate 
that,
although performance is unchanged for high lesion loads, 
for which segmentation is generally easier~\citep{Commowick2018},
the lesion shape prior 
clearly improves 
segmentations 
in subjects with small and medium lesion loads.
%
}



\arrayrulecolor{RevisionColor}
\begin{table*}[t]
\begin{center}
\resizebox{0.80\textwidth}{!}{%
\begin{tabular}{l|c|c|c|c|c|c}
\multicolumn{1}{c|}{\multirow{2}{*}{\revision{Lesion load}}} & \multicolumn{2}{c|}{\revision{Dice}}     & \multicolumn{2}{c|}{\revision{Precision}} & \multicolumn{2}{c}{\revision{Recall}}   \\
\multicolumn{1}{c|}{}                                      & \revision{Shape model} & \revision{No shape model} & \revision{Shape model}  & \revision{No shape model} & \revision{Shape model} & \revision{No shape model} \\
\hline
\revision{(0, 2{]} [ml]}  & \revision{0.42 ($\pm$0.10)} & \revision{0.38 ($\pm$0.10)} & \revision{0.32 ($\pm$0.12)} & \revision{0.24 ($\pm$0.07)} & \revision{0.28 ($\pm$0.09)} & \revision{0.24 ($\pm$0.07)} \\
\revision{(2, 10{]} [ml]} & \revision{0.50 ($\pm$0.13)} & \revision{0.47 ($\pm$0.13)} & \revision{0.37 ($\pm$0.13)} & \revision{0.33 ($\pm$0.11)} & \revision{0.34 ($\pm$0.12)} & \revision{0.32 ($\pm$0.12)} \\
\revision{{(}10, -) [ml]} & \revision{0.70 ($\pm$0.11)} & \revision{0.70 ($\pm$0.11)} & \revision{0.62 ($\pm$0.20)} & \revision{0.62 ($\pm$0.20)} & \revision{0.55 ($\pm$0.12)} & \revision{0.55 ($\pm$0.13)}
\\ \hline
\revision{(0, -) [ml]} & \revision{0.57 ($\pm$0.16)} & \revision{0.55 ($\pm$0.17)} & \revision{0.46 ($\pm$0.20)} & \revision{0.43 ($\pm$0.20)} & \revision{0.42 ($\pm$0.16)} & \revision{0.40 ($\pm$0.16)} \\
\end{tabular}%
}
\caption{
\revision{
Comparison in terms of lesion segmentation performance between the proposed method and a method where the lesion shape model was intentionally removed. Results are expressed in terms of mean 
$\pm$
standard deviation of Dice overlap, precision and recall for different ranges of lesion load. Lesion segmentations were computed across three different datasets (MSSeg, Trio and Achieva) on T1w-FLAIR input.    
}
}
\label{table:VAEvsNoVAE}
\end{center}
\end{table*}
\arrayrulecolor{black}

\begin{figure}[t]
	\begin{center}
		\includegraphics[width=.40\textwidth]{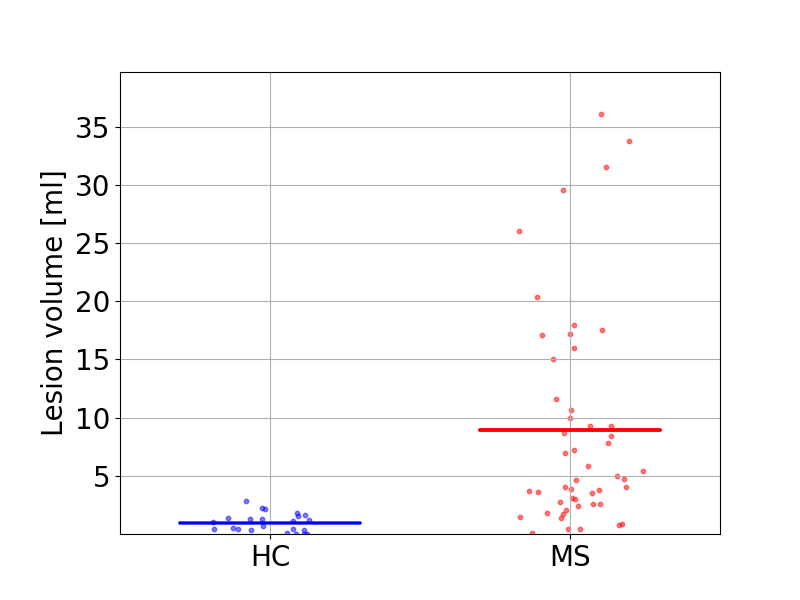}
	\end{center}
	\caption{
	Difference between healthy controls (HC) and MS subjects in lesion volume, as detected by the proposed method on the Achieva dataset (23 HC
	subjects, 50 MS subjects, T1w-FLAIR input).
	Lines indicate means across subjects. 
	}
   \label{fig:HCvsMSLesion}
\end{figure}

In order to 
demonstrate that the model 
also works robustly in control subjects (with no lesions at all),
and can therefore be safely applied in studies comparing MS subjects with controls,
we 
further
segmented T1w-FLAIR scans
\revision{of the Achieva dataset,}
and computed the total volume of the lesions in each subject. 
The results are shown in Fig.~\ref{fig:HCvsMSLesion}; 
the volumes were 8.95 $\pm$ 9.18 ml for MS subjects 
vs.~0.98 $\pm$ 0.77 ml for controls.  
Although the average lesion volume for controls was not 
exactly zero,
a visual inspection revealed 
that this was due to some controls having WM hyperintensities that were segmented by the method as MS lesions, which we find acceptable.

\subsection{Scanner and contrast adaptive segmentations}

\begin{figure*}
    \centering
    \includegraphics[width=.7\textwidth]{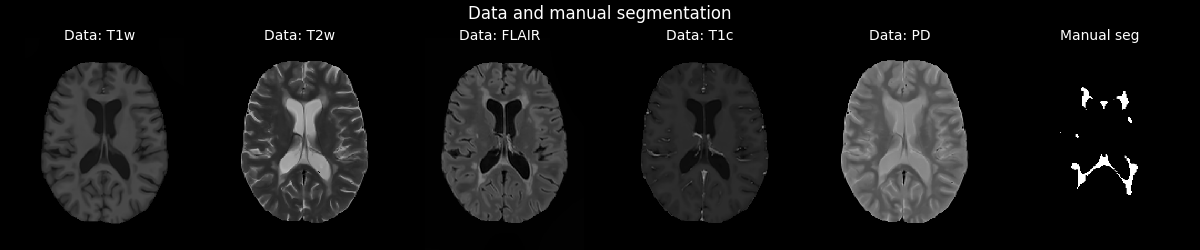} \\
    \includegraphics[width=.7\textwidth]{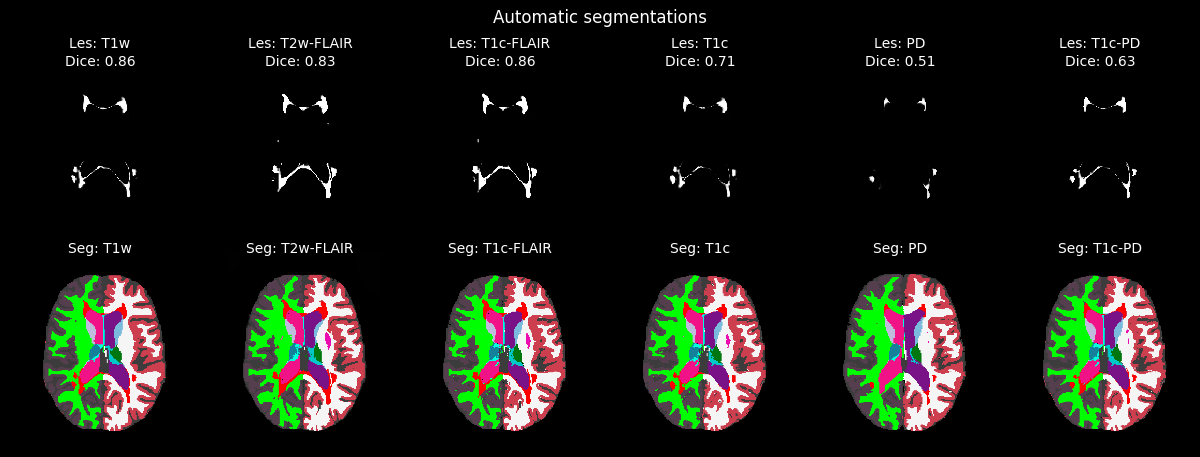} \\
    \vspace{0.05cm}
    \includegraphics[width=.7\textwidth]{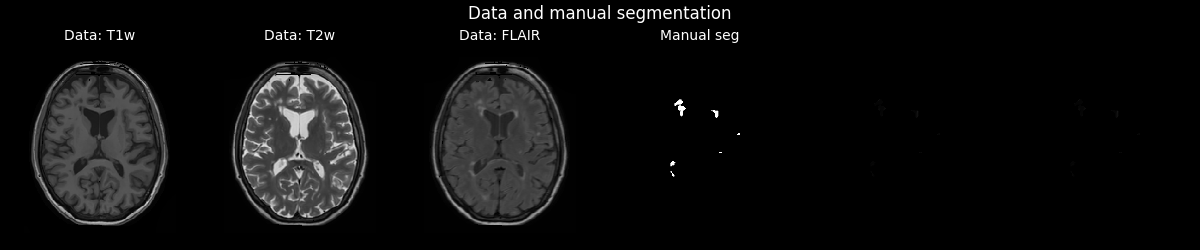} \\
    \includegraphics[width=.7\textwidth]{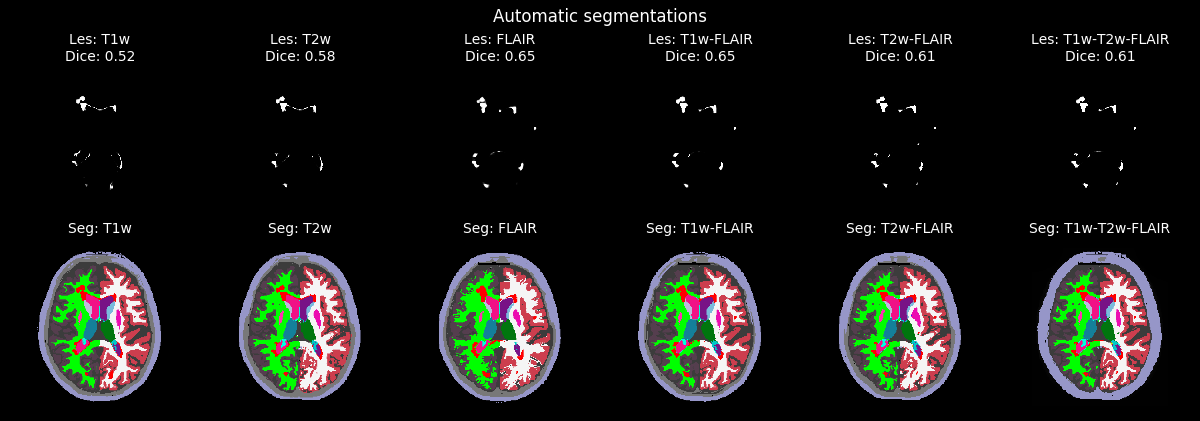} \\
    \vspace{0.05cm}
    \includegraphics[width=.7\textwidth]{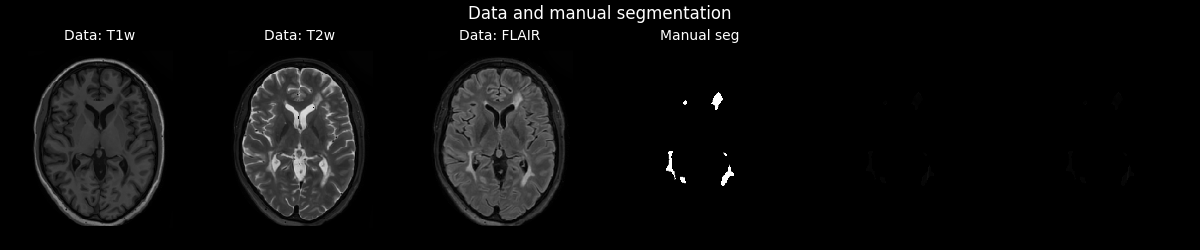} \\
    \includegraphics[width=.7\textwidth]{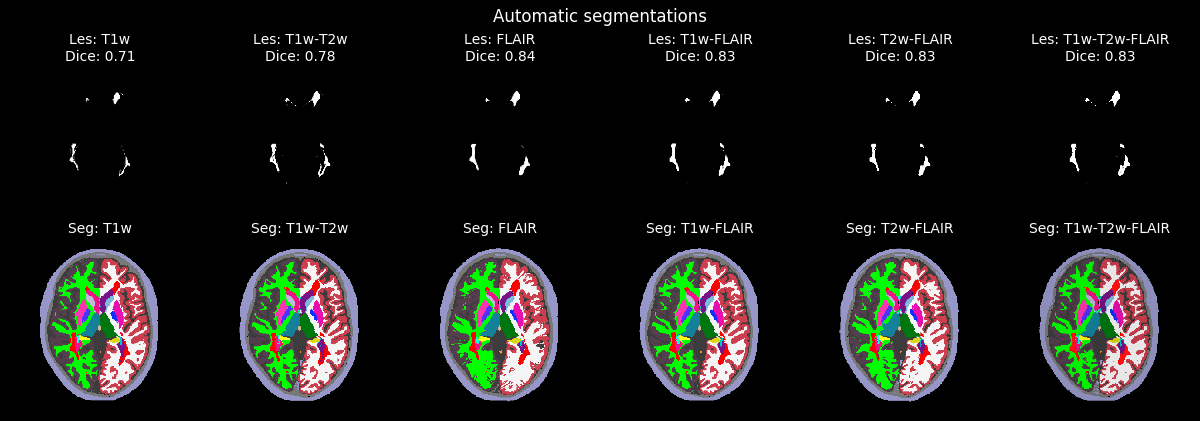} \\

   \caption{
   Contrast-adaptiveness of the proposed method to different combinations of input modalities.
   Segmentations are shown for one 
   subject of the MSSeg (top row),
   the Trio (middle row)
   and the Achieva MS (bottom row) dataset.
   For each subject the top row shows
   slices of the data
   and the manual lesion annotation;
   the middle row shows the lesion probability map 
   \revision{and Dice score}
   computed by the proposed method for specific input combinations;
   and 
   the bottom row 
   shows
   the 
   corresponding complete
   segmentations produced by the method. 
   \revision{Enlarged figures for each subject are available in the Supplementary Material Fig.~1-3.}
   %
   %
   %
   %
   }
   \label{fig:lesionSegmentationT1T2FLAIRImage}
\end{figure*}

In order to demonstrate the ability of our method to adapt to different types and combinations of MRI sequences acquired with different scanners, we show the method's segmentation results
along with the manual segmentations
for a representative subset of combinations for one subject in the MSSeg (consensus as manual segmentation), the Trio and the Achieva datasets in Fig.~\ref{fig:lesionSegmentationT1T2FLAIRImage}. 
It is not feasible to show all possible combinations.
For instance, mixing the 5 contrasts in the MSSeg dataset alone already yields 31 possible multi-contrast combinations.
Nonetheless, it is clear that the model is indeed able to adapt to the specific contrast properties of its input scans. 
A visual inspection of its whole-brain segmentation component seems to indicate that the method benefits from having access to the T1w contrast for best performance. This is especially clear when only the FLAIR contrast is provided, as this visually degrades the segmentation of the white-gray boundaries in the cortical regions due to the low contrast between white and gray matter in FLAIR. 

\begin{figure*}
    \centering
    \begin{subfigure}[t]{0.33\textwidth}
        \centering
        \includegraphics[width=\linewidth]{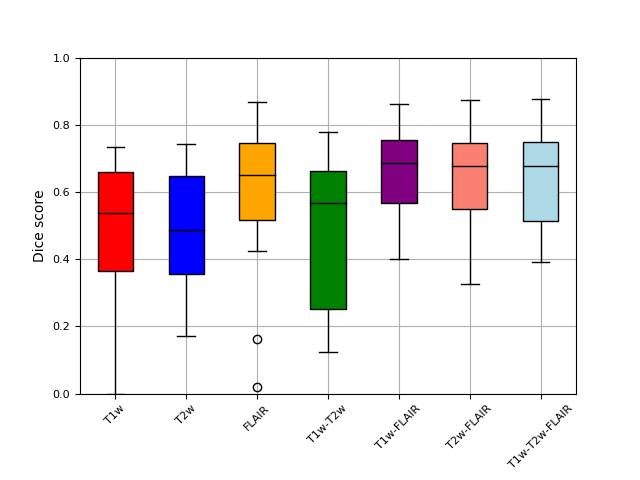} 
        \caption{MSSeg Dice scores.} \label{fig:lesMICCAIComb}
    \end{subfigure}
    \hfill
    \begin{subfigure}[t]{0.33\textwidth}
        \centering
        \includegraphics[width=\linewidth]{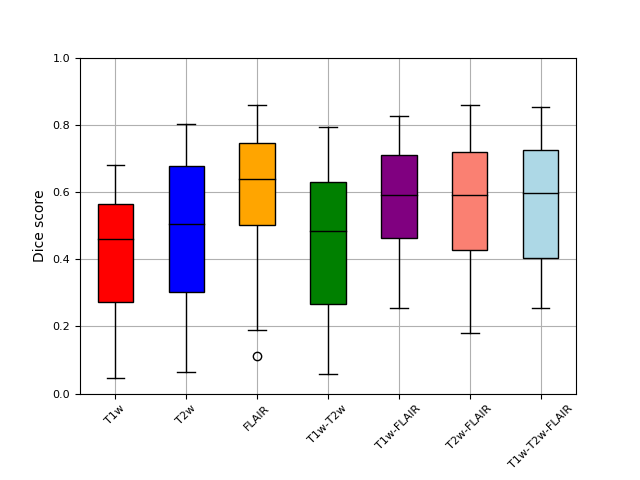} 
        \caption{Trio Dice scores.} \label{fig:lesTRIOComb}
    \end{subfigure}
    \begin{subfigure}[t]{0.33\textwidth}
        \centering
        \includegraphics[width=\linewidth]{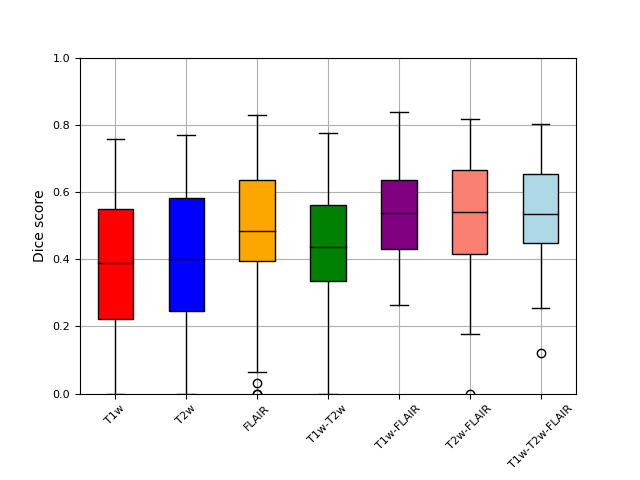} 
        \caption{Achieva Dice scores.} \label{fig:lesFATComb}
    \end{subfigure}

    \caption{
    Lesion segmentation performance 
    of the proposed method 
    in terms of Dice overlap with manual raters on three different datasets when different input contrasts are used (T1w, T2w, FLAIR, T1w-T2w, T1w-FLAIR, T2w-FLAIR, T1w-T2w-FLAIR). From left to right: Dice scores on MSSeg, Trio and Achieva MS data.
    }
    \label{fig:lesionSegmentationT1T2FLAIR}

\end{figure*}

When comparing the lesion probability maps produced by the method visually with the corresponding manual lesion segmentations, it seems that the method benefits from having access to the FLAIR contrast for 
\revisionTwo{the}
best lesion segmentation performance. 
This is confirmed by a quantitative analysis shown in Fig.~\ref{fig:lesionSegmentationT1T2FLAIR}, which plots the Dice overlap scores for each of the seven input combinations that all our three datasets have in common, namely  
T1w, T2w, FLAIR, T1w-T2w, T1w-FLAIR, T2w-FLAIR, and T1w-T2w-FLAIR.
Although the inclusion of additional contrasts does not hurt lesion segmentation performance, across all three datasets the best results are obtained whenever the FLAIR contrast is included as input to the model. This finding is perhaps not surprising, given that the manual delineations were all primarily based on the FLAIR image.

Considering both the whole-brain and lesion segmentation performance together, we conclude that the combination T1w-FLAIR is well-suited for obtaining good results with the proposed method, although it will also accept other and/or additional contrasts beyond T1w and FLAIR.

\subsection{Lesion segmentation}

In order to compare the lesion segmentation performance of our model against that of the two benchmark methods, and relate it to human inter-rater variability, we here present a number of results based on the T1w-FLAIR input combination (which is the combination required by the benchmark methods).
\revision{We also 
analyze
the lesion segmentation performance of our method on the public ISBI challenge.} 

\subsubsection*{Comparison with 
\revision{benchmark}
lesion segmentation methods}

\begin{figure*}
   \centering
   \includegraphics[width=.49\textwidth]{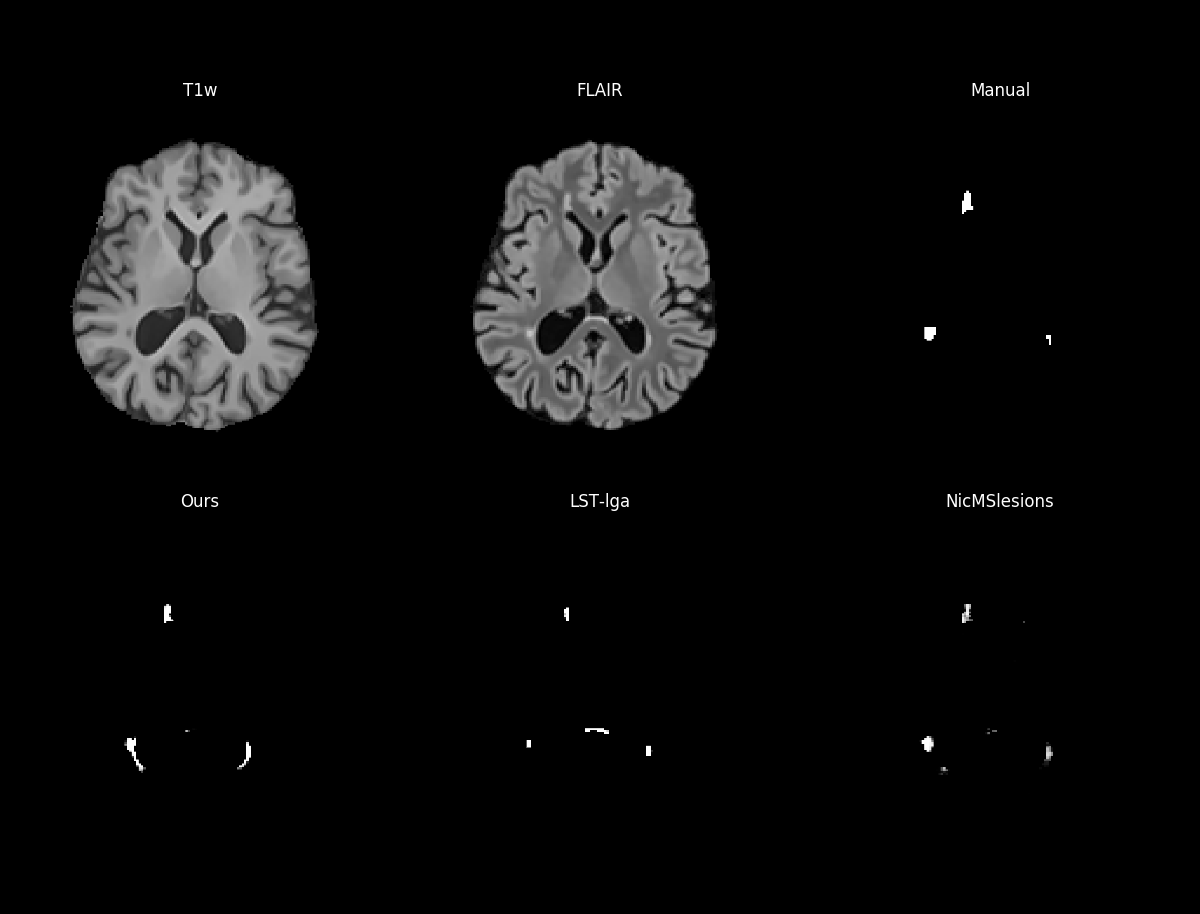}
   \includegraphics[width=.49\textwidth]{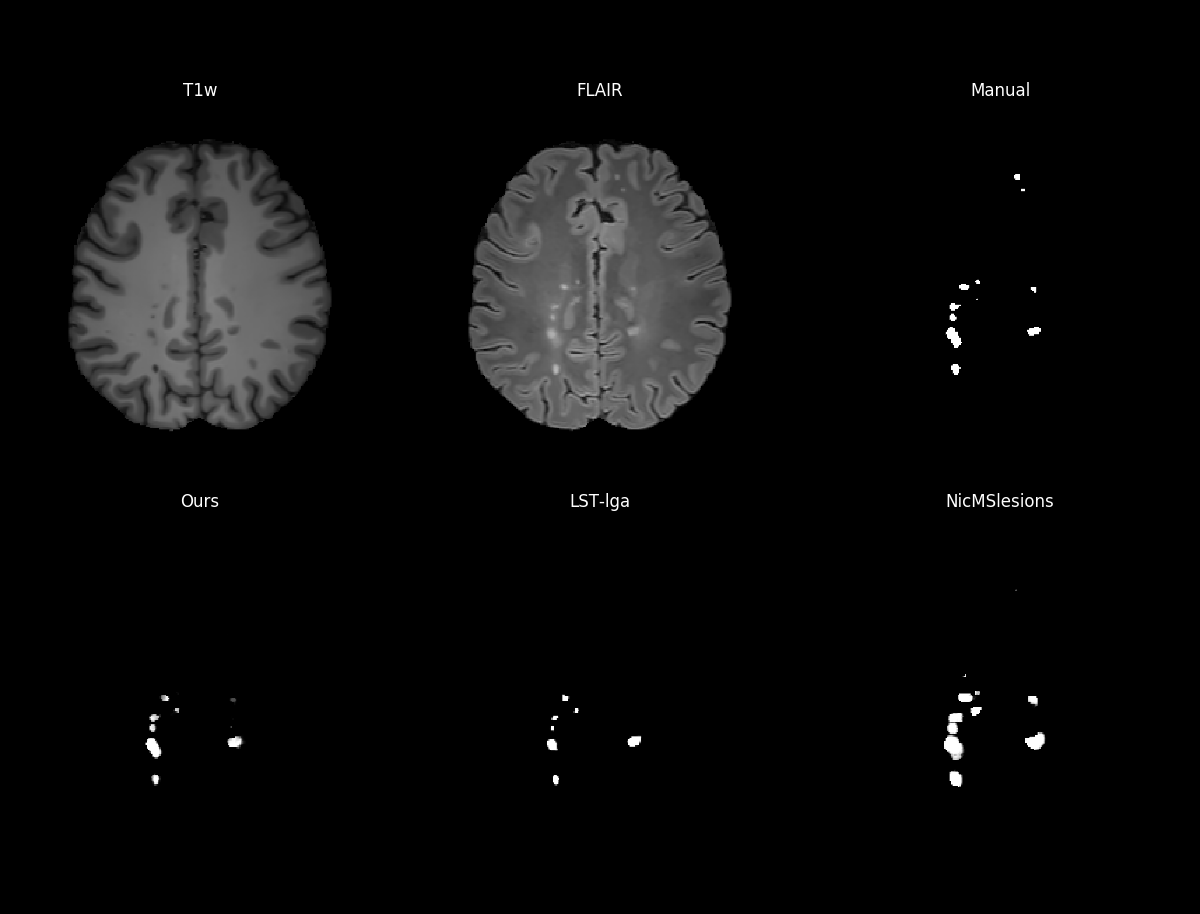}\\
   \vspace{0.05cm}
   \includegraphics[width=.49\textwidth]{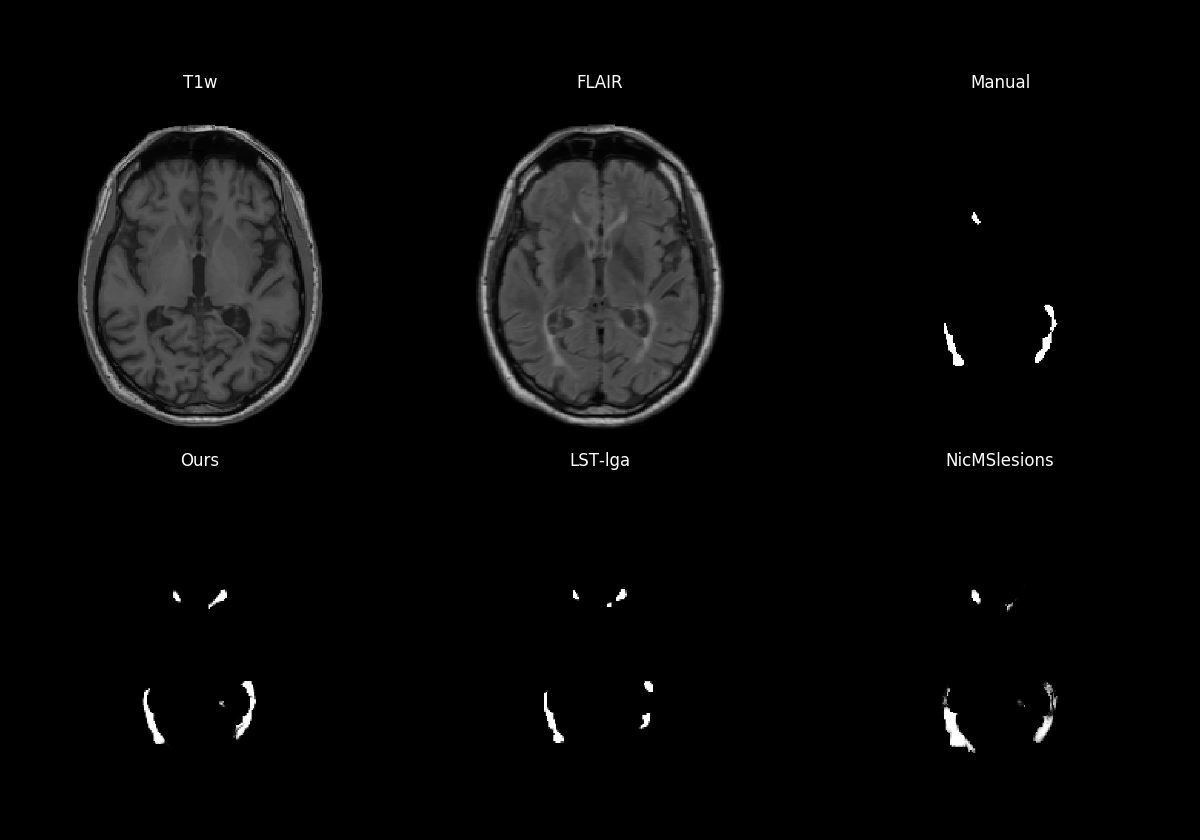}
   \includegraphics[width=.49\textwidth]{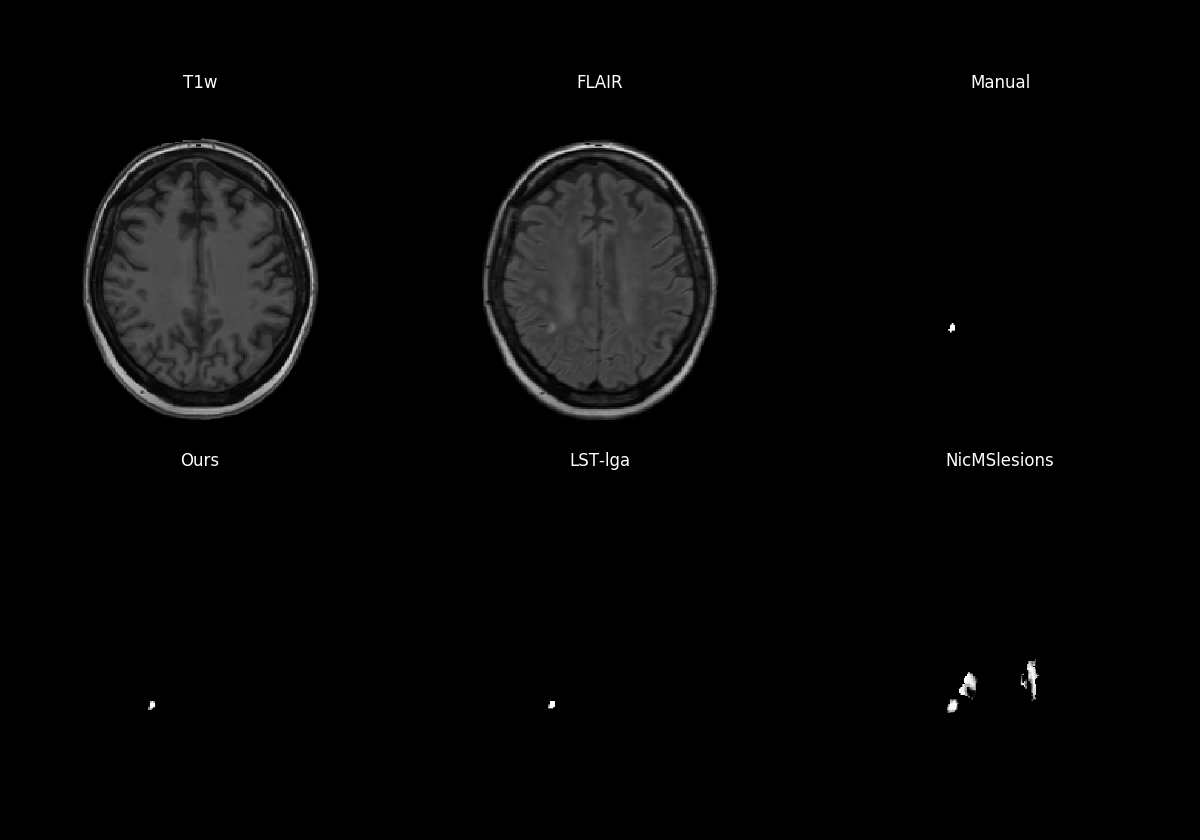}\\
   \vspace{0.05cm}
   \includegraphics[width=.49\textwidth]{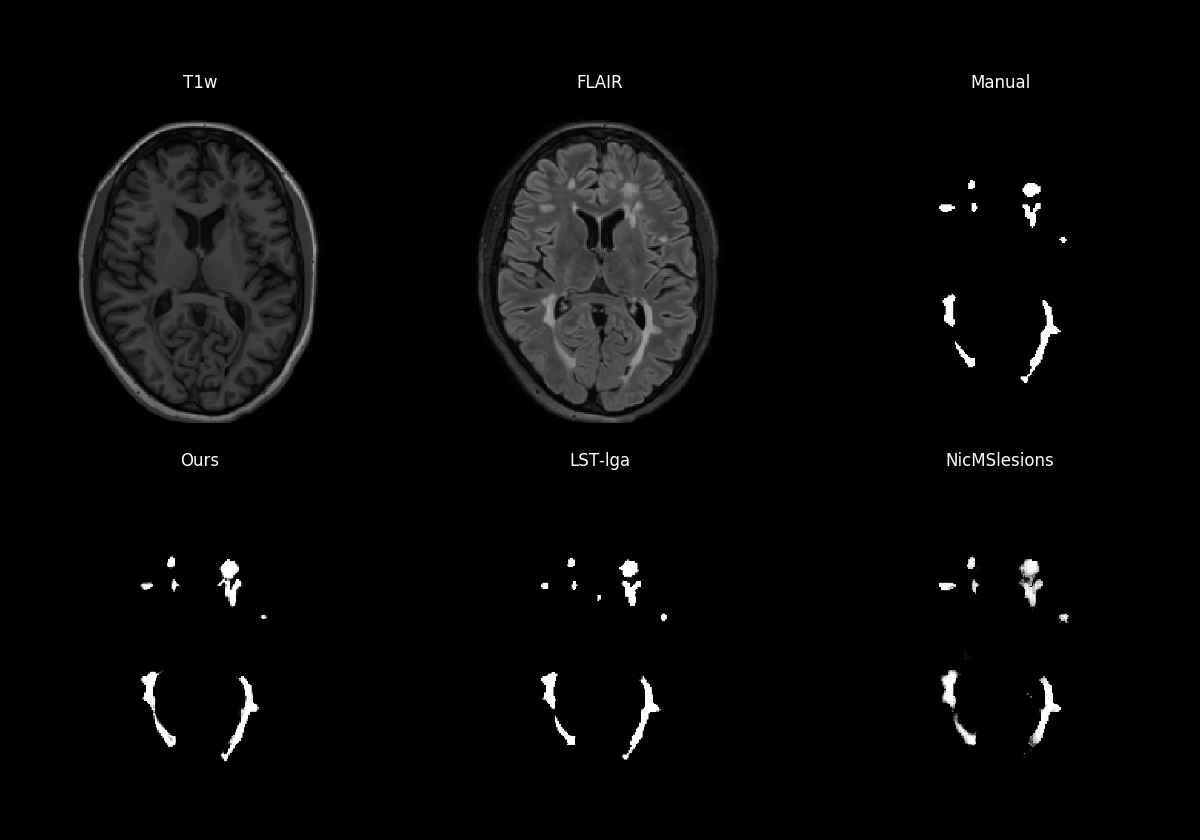}
   \includegraphics[width=.49\textwidth]{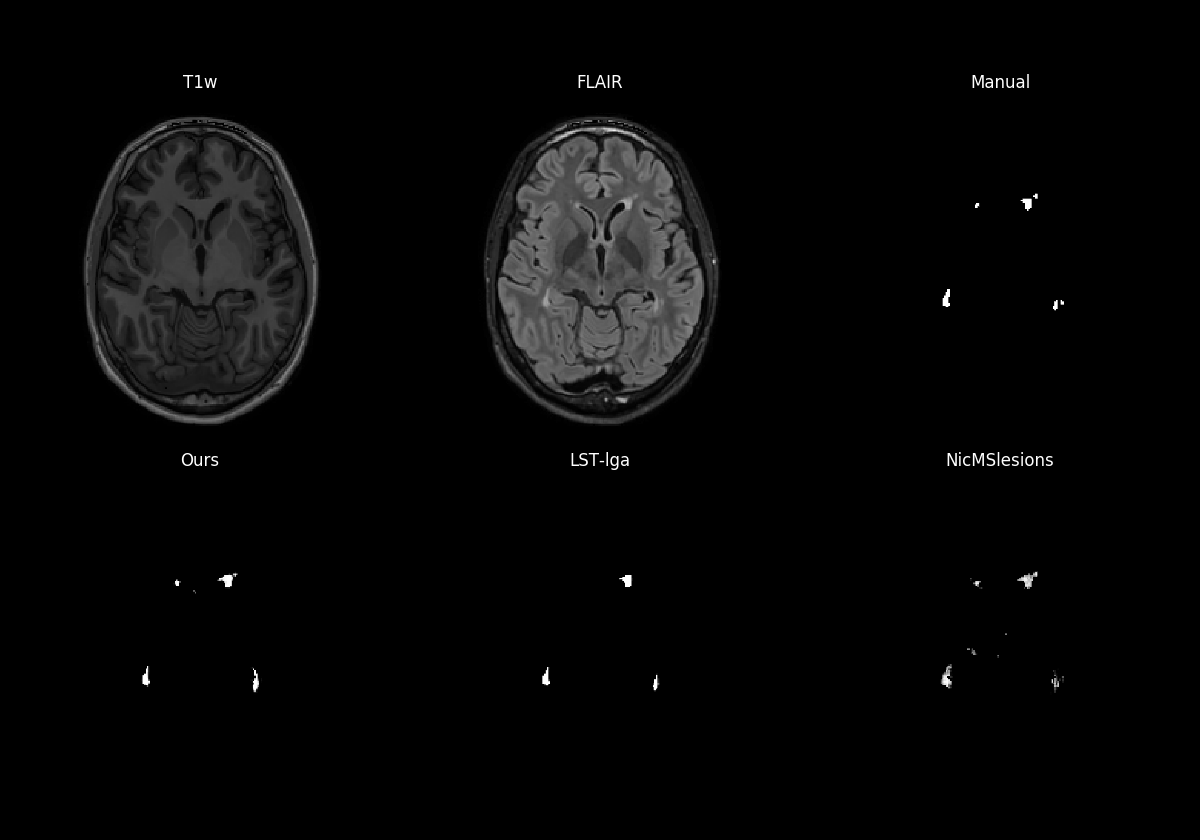}\\
   \caption{
   Visual comparison of lesion probability maps on three different datasets for the proposed method and two state-of-the-art lesion segmentation methods (LST-lga and NicMsLesions)
   on T1w-FLAIR input. 
   (Top) Two subjects from the MSSeg dataset;
   (Middle) Two subjects from the Trio dataset;
   (Bottom) Two subjects from the Achieva dataset.
   For each subject the top row shows slices of the data and the manual annotation while the bottom row shows the lesion probability maps for our model, LST-lga and NicMsLesions.
   }
   \label{fig:lesionSegmentationPerformanceImage}
  
\end{figure*}
\begin{figure*}
    \centering
    \begin{subfigure}[t]{0.33\textwidth}
        \centering
        \includegraphics[width=\linewidth]{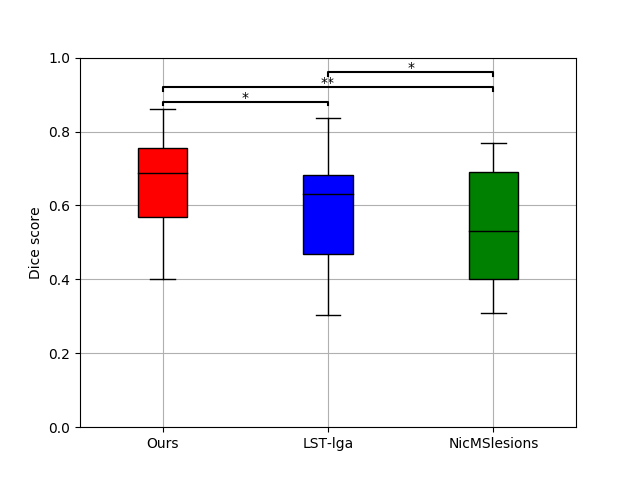} 
        \caption{MSSeg Dice Scores.} \label{fig:lesMICCAIVS}
    \end{subfigure}
    \hfill
    \begin{subfigure}[t]{0.33\textwidth}
        \centering
        \includegraphics[width=\linewidth]{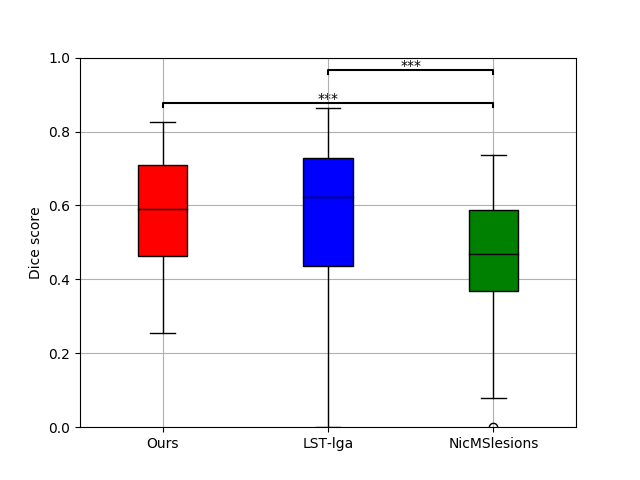} 
        \caption{Trio Dice Scores.} \label{fig:lesTRIOVS}
    \end{subfigure}
    \begin{subfigure}[t]{0.33\textwidth}
        \centering
        \includegraphics[width=\linewidth]{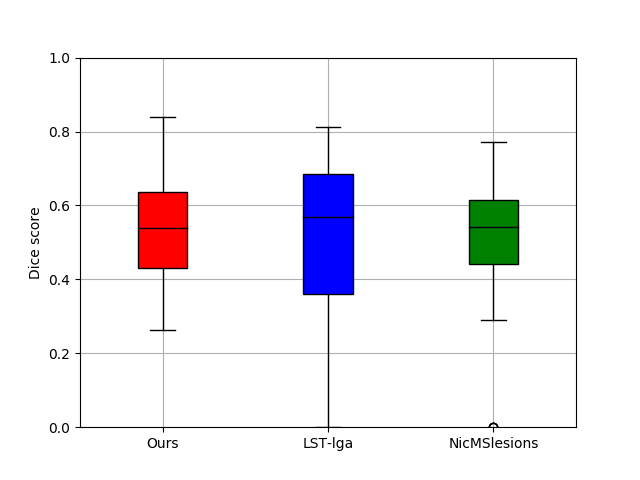} 
        \caption{Achieva Dice Scores.} \label{fig:lesFATVS}
    \end{subfigure}
   \caption{
   Lesion segmentation performance in terms of Dice overlap with manual raters 
   for the proposed method and two 
   benchmark
   methods (LST-lga and NicMsLesions) on T1w-FLAIR input.
   Statistically significant differences between two methods, computed with a two-tailed paired t-test, are indicated by asterisks (``***'' for p-value \textless~0.001, ``**'' for p-value \textless~0.01 and ``*'' for p-value \textless~0.05).
   From left to right: results on the MSSeg, the Trio and the Achieva dataset.
   }
    \label{fig:lesionSegmentationPerformance}

\end{figure*}

Fig~\ref{fig:lesionSegmentationPerformanceImage} shows automatic segmentations of two randomly selected subjects from 
\revision{the MSSeg, the Trio and the Achieva datasets,}
both for our method and for the two benchmark methods LST-lga and NicMSLesions, along with the corresponding manual segmentations 
 (consensus manual segmentations for MSSeg).
Visually, all three methods 
perform similarly on the Achieva MS data, but some of the results for NicMSLesions appear to be inferior to those obtained with the other two methods on MSSeg and Trio data. 
This qualitative observation is confirmed by the quantitative analysis shown in Fig.~\ref{fig:lesionSegmentationPerformance}, where the three methods' Dice overlap scores 
are compared on each dataset: similar performances are obtained 
for all methods
on 
\revisionTwo{the}
Achieva data, but NicMSLesions trails the other two methods on MSSeg and Trio data. 
Especially for MSSeg data this is a surprising result, since NicMSLesions was trained on this specific dataset, i.e., the subjects used for testing were part of the training data of this method, 
potentially biasing the results in favor of NicMSLesions.
Based on Dice scores, the proposed method outperforms LST-lga on MSSeg data, although there are no statistically significant differences between the two methods on the other datasets.


%

\iftrue
\revision{
\subsubsection*{Results on the ISBI data}
}
\fi



\revision{
We also evaluated the performance of the proposed method on the 
ISBI challenge data,
obtaining
a mean Dice score of 0.58
\revisionTwo{when T1w-FLAIR input is used}.
This 
score
is comparable to the 
ones we
obtained on the other three datasets analyzed in this paper (cf.~Fig.~\ref{fig:lesionSegmentationPerformance}) -- MSSeg: 0.65, Trio: 0.58 and Achieva: 0.54. 
A few example segmentation results on the ISBI data are available in the Supplementary Material, Fig.~4.
}

%


The ISBI challenge website\footnote{\url{https://smart-stats-tools.org/lesion-challenge}} ranks submissions according to an overall lesion segmentation performance score that takes into account Dice overlap, volume correlation, surface distance, and a few other metrics (see~\cite{Carass2017} for details). 
A score of 100 indicates perfect correspondence, while 
90 is meant to correspond to human inter-rater performance~\citep{MICCAI2008,Carass2017}.
%
We obtained a score of 87.87, 
which places us around half-way in the ranking of the original challenge~\citep{Carass2017},
although we note that the website currently lists methods with a much higher score.

\begin{figure}
    \centering
    {\includegraphics[width=0.33\textwidth]{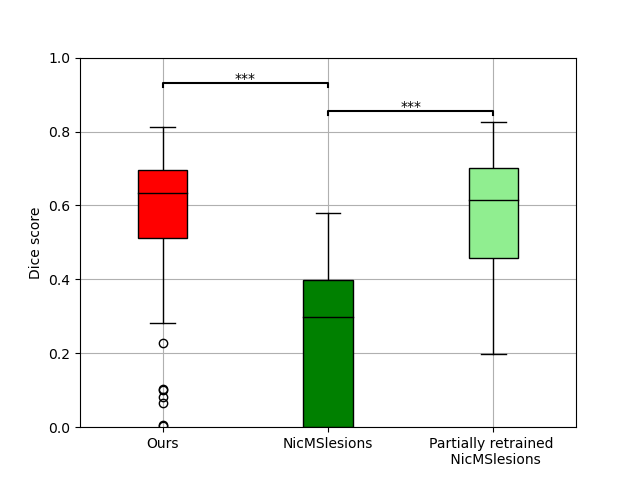}
    }
    \caption{
    \revisionTwo{
    Lesion segmentation performance in terms of Dice overlap with manual raters on the ISBI dataset 
    for the proposed method, 
    NicMsLesions, and 
    NicMsLesions with 
    partial
    retraining (see text for details).
    Statistically significant differences between two methods, computed with a two-tailed paired t-test, are indicated by asterisks (``***'' indicates p-value \textless~0.001).
    }
    }
    \label{fig:diceISBI}
\end{figure}

\revisionTwo{

In order to relate the performance of our method to the one obtained with the two benchmark methods, 
we also attempted to run LST-lga and NicMSLesions on this dataset.
However, the preprocessing applied to the ISBI challenge data proved problematic for LST-lga, and we were not able to get any results with this method. 
%
%
Results for NicMSLesions in terms of Dice overlap are shown in Fig.~\ref{fig:diceISBI}, together with those obtained with the proposed method. It is clear that NicMSLesions suffers strongly from the domain shift between 
its training data and the ISBI data, a fact that was already reported in~\citep{Valverde2019}. 
%
%
For completeness,
Fig.~\ref{fig:diceISBI} also includes results for NicMSLesions when
its 
network
was updated
on the ISBI training data as described in~\citep{Valverde2019}: different subsets of 
network parameters 
were retrained on the baseline scan of each of the five ISBI training subjects,
%
and the combination that performed best on all 21 training images 
was retained.
From the figure it can be seen that 
this 
partially retrained 
network has comparable performance to the proposed model, 
although the latter attains this performance without 
any 
retraining.
%
%

}

\subsubsection*{Inter-rater variability}
\label{subsec:interRaterLesion}


\revision{To evaluate}
the proposed method's lesion segmentation performance in the context of
human inter-rater variability, 
\revision{we took}
advantage of the availability of lesion segmentations by seven different 
raters in the MSSeg dataset. 
%
Table~\ref{table:meanDices} shows the lesion segmentation performance in terms of average Dice overlap between each pair of the seven raters, and between each rater and the proposed method.
On average, our method achieves a Dice overlap score of 0.57, which is slightly below the mean human raters' range of [0.59, 0.69].
\revision{We note that this result is in line with those obtained in the MSSeg challenge%
~\citep{Commowick2018}.}

%
%
\begin{table}
\resizebox{\columnwidth}{!}{%
\begin{tabular}{c|c|c|c|c|c|c|c||c}
 & R1 & R2 & R3 & R4 & R5 & R6 & R7 & Ours \\ \hline
R1 & - & 0.68 & 0.59 & 0.70 & 0.75 & 0.59 & 0.59 & 0.54 \\
R2 & 0.68 & - & 0.59 & 0.71 & 0.72 & 0.60 & 0.57 & 0.56 \\
R3 & 0.59 & 0.59 & - & 0.57 & 0.59 & 0.60 & 0.63 & 0.60 \\
R4 & 0.70 & 0.71 & 0.57 & - & 0.90 & 0.57 & 0.54 & 0.53 \\
R5 & 0.75 & 0.72 & 0.59 & 0.90 & - & 0.59 & 0.57 & 0.55 \\
R6 & 0.59 & 0.60 & 0.60 & 0.57 & 0.59 & - & 0.61 & 0.57 \\
R7 & 0.59 & 0.57 & 0.63 & 0.54 & 0.57 & 0.61 & - & 0.60 \\ \hline
Avg & \textbf{0.65} & \textbf{0.64} & \textbf{0.60} & \textbf{0.66} & \textbf{0.69} & \textbf{0.60} & \textbf{0.59} & \textbf{0.57}
\end{tabular}
}
\caption{
Comparison of lesion segmentation performance in terms of average Dice score between each pair of the seven raters of the MSSeg dataset, and between each rater and the proposed method (T1w-FLAIR input).
}
\label{table:meanDices}
\end{table}

\subsection{Whole-brain segmentation}
\label{subsec:wholebrainseg}
Since no ground truth segmentations are available for a direct evaluation of 
the whole-brain segmentation component of our method,
we performed an indirect validation, 
evaluating its potential for replacing lesion filling approaches that 
rely on
manually annotated lesions,
as well as its ability to 
replicate known
atrophy 
patterns 
in MS.
The results concentrate on the following 25 main neuroanatomical regions, segmented from T1w-FLAIR scans:
left and right
cerebral white matter, cerebellum white matter,  cerebral  cortex,  cerebellum cortex, lateral  ventricle, hippocampus, thalamus, putamen, pallidum, caudate, amygdala, nucleus accumbens and brain stem.
To avoid cluttering, the quantitative results for left and right structures are averaged. 
We 
note
that lesion segmentations are not merged into any 
of these brain structures (i.e., leaving ``holes'' in white matter), so that the results reflect performance only for the normal-appearing parts of structures.

\subsubsection*{Comparison with lesion filling}
\label{subsubsec:InterVarStructure}



It is well-known that white matter lesions can severely interfere with the quantification of normal-appearing structures when standard brain MRI segmentation techniques are used%
~\citep{chard2010reducing,battaglini2012evaluating,gelineau2012effect,ceccarelli2012impact,nakamura2009segmentation,vrenken2013recommendations}. 
A common strategy is therefore to use a lesion-filling%
~\citep{sdika2009nonrigid,chard2010reducing} 
procedure, in which lesions are first manually segmented, their original voxel intensities are replaced with normal-appearing white matter intensities, and standard tools are then used to segment the resulting, preprocessed images. Using such a procedure with SAMSEG would yield whole-brain segmentations that can serve as ``silver standard'' benchmarks against which the results of the proposed method (which works directly on the original scans) can be compared. In practice, however, we have noticed that replacing lesion intensities, which is typically done in T1w only, 
did
not work well in FLAIR in our experiments. Therefore, rather than explicitly replacing intensities, we 
obtained silver standard segmentations by simply masking out lesions
during the SAMSEG processing, 
effectively ignoring
lesion voxels during 
the 
model fitting.

We wished to
interpret segmentation vs.~silver standard discrepancies 
within the context of the human inter-rater variability associated with manually segmenting lesions.
Therefore, we
performed experiments on 
the MSSeg dataset, 
repeatedly re-computing
the silver standard using each of the seven raters' manual lesion annotations in turn. The results are shown in Tables~\ref{table:meanPearsonsStruct} and
~\ref{table:meanDicesStruct} for 
Pearson correlation coefficients between estimated volumes and
Dice segmentation overlap scores, respectively.
Each line in these tables corresponds to one structure, showing the average consistency between the silver standard of each rater compared to that of the six other raters, as well as the average consistency between the proposed method's segmentation and the silver standards of all raters. The results indicate that, in terms of Pearson correlation coefficient, the performance of our method falls within the range of inter-rater variability, albeit narrowly (average value 0.988 vs.~inter-rater range [0.988, 0.992]). In terms of Dice scores, however,
the method slightly underperforms compared to the inter-rater variability (average value 0.971 vs.~inter-rater range [0.978, 0.980]). 

\begin{table*}
\begin{minipage}[t]{0.48\textwidth}
\resizebox{\columnwidth}{!}{%
\begin{tabular}{l|c|c|c|c|c|c|c||c}
 & R1 & R2 & R3 & R4 & R5 & R6 & R7 & Ours \\ \hline
Cerebral White Matter & 0.992 & 0.992 & 0.991 & 0.993 & 0.993 & 0.993 & 0.987 & 0.989 \\
Cerebellum White Matter & 0.994 & 0.997 & 0.997 & 0.996 & 0.997 & 0.997 & 0.997 & 0.989 \\
Cerebral Cortex & 0.997 & 0.999 & 0.999 & 0.999 & 0.999 & 0.999 & 0.999 & 0.997 \\
Cerebellum Cortex & 0.999 & 0.999 & 0.997 & 0.999 & 0.997 & 0.999 & 0.999 & 0.999 \\
Lateral Ventricles & 0.996 & 0.995 & 0.996 & 0.997 & 0.998 & 0.994 & 0.996 & 0.992 \\ 
Hippocampus & 0.982 & 0.989 & 0.987 & 0.979 & 0.977 & 0.979 & 0.984 & 0.981 \\
Thalamus & 0.998 & 0.997 & 0.998 & 0.998 & 0.998 & 0.997 & 0.997 & 0.996 \\
Putamen & 0.999 & 0.999 & 0.999 & 0.999 & 0.999 & 0.999 & 0.999 & 0.996 \\
Pallidum & 0.988 & 0.993 & 0.993 & 0.994 & 0.993 & 0.994 & 0.990 & 0.989 \\
Caudate & 0.994 & 0.993 & 0.987 & 0.990 & 0.995 & 0.989 & 0.993 & 0.985 \\
Amygdala & 0.953 & 0.967 & 0.970 & 0.973 & 0.941 & 0.957 & 0.972 & 0.963 \\
Accumbens & 0.985 & 0.987 & 0.987 & 0.966 & 0.989 & 0.953 & 0.988 & 0.971 \\
Brain Stem & 0.991 & 0.994 & 0.990 & 0.992 & 0.992 & 0.988 & 0.992 & 0.989 \\
\hline
Average & \textbf{0.990} & \textbf{0.992} & \textbf{0.992} & \textbf{0.990} & \textbf{0.990} & \textbf{0.988} & \textbf{0.992} & \textbf{0.988}
\end{tabular}
}
\caption{
Average Pearson correlation coefficients of brain structure volume estimates between the silver standard of each rater compared to that of the six other raters in the MSSeg dataset, as well as the average consistency between the proposed method's segmentation and the silver standards of all raters (T1w-FLAIR input).
Each line shows an average across raters 
for a specific brain structure.
}
\label{table:meanPearsonsStruct}
\end{minipage}
\qquad
\begin{minipage}[t]{0.48\textwidth}
\resizebox{\columnwidth}{!}{%
\begin{tabular}{l|c|c|c|c|c|c|c||c}
 & R1 & R2 & R3 & R4 & R5 & R6 & R7 & Ours \\ \hline
Cerebral White Matter & 0.982 & 0.982 & 0.982 & 0.983 & 0.983 & 0.982 & 0.981 & 0.978 \\
Cerebellum White Matter & 0.987 & 0.987 & 0.987 & 0.987 & 0.988 & 0.987 & 0.987 & 0.983 \\
Cerebral Cortex & 0.989 & 0.990 & 0.989 & 0.990 & 0.989 & 0.989 & 0.989 & 0.986 \\
Cerebellum Cortex & 0.996 & 0.996 & 0.995 & 0.996 & 0.996 & 0.995 & 0.995 & 0.994 \\
Lateral Ventricles & 0.972 & 0.970 & 0.972 & 0.974 & 0.976 & 0.971 & 0.971 & 0.954 \\
Hippocampus & 0.975 & 0.972 & 0.971 & 0.973 & 0.974 & 0.972 & 0.972 & 0.965 \\
Thalamus & 0.980 & 0.981 & 0.981 & 0.982 & 0.981 & 0.982 & 0.981 & 0.975 \\
Putamen & 0.987 & 0.987 & 0.988 & 0.988 & 0.988 & 0.988 & 0.987 & 0.980 \\
Pallidum & 0.985 & 0.985 & 0.986 & 0.986 & 0.986 & 0.986 & 0.985 & 0.978 \\
Caudate & 0.961 & 0.957 & 0.956 & 0.961 & 0.964 & 0.957 & 0.954 & 0.937 \\
Amygdala & 0.973 & 0.972 & 0.972 & 0.973 & 0.972 & 0.972 & 0.972 & 0.967 \\
Accumbens & 0.957 & 0.958 & 0.960 & 0.960 & 0.960 & 0.943 & 0.960 & 0.945 \\
Brain Stem & 0.987 & 0.986 & 0.984 & 0.986 & 0.986 & 0.986 & 0.986 & 0.983 \\
\hline
Average & \textbf{0.979} & \textbf{0.979} & \textbf{0.979} & \textbf{0.980} & \textbf{0.980} & \textbf{0.978} & \textbf{0.978} & \textbf{0.971}
\end{tabular}
}
\caption{
Same as Table~\ref{table:meanPearsonsStruct}, but with Dice segmentation overlap scores.
Each line shows an average across raters -- similar to the last row of Table~\ref{table:meanDices} -- for a specific brain structure.
}
\label{table:meanDicesStruct}
\end{minipage}
\end{table*}

\subsubsection*{Detecting atrophy patterns in MS}

\begin{figure*}[t]
	\begin{center}
		\includegraphics[width=.195\textwidth]{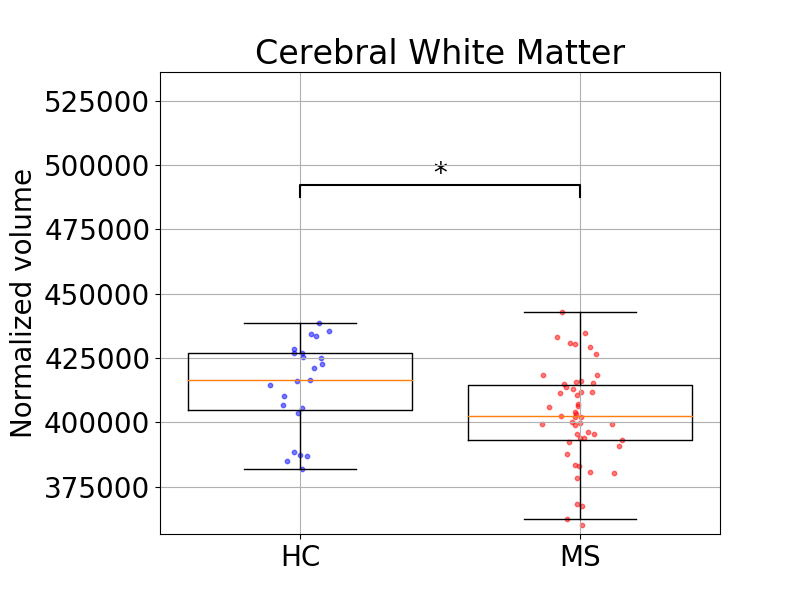}
		\includegraphics[width=.195\textwidth]{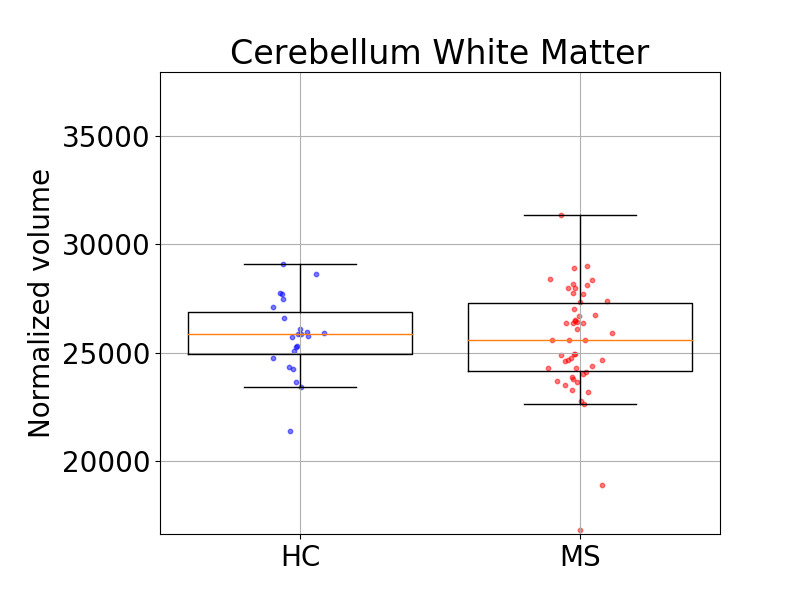}
		\includegraphics[width=.195\textwidth]{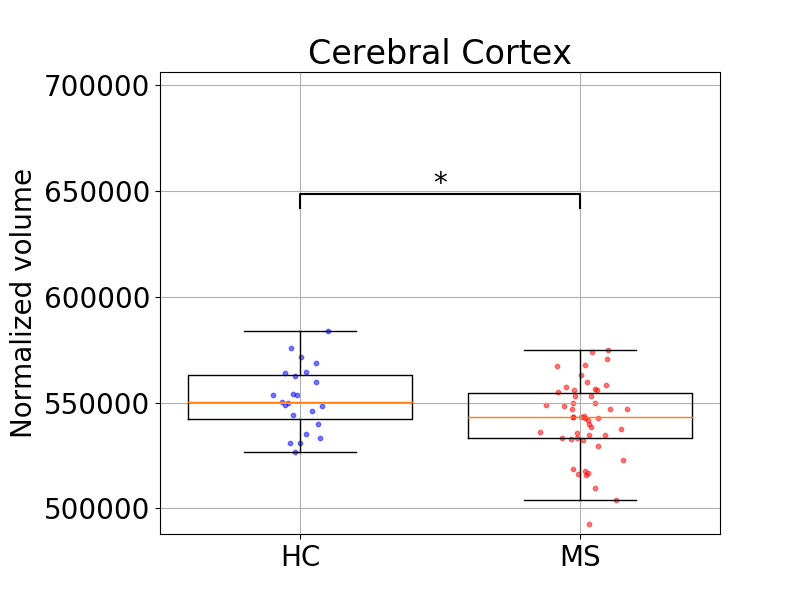}
		\includegraphics[width=.195\textwidth]{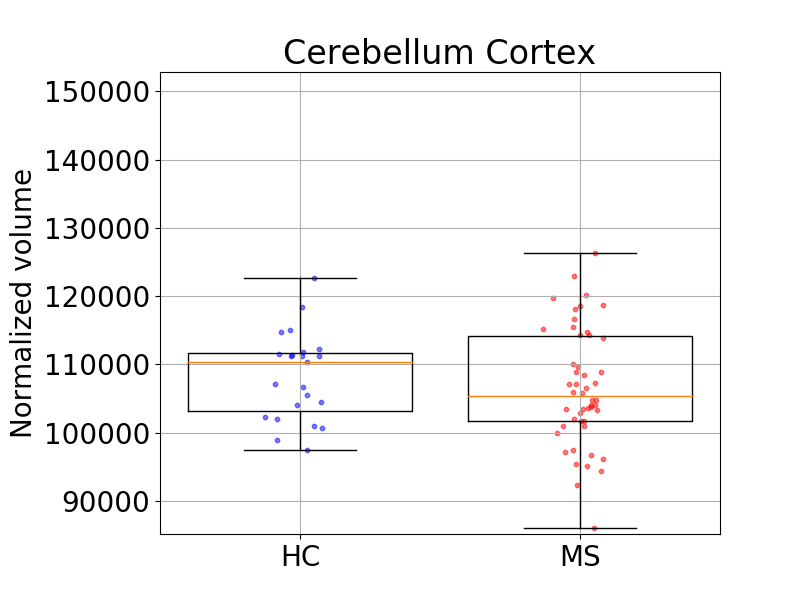}
		\includegraphics[width=.195\textwidth]{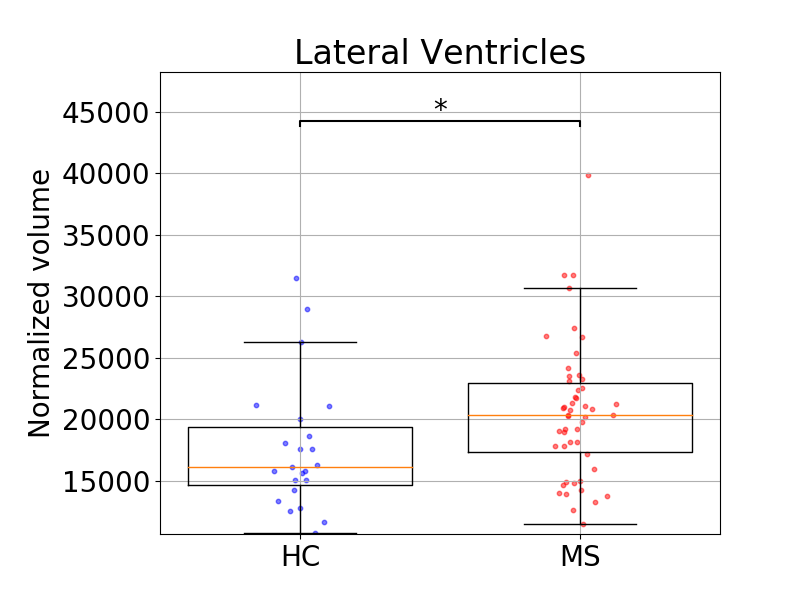}
		\includegraphics[width=.195\textwidth]{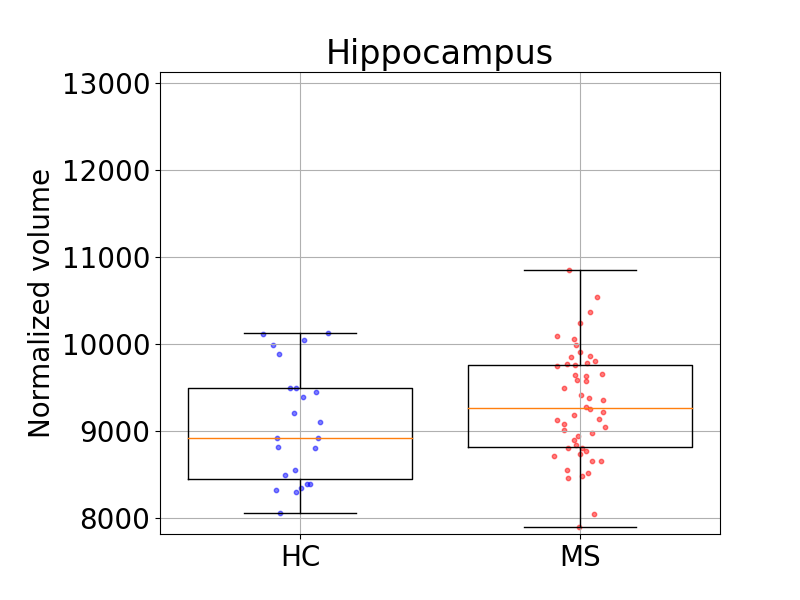}
		\includegraphics[width=.195\textwidth]{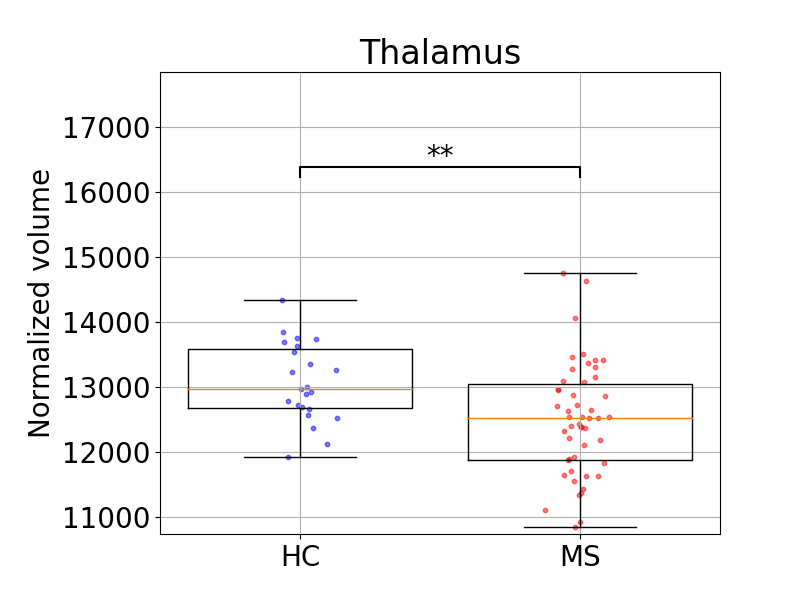}
		\includegraphics[width=.195\textwidth]{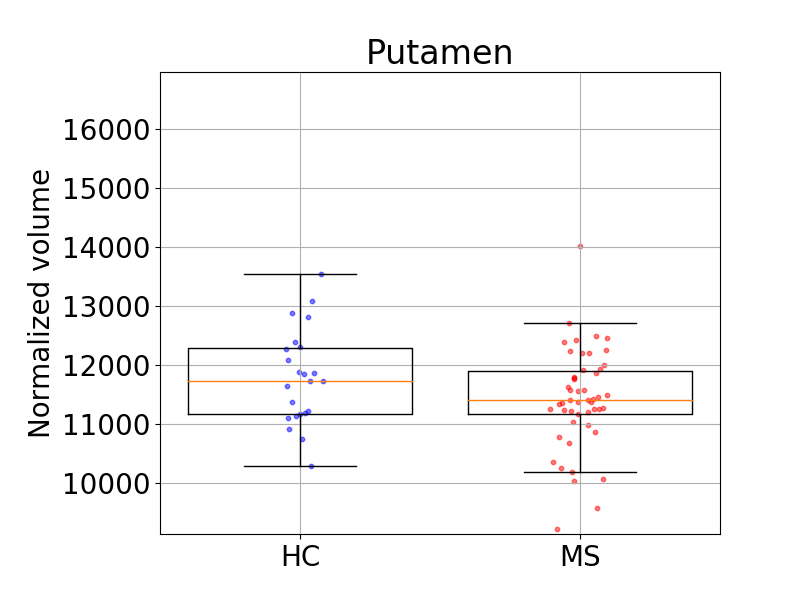}
		\includegraphics[width=.195\textwidth]{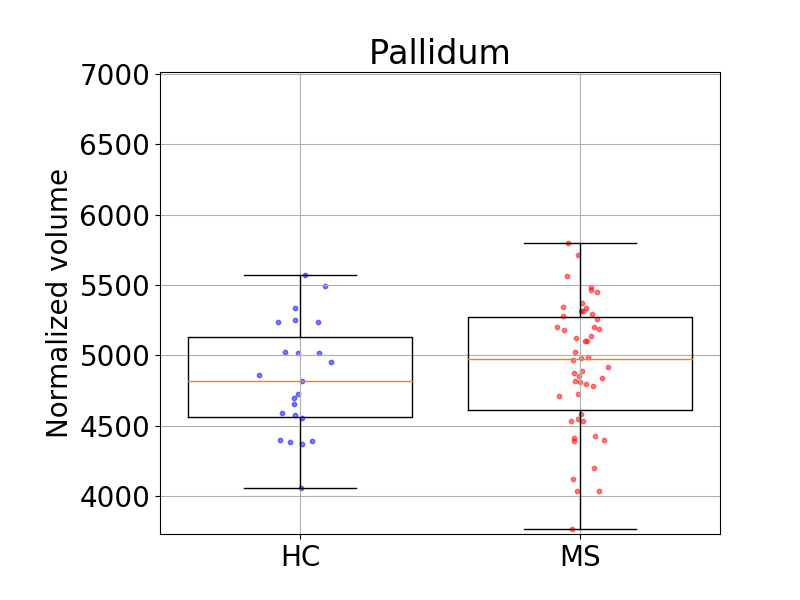}
		\includegraphics[width=.195\textwidth]{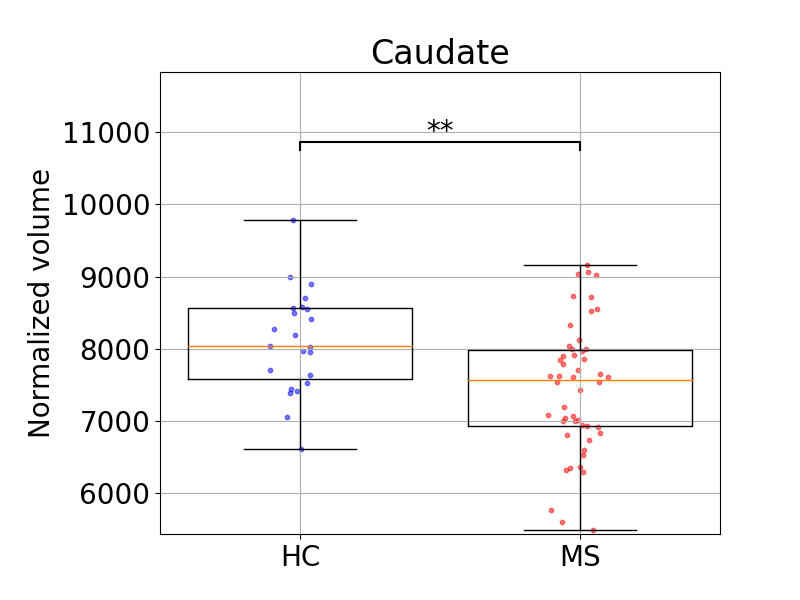}
		\includegraphics[width=.195\textwidth]{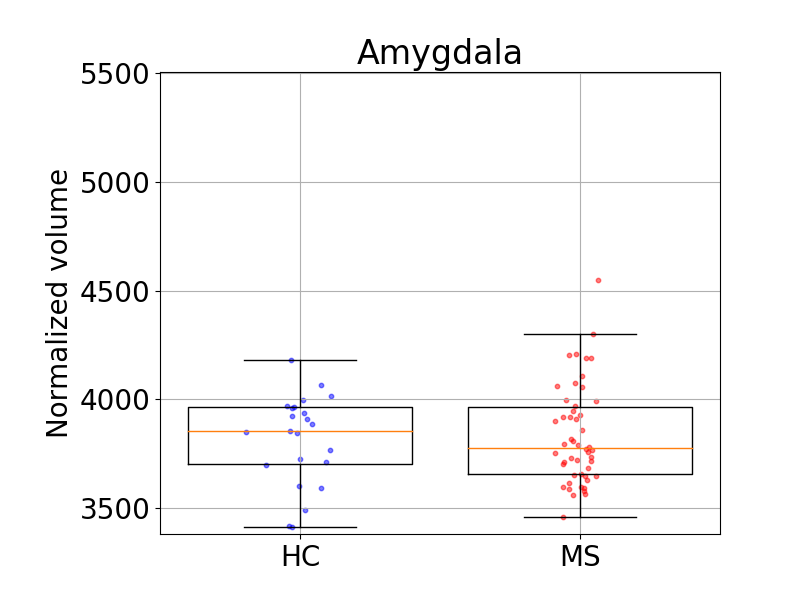}
		\includegraphics[width=.195\textwidth]{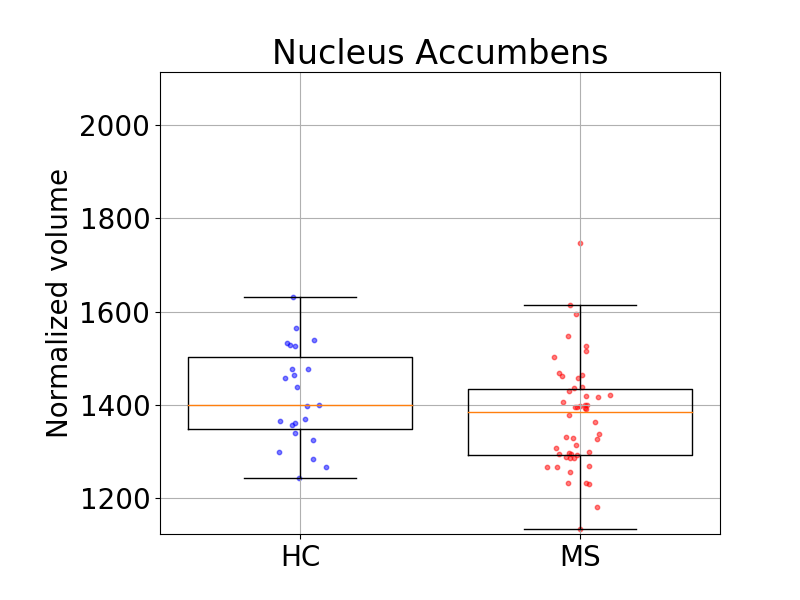}
		\includegraphics[width=.195\textwidth]{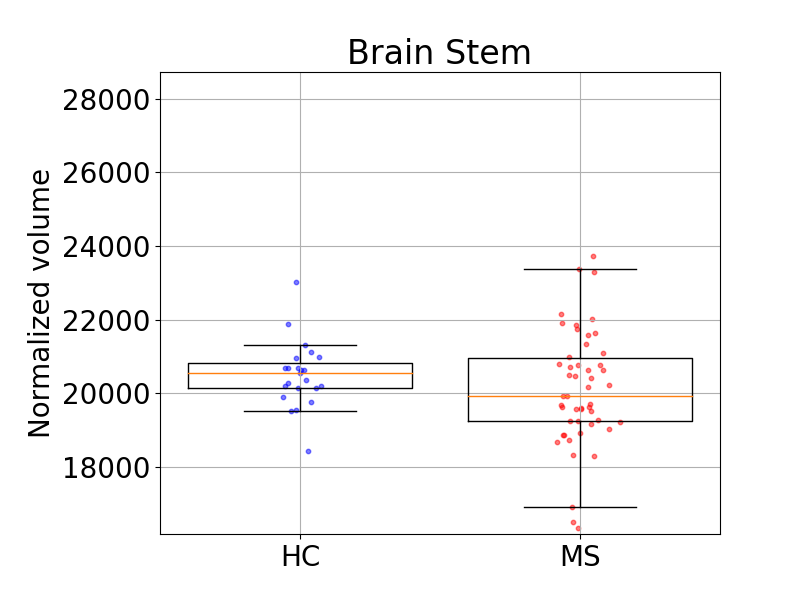}
	\end{center}
	\caption{
	Differences between healthy controls (HC) and MS subjects in 
	normalized volume estimates of various neuroanatomical structures, as detected by the proposed method on the Achieva dataset (23 HC subjects, 50 MS subjects, T1w-FLAIR input).
	Statistically significant differences between the two groups, computed with a Welch's t-test, are indicated by asterisks (``**'' for p-value \textless~0.01 and ``*'' for p-value \textless~0.05).
	}
	\label{fig:HCvsMS}
\end{figure*}

In a final analysis, we assessed whether previously reported volume reductions in specific brain structures in MS can 
automatically
be detected with the proposed method.
Towards this end, we segmented the 23 controls and the 50 MS subjects of the Achieva dataset,
and compared the volumes of various structures between the two groups.
Volumes were normalized for age, gender and total intracranial volume by regressing them out with a general linear model.
The intracranial volume used for the normalization was computed by summing the volumes of all the structures, as segmented by the method, within the intracranial vault.
%
%
%
%
%
The results are shown in Fig.~\ref{fig:HCvsMS}.
Although not all volumes showed significant difference between groups, well established differences were replicated. In particular, we demonstrated decreased volumes of cerebral white matter, cerebral cortex, thalamus and caudate \citep{Chard2002,Houtchens2007,Azevedo2018} as well as an increased volume of the lateral ventricles \citep{Zivadinov2016}.

\section{Discussion and conclusion}
\label{sec:Discussion}

In this paper, we have proposed a method for the simultaneous segmentation of white matter lesions and normal-appearing neuroanatomical structures from multi-contrast brain MRI scans of 
MS
patients.
The method 
integrates 
a 
novel 
model for white matter lesions
into 
a 
previously validated
generative 
model for whole-brain segmentation.
By using separate models 
for the shape of anatomical structures and their appearance in MRI,
the algorithm
is able to
adapt to data acquired with different scanners and imaging protocols
without 
needing to be retrained.
We validated the method using 
\revision{four}
disparate
datasets, 
showing 
\revision{robust}
performance in white matter lesion segmentation while simultaneously segmenting
dozens of
other brain structures.
We further demonstrated that 
it
can
also be 
\revision{safely}
applied 
to MRI scans of healthy controls,
and replicate previously documented atrophy patterns in 
deep gray matter structures in MS.
The proposed algorithm is publicly available as part of the open-source neuroimaging package FreeSurfer.
By performing 
both
whole-brain and white matter lesion segmentation 
at the same time%
, 
the 
method we propose 
aims to supplant 
the two-stage ``lesion filling'' procedure that is commonly used in 
morphometric studies
in MS,
in which lesions segmented in a first step are 
used to 
avoid biasing 
a subsequent 
analysis of 
normal-appearing
structures 
with 
software tools developed for 
healthy brain scans.
In order to 
evaluate whether 
our
method 
is successful in this regard%
,
we compared its whole-brain segmentation performance 
against the results obtained when lesions are segmented \emph{a priori}
by 
seven
different human raters
instead of automatically
by the method itself%
.
Our results show that 
the volumes of various neuroanatomical structures obtained 
when lesions are segmented 
automatically
fall within the range of inter-rater variability%
,
indicating that the proposed method may be used 
instead of lesion filling
with manual lesion segmentations
in large volumetric studies of brain atrophy in MS.
When detailed spatial overlap is analyzed, however, 
we found that the 
automatic segmentation does not 
fully 
reach the 
performance obtained with human lesion annotation
as measured by Dice overlap.

%
%
Like many other methods for MS lesion segmentation, the method proposed here 
produces
a spatial 
map indicating in each voxel its probability of belonging to a lesion,
which can then be thresholded to obtain
a final 
lesion segmentation.
Although in our experience good results 
can be
obtained 
by using 
the same
threshold value 
across datasets 
(e.g., $\gamma=0.5$),
%
changing
this value
allows 
one
to 
adjust
the
trade-off
between false positive and false negative 
lesion detections.
Since some MRI sequences and scanners will depict lesions with a higher contrast than others, and 
because there is often considerable disagreement between human experts regarding the exact extent of lesions~\citep{Zijdenbos_MICCAI},
in our implementation we therefore expose 
this threshold value
as an optional, tunable parameter to the end-user.
%
Suitable threshold values can be found
by visually inspecting the lesion segmentations of a few cases or, in large-scale studies, using cross-validation 
as we did in 
our experiments.

By providing 
the ability to 
robustly and
efficiently
segment
multi-contrasts scans of MS patients across a wide range of imaging 
equipment
and protocols,
the software tool presented here
may help facilitate 
large 
cohort studies
aiming to 
elucidate
the morphological and temporal dynamics underlying disease progression and accumulation of disability in MS.
Furthermore, in current clinical practice, high-resolution multi-contrast images, which can be used to increase the accuracy of lesion segmentation, represent a significantly increased burden for the neuroradiologist to read, and are hence frequently not acquired. The 
emergence of robust, multi-contrast segmentation tools such as ours
may help break the link between the resolution and number of contrasts of the acquired data and the human time needed to evaluate it, thus potentially 
increasing the accuracy of the resulting measures.

The ability of the proposed method to automatically 
tailor
its appearance models 
for 
specific datasets
makes it 
very flexible,
allowing it to seamlessly take advantage of novel, potentially more sensitive and specific MRI acquisitions as they are developed.
Although not extensively tested, the proposed method should make it possible 
to, with minimal adjustments, segment 
data acquired 
with advanced research sequences
such as MP2RAGE \citep{MP2RAGEOrig}, DIR \citep{DIROrig}, FLAIR$^2$ \citep{FLAIR2} or T2* \citep{T2StartOrig}
, both at 
conventional and at
ultra-high 
magnetic 
field strengths.
%
%
%
We are currently
pursuing several
extensions of the proposed 
method,
including
the ability to 
go on and
create 
cortical surfaces 
and parcellations 
in FreeSurfer,
as well as 
a dedicated version 
for
longitudinal data~\revision{\citep{cerri2020}}.

\section{Acknowledgments}
\label{sec:Acknowledgments}

This project has received funding from the European Union's Horizon 2020 research and innovation program under the Marie Sklodowska-Curie grant agreement No 765148,
as well as from the National Institute Of Neurological Disorders and Stroke under project number R01NS112161.
Hartwig R. Siebner holds a 5-year professorship in precision medicine at the Faculty of Health Sciences and Medicine, University of Copenhagen which is sponsored by the Lundbeck Foundation (Grant Nr. R186-2015-2138). 
Mark M\"{u}hlau was supported by the German Research Foundation (Priority Program SPP2177, Radiomics: Next Generation of Biomedical Imaging) -- project number 428223038.

\section{Conflicts of interest}
\label{sec:conflictsOfInterest}
Hartwig R. Siebner has received honoraria as speaker from Sanofi Genzyme, Denmark and Novartis, Denmark, as consultant from Sanofi Genzyme, Denmark and as senior editor (NeuroImage) from Elsevier Publishers, Amsterdam, The Netherlands. He has received royalties as book editor from Springer Publishers, Stuttgart, Germany.

\appendix


\section{Parameter optimization in SAMSEG}
\label{app:GEMS}


We here describe how we perform the optimization of $p(\bldgr{\theta} | \fat{D} )$ with respect to $\bldgr{\theta}$ in the
original
SAMSEG model%
. 
We follow 
a coordinate ascent approach, in which a
limited-memory BFGS optimization of $\bldgr{\theta}_{\fat{l}}$ 
is interleaved with a generalized EM (GEM) optimization of the remaining parameters 
$\bldgr{\theta}_{\fat{d}}$.
The GEM algorithm was derived in~\citep{VanLeemput1999} based on~\citep{Wells1996}, and is repeated here for the sake of completeness.
It iteratively constructs a tight lower bound to the objective function by computing the 
soft label assignments $w_{i,k}$ 
based on the current estimate of $\bldgr{\theta}_{\fat{d}}$
(Eq.~\eqref{eq:healthySeg}),
and 
subsequently
improves the lower bound (and therefore the objective function) 
using the following set of analytical update equations
for 
these parameters%
:
\begin{linenomath}
\begin{align*}
  \bldgr{\mu}_k 
  & \gets 
  \fat{m}_k
  \quad \mathrm{and} \quad
  \bldgr{\Sigma}_k 
  \gets
  \fat{V}_k, 
  \quad \forall k
  \\
\left(
  \begin{array}{@{}c@{}}
    \fat{c}_1 \\
    \vdots \\
    \fat{c}_N
  \end{array}
\right)
  & 
  \gets \left(\begin{array}{@{}ccc@{}}
                \fat{A}\transp\fat{S}_{1,1}\fat{A} & \hdots & \fat{A}\transp\fat{S}_{1,N}\fat{A}\\
                \vdots & \ddots & \vdots\\
                \fat{A}\transp\fat{S}_{N,1}\fat{A} & \hdots & \fat{A}\transp\fat{S}_{N,N}\fat{A}
                \end{array}\right)^{-1} 
                \left(\begin{array}{@{}c@{}} %
                \fat{A}\transp\left( \sum_{n=1}^N \fat{S}_{1,n}\fat{r}_{1,n}\right)\\
                \vdots\\
                \fat{A}\transp\left( \sum_{n=1}^N \fat{S}_{N,n}\fat{r}_{N,n}\right)
                \end{array}\right),
\end{align*}
\end{linenomath}
where 
\begin{linenomath}
\begin{align*}
  & 
  \fat{m}_k 
  = \frac{\sum_{i=1}^I w_{i,k} ( \fat{d}_i - \fat{C}\bldgr{\phi}_i ) }{N_k}
  \quad \mathrm{with} \quad
  N_k = \sum_{i=1}^I w_{i,k},
   \\
  & \fat{V}_k 
  = \frac{\sum_{i=1}^I w_{i,k} ( \fat{d}_i - \fat{C}\bldgr{\phi}_i - \fat{m}_k )
                           ( \fat{d}_i - \fat{C}\bldgr{\phi}_i - \fat{m}_k )\transp }
                           {N_k},
   \\
& \fat{A} = \left(\begin{array}{@{}ccc@{}}
                                        \phi_{1}^1 & \hdots & \phi_{P}^1\\
                                        \vdots & \ddots & \vdots\\
                                        \phi_{1}^I & \hdots & \phi_{P}^I
                                        \end{array}\right), \quad
\fat{S}_{m,n} = \text{diag}\left(s_{i}^{m,n}\right)
, \quad 
\fat{r}_{m,n} = \left( 
                     \begin{array}{@{}c@{}}
                       r_{1}^{m,n}\\
                      \vdots \\
                      r_{I}^{m,n} 
                     \end{array}
                   \right)
\\
& \textrm{and} 
\\ 
& 
s_{i}^{m,n} = \sum_{k=1}^K s_{i,k}^{m,n}
, \,\,\,  
s_{i,k}^{m,n} =  w_{i,k}\left(\fat{\Sigma}_k^{-1}\right)_{m,n} 
, \,\,\,
r_{i}^{m,n}=d_{i}^{n} - \frac{\sum_{l=1}^{K} s_{i,k}^{m,n}\left(\bldgr{\mu}_k\right)_{n}}{\sum_{l=1}^K s_{i,k}^{m,n}}.
\end{align*}
\end{linenomath}

\section{Parameter optimization}
\label{app:optimizationWithLesionPrior}

Here we describe how we perform the
optimization of $p( \bldgr{\theta}, \bldgr{\theta}_{les} | \fat{D} )$ 
with respect to $\bldgr{\theta}$ and $\bldgr{\theta}_{les}$
in the 
augmented
model of Sec.~\ref{sec:modelingLesion}
with the decoder outputs $f_i(\fat{h})$ 
all
clamped to value $1$.
In that case, 
the model can be 
reformulated in
the same form as 
the original SAMSEG model,
so that the same optimization 
strategy 
can be used.
In particular, 
lesions
can be considered to 
form an extra class 
(with index $K+1$)
in a SAMSEG model with $K+1$ labels,
provided that 
the mesh vertex label probabilities 
\begin{linenomath}
\begin{align*}
  \tilde{\alpha}_j^k 
  =
  \begin{cases}
    \beta_j                   & \text{if } k=K+1 \text{ (lesion)}, \\
    \alpha_j^k (\beta_j-1)    & \text{otherwise} .
  \end{cases}
\end{align*}
\end{linenomath}
are used instead of the original $\alpha_j^k$'s in the atlas interpolation model of Eq.~\eqref{eq:atlasPrior}.

The 
optimization 
described in~\ref{app:GEMS}
does require
one modification
because of
the 
prior $p( \bldgr{\theta}_{les} | \bldgr{\theta}_{\fat{d}})$ binding 
the means and variances of the WM and lesion classes
together.
The following altered update equations for these parameters guarantee that the 
EM lower bound,
and therefore the objective function, is improved in each iteration
of the GEM algorithm:
\begin{linenomath}
\begin{align*}
  \bldgr{\mu}_{WM} \quad
  \gets & \quad
  \left( 
    N_{WM} \fat{I}
    +
    \frac{ \nu N_{WM} }{ \nu + N_{WM} } \bldgr{\Sigma}_{WM} \bldgr{\Sigma}_{les}^{-1}
  \right)^{-1}
   \\
  & \quad \quad
  \left( 
    N_{WM} 
    \fat{m}_{WM}
    +
    \frac{ \nu N_{WM} }{ \nu + N_{WM} } \bldgr{\Sigma}_{WM} \bldgr{\Sigma}_{les}^{-1}
    \fat{m}_{les}
  \right)
  , \\
  \bldgr{\Sigma}_{WM}  \quad
  \gets & \quad
  \frac{
    N_{WM}
    \fat{V}_{WM}
    + \bldgr{\Sigma}_{les} \bldgr{\Sigma}_{WM}^{-1} \bldgr{\Psi}_{les}    
  }{N_{WM} + N_{les} + N + 2 }
  ,
   \\
  \bldgr{\mu}_{les} \quad
  \gets & \quad
  \frac{ N_{les} \fat{m}_{les} + \nu \bldgr{\mu}_{WM} }%
       { N_{les} + \nu}
  ,
  \\
  \bldgr{\Sigma}_{les} \quad 
  \gets & \quad
  \frac{
        \bldgr{\Psi}_{les}        
        +
        \nu \kappa \bldgr{\Sigma}_{WM}  
  }
  {
  N_{les} + \nu
  },
  \\
  \textrm{where} 
  \quad 
  \bldgr{\Psi}_{les} & = \frac{N_{les} \nu}{N_{les} + \nu}(\fat{m}_{les} - \bldgr{\mu}_{WM} ) 
                                        (\fat{m}_{les} - \bldgr{\mu}_{WM} )\transp
               +
               N_{les} \fat{V}_{les} 
  .               
\end{align*}
\end{linenomath}

\section{Estimating lesion probabilities}
\label{app:MCMC}

We here describe how we we approximate $p( z_i = 1 | \fat{d}_i, \bldgr{\hat{\theta}} )$ using Monte Carlo sampling. 
We use a Markov chain Monte Carlo (MCMC) approach to sample 
triplets
$\{ \bldgr{\theta}_{les}^{(s)}, \fat{z}^{(s)}, \fat{h}^{(s)} \}$
from the distribution 
$p( \bldgr{\theta}_{les}, \fat{z}, \fat{h} | \fat{D}, \bldgr{\hat{\theta}} )$:
Starting from an initial 
lesion segmentation $\fat{z}^{(0)}$ 
obtained 
from 
the parameter estimation procedure described in~\ref{app:optimizationWithLesionPrior},
we use a blocked Gibbs sampler in which each 
variable
is updated conditioned on the other 
ones:
\begin{linenomath}
\begin{align*}
  \bldgr{\Sigma}_{les}^{(s+1)}
  & \sim
  p( \bldgr{\Sigma}_{les} | \fat{D},  \bldgr{\hat{\theta}}, \fat{z}^{(s)} )
  = \\
  & \quad
  \mathrm{IW}\left( \bldgr{\Sigma}_{les} \, \big| \, \bldgr{\Psi}_{les}^{(s)} + \nu \kappa \bldgr{\hat{\Sigma}}_{WM}, 
  \,\,
  N_{les}^{(s)} + \nu - N - 2 \right)
  \\
  \bldgr{\mu}_{les}^{(s+1)}
  & \sim
  p( \bldgr{\mu}_{les} | \fat{D},  \bldgr{\hat{\theta}}, \fat{z}^{(s)}, \bldgr{\Sigma}_{les}^{(s+1)} )
  = \\
  & \quad
  \mathcal{N}\left( \bldgr{\mu}_{les} \, \bigg| \, \frac{N_{les}^{(s)} \fat{m}_{les}^{(s)} + \nu \bldgr{\hat{\mu}}_{WM}}{N_{les}^{(s)} + \nu}, \,\, \frac{\bldgr{\Sigma}_{les}^{(s+1)}}{N_{les}^{(s)} + \nu}  \right)
  \\
  \fat{h}^{(s+1)}
  & \sim
  p( \fat{h} | \fat{z}^{(s)} )
  \simeq 
  \mathcal{N}\left( 
    \,
    \fat{h}
    \,
    \big| 
    \,
    \bldgr{\mu}_{\upsilon}(\fat{z}^{(s)}), 
    \,
    \mathrm{diag}( \bldgr{\sigma}^2_{\upsilon}(\fat{z}^{(s)}) ) 
  \right)
  \\
  \fat{z}^{(s+1)}
  & \sim
  p( \fat{z} | \fat{D}, \bldgr{\hat{\theta}}, \fat{h}^{(s+1)}, \bldgr{\theta}_{les}^{(s+1)} )
  = \prod_{i=1}^I p( z_i | \fat{d}_i, \bldgr{\hat{\theta}}, \fat{h}^{(s+1)}, \bldgr{\theta}_{les}^{(s+1)} )
  ,
\end{align*}
\end{linenomath}
where we use the
encoder
variational approximation obtained during the training of the lesion shape model
(see Sec.~\ref{sec:VAE})
to sample from $\fat{h}$
in the next-to-last step,
and 
\begin{linenomath}
\begin{align*}
  p( z_i = 1 | \fat{d}_i, \bldgr{\hat{\theta}}, \fat{h},\bldgr{\theta}_{les} )
  =
  \frac{ \mathcal{N}( \fat{d}_i | \bldgr{\mu}_{les} + \fat{C} \bldgr{\phi}_i, \bldgr{\Sigma}_{les} ) 
          f_i( \fat{h} ) 
          \rho_i( \bldgr{\hat{\theta}}_{\fat{l}} ) 
      }
      {
      \sum_{l_i=1}^K \sum_{z'_i=0}^1
      p( \fat{d}_i | l_i, z'_i, \bldgr{\hat{\theta}}_{\fat{l}}, \bldgr{\theta}_{les} )
      p ( z'_i | \bldgr{\hat{\theta}}_{\fat{l}}, \fat{h} )
      p ( l_i | \bldgr{\hat{\theta}}_{\fat{l}} )
      }  
\end{align*}
\end{linenomath}
in the last step. 
In these equations, the variables 
$N_{les}^{(s)}$,
$\fat{m}_{les}^{(s)}$,  
$\fat{V}_{les}^{(s)}$
and
$\bldgr{\Psi}_{les}^{(s)}$
are as defined before, but using voxel assignments $w_{i,les} = z_i^{(s)}$.
Once $S$ samples are obtained, we approximate $p( z_i = 1 | \fat{d}_i, \bldgr{\hat{\theta}} )$
as
\begin{linenomath}
\begin{align*}
p( z_i = 1 | \fat{d}_i, \bldgr{\hat{\theta}} ) 
  \simeq
  \frac{1}{S}
  \sum_{s=1}^S
  p( z_i = 1 | \fat{d}_i, \bldgr{\hat{\theta}}, \fat{h}^{(s)},\bldgr{\theta}_{les}^{(s)} )
  .
\end{align*}
\end{linenomath}

In our implementation, we use $S=50$ samples, obtained after discarding the first 50 sweeps of the sampler (so-called ``burn-in'' phase).
The algorithm repeatedly invokes the decoder and encoder networks of the lesion shape model described in Sec.~\ref{sec:VAE}. 
Since 
this
shape model was trained in a specific 
isotropic space, 
the algorithm requires 
transitioning between
this training space and subject space
using an affine transformation.
This is 
accomplished
by 
resampling 
the input and output of the encoder and decoder, respectively, using linear interpolation.

\section*{Bibliography}
\bibliography{bib} 

@article{Houtchens2007,
author = {Houtchens, M. K. and Benedict, R. H.B. and Killiany, R. and Sharma, J. and Jaisani, Z. and Singh, B. and Weinstock-Guttman, B. and Guttmann, C. R.G. and Bakshi, R.},
journal = {Neurology},
title = {{Thalamic atrophy and cognition in multiple sclerosis}},
year = {2007},
volume={69},
number={12},
pages={1213--1223}
}

@article{Azevedo2018,
author = {Azevedo, Christina J. and Cen, Steven Y. and Khadka, Sankalpa and Liu, Shuang and Kornak, John and Shi, Yonggang and Zheng, Ling and Hauser, Stephen L. and Pelletier, Daniel},
journal = {Annals of Neurology},
title = {{Thalamic atrophy in multiple sclerosis: A magnetic resonance imaging marker of neurodegeneration throughout disease}},
year = {2018},
volume={83},
number={2},
pages={223--234}
}

@article{Valverde2019,
author = {Valverde, Sergi and Salem, Mostafa and Cabezas, Mariano and Pareto, Deborah and Vilanova, Joan C. and Rami{\'{o}}-Torrent{\`{a}}, Llu{\'{i}}s and Rovira, {\`{A}}lex and Salvi, Joaquim and Oliver, Arnau and Llad{\'{o}}, Xavier},
journal = {NeuroImage: Clinical},
title = {{One-shot domain adaptation in multiple sclerosis lesion segmentation using convolutional neural networks}},
year = {2019},
volume = {21},
pages = {101638},
}

@article{Ashburner2000,
author = {Ashburner, John and Andersson, Jesper L.R. and Fristen, Karl J.},
journal = {Human Brain Mapping},
title = {{Image registration using a symmetric prior - In three dimensions}},
year = {2000},
volume={9},
number={4},
pages={212--225}
}

@article{Valverde2017,
author = {Valverde, Sergi and Cabezas, Mariano and Roura, Eloy and Gonz{\'{a}}lez-Vill{\`{a}}, Sandra and Pareto, Deborah and Vilanova, Joan C and Rami{\'{o}}-Torrent{\`{a}}, Llu{\'{i}}s and Rovira, {\`{A}}lex and Oliver, Arnau and Llad{\'{o}}, Xavier},
doi = {10.1016/j.neuroimage.2017.04.034},
issn = {1053-8119},
journal = {NeuroImage},
pages = {159--168},
title = {{Improving automated multiple sclerosis lesion segmentation with a cascaded 3D convolutional neural network approach}},
volume = {155},
year = {2017}
}

@misc{Abadi2016,
    title={TensorFlow: Large-Scale Machine Learning on Heterogeneous Distributed Systems},
    author={Martín Abadi and Ashish Agarwal and Paul Barham and Eugene Brevdo and Zhifeng Chen and Craig Citro and Greg S. Corrado and Andy Davis and Jeffrey Dean and Matthieu Devin and Sanjay Ghemawat and Ian Goodfellow and Andrew Harp and Geoffrey Irving and Michael Isard and Yangqing Jia and Rafal Jozefowicz and Lukasz Kaiser and Manjunath Kudlur and Josh Levenberg and Dan Mane and Rajat Monga and Sherry Moore and Derek Murray and Chris Olah and Mike Schuster and Jonathon Shlens and Benoit Steiner and Ilya Sutskever and Kunal Talwar and Paul Tucker and Vincent Vanhoucke and Vijay Vasudevan and Fernanda Viegas and Oriol Vinyals and Pete Warden and Martin Wattenberg and Martin Wicke and Yuan Yu and Xiaoqiang Zheng},
    year={2015},
    eprint={1603.04467},
    archivePrefix={arXiv},
    primaryClass={cs.DC}
}

@article{Cotton2015,
author = {Cotton, F. and Kremer, S. and Hannoun, S. and Vukusic, S. and Dousset, V. and Roxana, Am{\'{e}}lie and Ren{\'{e}}, Anxionnat and Jean-Paul, Armspach and Bertrand, Audoin and Christian, Barillot and Isabelle, Berry and Fabrice, Bonneville and Claire, Boutet and Giovanni, Castelnovo and Fr{\'{e}}d{\'{e}}ric, Cervenanski and Mikael, Cohen and Olivier, Commowick and Fran{\c{c}}ois, Cotton and J{\'{e}}r{\^{o}}me, De Seze and Vincent, Dousset and Fran{\c{c}}oise, Durand Dubief and Gilles, Edan and Jean-Christophe, Ferre and Damien, Galanaud and Tristan, Glattard and Sylvie, Grand and Justine, Guillaumont and R{\'{e}}my, Guillevin and Charles, Guttmann and Salem, Hannoun and Fabrice, Heitz and Alexandre, Krainik and St{\'{e}}phane, Kremer and Pierre, Labauge and {de Champfleur Nicolas}, Menjot and Jean-Philippe, Ranjeva and Jean-Am{\'{e}}d{\'{e}}e, Roch and Dominique, Sappey Marinier and Julien, Savatovsky and Bruno, Stankoff and Ayman, Tourbah and Thomas, Tourdias and Sandra, Vukusic},
booktitle = {Journal of Neuroradiology},
title = {{OFSEP, a nationwide cohort of people with multiple sclerosis: Consensus minimal MRI protocol}},
year = {2015},
volume={42},
numbers={3},
pages={133--140}
}

@article{Dempster1977JRSS,
 author = {A. P. Dempster and N. M. Laird and D. B. Rubin},
 journal = {Journal of the Royal Statistical Society. Series B (Methodological)},
 number = {1},
 pages = {1--38},
 publisher = {[Royal Statistical Society, Wiley]},
 title = {Maximum Likelihood from Incomplete Data via the EM Algorithm},
 volume = {39},
 year = {1977}
}

@article{VanLeemput1999,
author = {{Van Leemput}, Koen and Maes, Frederik and Vandermeulen, Dirk and Suetens, Paul},
journal = {IEEE Transactions on Medical Imaging},
title = {{Automated model-based bias field correction of MR images of the brain}},
year = {1999},
volume={18},
number={10},
pages={885-896}
}

@article{Wells1996,
author = {Wells, W. M. and Grimson, W. E.L. and Kikinis, R. and Jolesz, F. A.},
journal = {IEEE Transactions on Medical Imaging},
title = {{Adaptive segmentation of {MRI} data}},
year = {1996},
volume={15},
number={4},
pages={429-442}
}

@article{Barkhof2009,
author = {Barkhof, Frederik and Calabresi, Peter A. and Miller, David H. and Reingold, Stephen C.},
journal = {Nature Reviews Neurology},
month = {may},
number = {5},
pages = {256--266},
publisher = {Nature Publishing Group},
title = {{Imaging outcomes for neuroprotection and repair in multiple sclerosis trials}},
volume = {5},
year = {2009}
}

@article{Schmidt2012,
author = {Schmidt, Paul and Gaser, Christian and Arsic, Milan and Buck, Dorothea and F{\"{o}}rschler, Annette and Berthele, Achim and Hoshi, Muna and Ilg, R{\"{u}}diger and Schmid, Volker J and Zimmer, Claus and Hemmer, Bernhard and M{\"{u}}hlau, Mark},
journal = {NeuroImage},
number = {4},
pages = {3774--3783},
title = {{An automated tool for detection of FLAIR-hyperintense white-matter lesions in Multiple Sclerosis}},
volume = {59},
year = {2012}
}

@article{Shiee2010,
author = {Shiee, Navid and Bazin, Pierre-Louis and Ozturk, Arzu and Reich, Daniel S and Calabresi, Peter A and Pham, Dzung L},
journal = {NeuroImage},
number = {2},
pages = {1524--1535},
title = {{A topology-preserving approach to the segmentation of brain images with multiple sclerosis lesions}},
volume = {49},
year = {2010}
}

@article{Bazin,
author = {Bazin, Pierre-Louis and Pham, Dzung L},
journal = {Medical Image Analysis},
number = {5},
pages = {616--625},
title = {{Homeomorphic brain image segmentation with topological and statistical atlases}},
volume = {12},
year = {2008}
}

@article{Carass2017,
author = {Carass, Aaron and Roy, Snehashis and Jog, Amod and Cuzzocreo, Jennifer L and Magrath, Elizabeth and Gherman, Adrian and Button, Julia and Nguyen, James and Prados, Ferran and Sudre, Carole H and Cardoso, Manuel Jorge and Cawley, Niamh and Ciccarelli, Olga and Wheeler-Kingshott, Claudia A M and Ourselin, S{\'{e}}bastien and Catanese, Laurence and Deshpande, Hrishikesh and Maurel, Pierre and Commowick, Olivier and Barillot, Christian and Tomas-Fernandez, Xavier and Warfield, Simon K and Vaidya, Suthirth and Chunduru, Abhijith and Muthuganapathy, Ramanathan and Krishnamurthi, Ganapathy and Jesson, Andrew and Arbel, Tal and Maier, Oskar and Handels, Heinz and Iheme, Leonardo O and Unay, Devrim and Jain, Saurabh and Sima, Diana M and Smeets, Dirk and Ghafoorian, Mohsen and Platel, Bram and Birenbaum, Ariel and Greenspan, Hayit and Bazin, Pierre-Louis and Calabresi, Peter A and Crainiceanu, Ciprian M and Ellingsen, Lotta M and Reich, Daniel S and Prince, Jerry L and Pham, Dzung L},
journal = {NeuroImage},
pages = {77--102},
title = {{Longitudinal Multiple Sclerosis Lesion Segmentation: Resource {\&} Challenge HHS Public Access}},
volume = {148},
year = {2017}
}

@article{Commowick2018,
author = {Commowick, Olivier and Istace, Audrey and Kain, Micha{\"{e}}l and Laurent, Baptiste and Leray, Florent and Simon, Mathieu and Pop, Sorina Camarasu and Girard, Pascal and Am{\'{e}}li, Roxana and Ferr{\'{e}}, Jean-Christophe and Kerbrat, Anne and Tourdias, Thomas and Cervenansky, Fr{\'{e}}d{\'{e}}ric and Glatard, Tristan and Beaumont, J{\'{e}}r{\'{e}}my and Doyle, Senan and Forbes, Florence and Knight, Jesse and Khademi, April and Mahbod, Amirreza and Wang, Chunliang and Mckinley, Richard and Wagner, Franca and Muschelli, John and Sweeney, Elizabeth and Roura, Eloy and Llad{\'{o}}, Xavier and Santos, Michel M and Santos, Wellington P and Silva-Filho, Abel G and Tomas-Fernandez, Xavier and Urien, H{\'{e}}l{\`{e}}ne and Bloch, Isabelle and Valverde, Sergi and Cabezas, Mariano and Vera-Olmos, Francisco Javier and Malpica, Norberto and Guttmann, Charles and Vukusic, Sandra and Edan, Gilles and Dojat, Michel and Styner, Martin and Warfield, Simon K and Cotton, Fran{\c{c}}ois and Barillot, Christian},
title = {Objective Evaluation of Multiple Sclerosis Lesion Segmentation using a Data Management and Processing Infrastructure},
volume = {8},
number={13650},
year = {2018},
journal={Scientific Reports}
}

@article{Danelakis2018,
author = {Danelakis, Antonios and Theoharis, Theoharis and Verganelakis, Dimitrios A.},
journal = {Computerized Medical Imaging and Graphics},
month = {dec},
pages = {83--100},
publisher = {Pergamon},
title = {{Survey of automated multiple sclerosis lesion segmentation techniques on magnetic resonance imaging}},
volume = {70},
year = {2018}
}

@misc{Mckinley2019,
    title={Simultaneous lesion and neuroanatomy segmentation in Multiple Sclerosis using deep neural networks},
    author={Richard McKinley and Rik Wepfer and Fabian Aschwanden and Lorenz Grunder and Raphaela Muri and Christian Rummel and Rajeev Verma and Christian Weisstanner and Mauricio Reyes and Anke Salmen and Andrew Chan and Franca Wagner and Roland Wiest},
    year={2019},
    eprint={1901.07419},
    archivePrefix={arXiv},
    primaryClass={cs.CV}
}

@misc{Kingma,
    title={Adam: A Method for Stochastic Optimization},
    author={Diederik P. Kingma and Jimmy Ba},
    year={2014},
    eprint={1412.6980},
    archivePrefix={arXiv},
    primaryClass={cs.LG}
}

@inproceedings{Rezende2014,
archivePrefix = {arXiv},
arxivId = {1401.4082},
author = {Rezende, Danilo Jimenez and Mohamed, Shakir and Wierstra, Daan},
title = {{Stochastic Backpropagation and Approximate Inference in Deep Generative Models}},
year = {2014},
booktitle = {Proceedings of the 31st International Conference on International Conference on Machine Learning},
volume={32},
pages = {1278--1286},
numpages = {9},
}

@article{VanLeemput2001,
author = {{Van Leemput}, K and Maes, F and Vandermeulen, D and Colchester, A and Suetens, P},
journal = {IEEE transactions on medical imaging},
number = {8},
pages = {677--88},
title = {{Automated segmentation of multiple sclerosis lesions by model outlier detection.}},
volume = {20},
year = {2001}
}

@article{Garcia-Lorenzo2013,
author = {Garc{\'{i}}a-Lorenzo, Daniel and Francis, Simon and Narayanan, Sridar and Arnold, Douglas L. and Collins, D. Louis},
journal = {Medical Image Analysis},
number = {1},
pages = {1--18},
title = {{Review of automatic segmentation methods of multiple sclerosis white matter lesions on conventional magnetic resonance imaging}},
volume = {17},
year = {2013}
}

@misc{Kingma2013,
    title={Auto-Encoding Variational Bayes},
    author={Diederik P Kingma and Max Welling},
    year={2013},
    eprint={1312.6114},
    archivePrefix={arXiv},
    primaryClass={stat.ML}
}

@article{VanLeemput2009,
author = {{Van Leemput}, Koen},
journal = {IEEE Transactions on Medical Imaging},
number = {6},
pages = {822--837},
title = {{Encoding probabilistic brain atlases using Bayesian inference}},
volume = {28},
year = {2009}
}

@inproceedings{Puonti2016a,
author = {Puonti, Oula and Van Leemput, Koen},
booktitle = {Lecture Notes in Computer Science (including subseries Lecture Notes in Artificial Intelligence and Lecture Notes in Bioinformatics)},
pages = {9--20},
title = {{Simultaneous whole-brain segmentation and white matter lesion detection using contrast-adaptive probabilistic models}},
volume = {9556},
year = {2016}
}

@article{Puonti2016,
author = {Puonti, Oula and Iglesias, Juan Eugenio and {Van Leemput}, Koen},
journal = {NeuroImage},
pages = {235--249},
title = {{Fast and sequence-adaptive whole-brain segmentation using parametric Bayesian modeling}},
volume = {143},
year = {2016}
}

@article{Rissanen2014,
journal = {Journal of Nuclear Medicine},
author = {Rissanen, Eero and Tuisku, Jouni and Rokka, Johanna and Paavilainen, Teemu and Parkkola, Riitta and Rinne, Juha O. and Airas, Laura},
title = {{In vivo detection of diffuse inflammation in secondary progressive multiple sclerosis using PET imaging and the radioligand11C-PK11195}},
year = {2014},
volume={55},
number={6},
pages={939--944}
}

@article{Muhlau2013,
author = {M{\"{u}}hlau, Mark and Buck, Dorothea and F{\"{o}}rschler, Annette and Boucard, Christine C. and Arsic, Milan and Schmidt, Paul and Gaser, Christian and Berthele, Achim and Hoshi, Muna and Jochim, Angela and Kronsbein, Helena and Zimmer, Claus and Hemmer, Bernhard and Ilg, R{\"{u}}diger},
journal = {Multiple Sclerosis Journal},
title = {{White-matter lesions drive deep gray-matter atrophy in early multiple sclerosis: Support from structural MRI}},
year = {2013},
volume={19},
number={11},
pages={1485--1492}
}

@article{MICCAI2008,
author = {Styner, Martin and Lee, Joohwi and Chin, Brian and Chin, Matthew and Commowick, Olivier and Tran, Hoai-Huong and Jewells, Valerie and Warfield, Simon},
year = {2008},
month = {11},
pages = {1--6},
title = {3{D} segmentation in the clinic: A grand challenge {II}: {MS} lesion segmentation},
journal = {MIDAS Journal}
}

@inproceedings{Ioffe2015,
author = {Ioffe, Sergey and Szegedy, Christian},
title = {Batch Normalization: Accelerating Deep Network Training by Reducing Internal Covariate Shift},
year = {2015},
booktitle = {Proceedings of the 32nd International Conference on International Conference on Machine Learning},
volume={37},
pages = {448--456},
numpages = {9},
location = {Lille, France},
}

@inproceedings{Ait-Ali2005,
author = {A{\"{i}}t-Ali, L. S. and Prima, S. and Hellier, P. and Carsin, B. and Edan, G. and Barillot, C.},
booktitle = {Lecture Notes in Computer Science (including subseries Lecture Notes in Artificial Intelligence and Lecture Notes in Bioinformatics)},
title = {{STREM: A robust multidimensional parametric method to segment MS lesions in MRI}},
year = {2005},
volume={3749},
pages={409-416}
}

@inproceedings{Bricq2008,
author = {Bricq, S. and Collet, Ch and Armspach, J. P.},
booktitle = {2008 5th IEEE International Symposium on Biomedical Imaging: From Nano to Macro, Proceedings, ISBI},
title = {{Lesions detection on 3D brain MRI using trimmmed likelihood estimator and probabilistic atlas}},
year = {2008},
volume={},
number={},
pages={93--96},
}

@inproceedings{Rousseau2008,
author = {Rousseau, F. and Blanc, F. and {De S{\`{e}}ze}, J. and Rumbach, L. and Armspach, J. P.},
booktitle={2008 5th IEEE International Symposium on Biomedical Imaging: From Nano to Macro},
title={An a contrario approach for outliers segmentation: Application to Multiple Sclerosis in {MRI}}, year={2008},
volume={},
number={},
pages={9-12}}

@article{Prastawa2008,
author = {Prastawa, Marcel and Gerig, Guido},
title = {{Automatic MS Lesion Segmentation By Outlier Detection and Information Theoretic Region Partitioning}},
year = {2008},
journal = {MIDAS Journal}
}

@article{Jain2015,
author = {Jain, Saurabh and Sima, Diana M and Ribbens, Annemie and Cambron, Melissa and Maertens, Anke and Hecke, Wim Van and Mey, Johan De and Barkhof, Frederik and Steenwijk, Martijn D and Daams, Marita and Maes, Frederik and Huffel, Sabine Van and Vrenken, Hugo and Smeets, Dirk},
journal = {NeuroImage: Clinical},
pages = {367--375},
title = {{Automatic segmentation and volumetry of multiple sclerosis brain lesions from MR images}},
volume = {8},
year = {2015}
}

@inproceedings{Liu2009,
author = {Liu, Jundong and Smith, Charles D. and Chebrolu, Himachandra},
booktitle={2009 IEEE Computer Society Conference on Computer Vision and Pattern Recognition Workshops},
title = {{Automatic multiple sclerosis detection based on integrated square estimation}},
year = {2009},
volume={},
number={},
pages={31--38},
}

@article{Garcia-Lorenzo2011,
author = {Garc{\'{i}}a-Lorenzo, Daniel and Prima, Sylvain and Arnold, Douglas L. and Collins, D. Louis and Barillot, Christian},
journal = {IEEE Transactions on Medical Imaging},
title = {{Trimmed-likelihood estimation for focal lesions and tissue segmentation in multisequence MRI for multiple sclerosis}},
year = {2011},
volume={30},
number={8},
pages={1455--1467}
}

@article{Guttmann1999,
author = {Guttmann, Charles R.G. and Kikinis, Ron and Anderson, Mark C. and Jakab, Marianna and Warfield, Simon K. and Killiany, Ron J. and Weiner, Howard L. and Jolesz, Ferenc A.},
journal = {Journal of Magnetic Resonance Imaging},
title = {{Quantitative follow-up of patients with multiple sclerosis using MRI: Reproducibility}},
year = {1999},
volume={4},
number={9},
pages={509--518}
}

@article{Kikinis1999,
author = {Kikinis, Ron and Guttmann, Charles R.G. and Metcalf, David and Wells, William M. and Ettinger, Gil J. and Weiner, Howard L. and Jolesz, Ferenc A.},
journal = {Journal of Magnetic Resonance Imaging},
title = {{Quantitative follow-up of patients with multiple sclerosis using MRI: Technical aspects}},
year = {1999},
volume={9},
number={4},
pages={519--530}
}

@article{Sudre2015,
author = {Sudre, Carole H. and Cardoso, M. Jorge and Bouvy, Willem H. and Biessels, Geert Jan and Barnes, Josephine and Ourselin, Sebastien},
journal = {IEEE Transactions on Medical Imaging},
title = {{Bayesian Model Selection for Pathological Neuroimaging Data Applied to White Matter Lesion Segmentation}},
year = {2015},
volume = {34},
number = {10},
pages = {2079--2102}
}

@article{FLAIR2,
author = {Wiggermann, V. and Hernandez-Torres, E. and Traboulsee, A. and Li, D. K.B. and Rauscher, A.},
journal = {American Journal of Neuroradiology},
title = {{FLAIR2: A combination of FLAIR and T2 for improved MS lesion detection}},
year = {2016},
number = {2},
pages = {259--265},
volume = {37}
}

@article{rosati2001prevalence,
  title={The prevalence of multiple sclerosis in the world: an update},
  author={Rosati, G},
  journal={Neurological Sciences},
  volume={22},
  number={2},
  pages={117--139},
  year={2001},
  publisher={Springer}
}

@article{goldenberg2012multiple,
  title={Multiple sclerosis review},
  author={Goldenberg, Marvin M},
  journal={Pharmacy and Therapeutics},
  volume={37},
  number={3},
  pages={175},
  year={2012},
  publisher={MediMedia, USA}
}

@article{adelman2013cost,
  title={The cost burden of multiple sclerosis in the {U}nited {S}tates: a systematic review of the literature},
  author={Adelman, Gabriel and Rane, Stanley G and Villa, Kathleen F},
  journal={Journal of Medical Economics},
  volume={16},
  number={5},
  pages={639--647},
  year={2013},
  publisher={Taylor \& Francis}
}

@article{bakshi2008mri,
  title={{MRI} in multiple sclerosis: current status and future prospects},
  author={Bakshi, Rohit and Thompson, Alan J and Rocca, Maria A and Pelletier, Daniel and Dousset, Vincent and Barkhof, Frederik and Inglese, Matilde and Guttmann, Charles RG and Horsfield, Mark A and Filippi, Massimo},
  journal={The Lancet Neurology},
  volume={7},
  number={7},
  pages={615--625},
  year={2008},
  publisher={Elsevier}
}

@article{lovblad2010mr,
  title={{MR} imaging in multiple sclerosis: review and recommendations for current practice},
  author={L{\"o}vblad, K-O and Anzalone, N and D{\"o}rfler, A and Essig, M and Hurwitz, B and Kappos, L and Lee, S-K and Filippi, M},
  journal={American Journal of Neuroradiology},
  volume={31},
  number={6},
  pages={983--989},
  year={2010},
  publisher={Am Soc Neuroradiology}
}

@article{blystad2015quantitative,
  title={Quantitative {MRI} for analysis of active multiple sclerosis lesions without gadolinium-based contrast agent},
  author={Blystad, Ida and H{\aa}kansson, Irene and Tisell, Anders and Ernerudh, Jan and Smedby, {\"O}rjan and Lundberg, Peter and Larsson, E-M},
  journal={American Journal of Neuroradiology},
  year={2015},
  publisher={American Society Neuroradiology},
  number = {1},
  pages = {94--100},
  volume = {37},
}

@article{filippi2006efns,
  title={{EFNS} guidelines on the use of neuroimaging in the management of multiple sclerosis},
  author={Filippi, M and Rocca, MA and Arnold, DL and Bakshi, R and Barkhof, F and De Stefano, N and Fazekas, F and Frohman, E and Wolinsky, JS},
  journal={European Journal of Neurology},
  volume={13},
  number={4},
  pages={313--325},
  year={2006},
  publisher={Wiley Online Library}
}

@article{geurts2012measurement,
  title={Measurement and clinical effect of grey matter pathology in multiple sclerosis},
  author={Geurts, Jeroen JG and Calabrese, Massimiliano and Fisher, Elizabeth and Rudick, Richard A},
  journal={The Lancet Neurology},
  volume={11},
  number={12},
  pages={1082--1092},
  year={2012},
  publisher={Elsevier}
}

@InProceedings{Zijdenbos_MICCAI,
  author =       {A. Zijdenbos and R. Forghani and A. Evans},
  title =        {Automatic Quantification of {MS} Lesions in {3D} {MRI} Brain Data Sets: Validation of {INSECT}},
  booktitle =    {Proceedings of Medical Image Computing and Computer-Assisted Intervention -- {MICCAI'98}},
  OPTcrossref =  {},
  OPTkey =       {},
  pages =     {439--448},
  year =      {1998},
  OPTeditor =    {},
  volume =    {1496},
  OPTnumber =    {},
  series =    {Lecture Notes in Computer Science},
  OPTaddress =   {},
  OPTmonth =     {},
  OPTorganization = {},
  publisher = {Springer},
  OPTnote =      {},
  OPTannote =    {}
}

@article{smith2002accurate,
  title={Accurate, robust, and automated longitudinal and cross-sectional brain change analysis},
  author={Smith, Stephen M and Zhang, Yongyue and Jenkinson, Mark and Chen, Jacqueline and Matthews, PM and Federico, Antonio and De Stefano, Nicola},
  journal={NeuroImage},
  volume={17},
  number={1},
  pages={479--489},
  year={2002},
  publisher={Elsevier}
}

@article{smeets2016reliable,
  title={Reliable measurements of brain atrophy in individual patients with multiple sclerosis},
  author={Smeets, Dirk and Ribbens, Annemie and Sima, Diana M. and Cambron, Melissa and Horakova, Dana and Jain, Saurabh and Maertens, Anke and Van Vlierberghe, Eline and Terzopoulos, Vasilis and Van Binst, Anne-Marie and Vaneckova, Manuela and Krasensky, Jan and Uher, Tomas and Seidl, Zdenek and De Keyser, Jacques and Nagels, Guy and De Mey, Johan and Havrdova, Eva and Van Hecke, Wim},
  journal={Brain and Behavior},
  volume={6},
  number={9},
  year={2016},
  pages={e00518},
  publisher={Wiley Online Library}
}

@article{gelineau2012effect,
  title={The effect of hypointense white matter lesions on automated gray matter segmentation in multiple sclerosis},
  author={Gelineau-Morel, Rose and Tomassini, Valentina and Jenkinson, Mark and Johansen-Berg, Heidi and Matthews, Paul M and Palace, Jacqueline},
  journal={Human Brain Mapping},
  volume={33},
  number={12},
  pages={2802--2814},
  year={2012},
  publisher={Wiley Online Library}
}

@article{chard2010reducing,
  title={Reducing the impact of white matter lesions on automated measures of brain gray and white matter volumes},
  author={Chard, Declan T and Jackson, Jonathan S and Miller, David H and Wheeler-Kingshott, Claudia AM},
  journal={Journal of Magnetic Resonance Imaging},
  volume={32},
  number={1},
  pages={223--228},
  year={2010},
  publisher={Wiley Online Library}
}

@article{sdika2009nonrigid,
  title={Nonrigid registration of multiple sclerosis brain images using lesion inpainting for morphometry or lesion mapping},
  author={Sdika, Micha{\"e}l and Pelletier, Daniel},
  journal={Human Brain Mapping},
  volume={30},
  number={4},
  pages={1060--1067},
  year={2009},
  publisher={Wiley Online Library}
}

@article{ceccarelli2012impact,
  title={The impact of lesion in-painting and registration methods on voxel-based morphometry in detecting regional cerebral gray matter atrophy in multiple sclerosis},
  author={Ceccarelli, A and Jackson, JS and Tauhid, S and Arora, A and Gorky, J and Dell'Oglio, E and Bakshi, A and Chitnis, Tanuja and Khoury, Samia Joseph and Weiner, Howard Lee and others},
  journal={American Journal of Neuroradiology},
  year={2012},
  volume={33},
  number={8},
  pages={1579--1585},
}

@article{battaglini2012evaluating,
  title={Evaluating and reducing the impact of white matter lesions on brain volume measurements},
  author={Battaglini, Marco and Jenkinson, Mark and De Stefano, Nicola},
  journal={Human Brain Mapping},
  volume={33},
  number={9},
  pages={2062--2071},
  year={2012},
  publisher={Wiley Online Library}
}

@article{FreeSurfer,
  title={{FreeSurfer}},
  author={Fischl, Bruce},
  journal={NeuroImage},
  volume={62},
  number={2},
  pages={774--781},
  year={2012},
  publisher={Elsevier}
}

@article{vrenken2013recommendations,
  title={Recommendations to improve imaging and analysis of brain lesion load and atrophy in longitudinal studies of multiple sclerosis},
  author={Vrenken, H and Jenkinson, M and Horsfield, MA and Battaglini, M and Van Schijndel, RA and Rostrup, E and Geurts, JJG and Fisher, E and Zijdenbos, A and Ashburner, J and others},
  journal={Journal of Neurology},
  volume={260},
  number={10},
  pages={2458--2471},
  year={2013},
  publisher={Springer}
}

@article{nakamura2009segmentation,
  title={Segmentation of brain magnetic resonance images for measurement of gray matter atrophy in multiple sclerosis patients},
  author={Nakamura, Kunio and Fisher, Elizabeth},
  journal={Neuroimage},
  volume={44},
  number={3},
  pages={769--776},
  year={2009},
  publisher={Elsevier}
}

@article{Thompson2018,
author = {Thompson, Alan J. and Banwell, Brenda L. and Barkhof, Frederik and Carroll, William M. and Coetzee, Timothy and Comi, Giancarlo and Correale, Jorge and Fazekas, Franz and Filippi, Massimo and Freedman, Mark S. and Fujihara, Kazuo and Galetta, Steven L. and Hartung, Hans Peter and Kappos, Ludwig and Lublin, Fred D. and Marrie, Ruth Ann and Miller, Aaron E. and Miller, David H. and Montalban, Xavier and Mowry, Ellen M. and Sorensen, Per Soelberg and Tintor{\'{e}}, Mar and Traboulsee, Anthony L. and Trojano, Maria and Uitdehaag, Bernard M.J. and Vukusic, Sandra and Waubant, Emmanuelle and Weinshenker, Brian G. and Reingold, Stephen C. and Cohen, Jeffrey A.},
journal = {The Lancet Neurology},
title = {{Diagnosis of multiple sclerosis: 2017 revisions of the McDonald criteria}},
year = {2018},
volume={17},
number={2},
pages={162--173}
}

@article{Sormani,
author = {Sormani, Maria Pia. Bruzzi, Paolo},
journal={The Lancet Neurology},
title = {{MRI lesions as a surrogate for relapses in multiple sclerosis: a meta-analysis of randomised trials}},
year = {2013},
volume = {12},
number = {7},
pages = {669--676}
}

@article{Bianca,
author = {Griffanti, Ludovica and Zamboni, Giovanna and Khan, Aamira and Li, Linxin and Bonifacio, Guendalina and Sundaresan, Vaanathi and Schulz, Ursula G. and Kuker, Wilhelm and Battaglini, Marco and Rothwell, Peter M. and Jenkinson, Mark},
journal = {NeuroImage},
pages = {191--205},
title = {{BIANCA (Brain Intensity AbNormality Classification Algorithm): A new tool for automated segmentation of white matter hyperintensities}},
volume = {141},
year = {2016},
number = {1}
}

@article{MP2RAGEOrig,
author = {Marques, Jos{\'{e}} P and Kober, Tobias and Krueger, Gunnar and der Zwaag, Wietske van and de Moortele, Pierre-Fran{\c{c}}ois Van and Gruetter, Rolf},
journal = {NeuroImage},
number = {2},
pages = {1271--1281},
title = {{MP2RAGE, a self bias-field corrected sequence for improved segmentation and T1-mapping at high field}},
volume = {49},
year = {2010}
}

@article{DIROrig,
author = {Redpath, T. W. and Smith, F. W.},
journal = {{British Journal of Radiology}},
title = {{Technical note: Use of a double inversion recovery pulse sequence to image selectively grey or white brain matter}},
year = {1994},
volume={67},
pages={1258--1263},
number={804}
}

@article{T2StartOrig,
    author = {Anderson, L.J. and Holden, S. and Davis, B. and Prescott, E. and Charrier, C.C. and Bunce, N.H. and Firmin, D.N. and Wonke, B. and Porter, J. and Walker, J.M. and Pennell, D.J.},
    title = {{Cardiovascular T2-star (T2*) magnetic resonance for the early diagnosis of myocardial iron overload}},
    journal = {European Heart Journal},
    volume = {22},
    number = {23},
    pages = {2171-2179},
    year = {2001},
}

@article{Chard2002,
    author = {Chard, D. T. and Griffin, C. M. and Parker, G. J. M. and Kapoor, R. and Thompson, A. J. and Miller, D. H.},
    title = {Brain atrophy in clinically early relapsing-remitting multiple sclerosis},
    journal = {Brain},
    volume = {125},
    number = {2},
    pages = {327--337},
    year = {2002},
}

@article{Zivadinov2016,
author = {Zivadinov, Robert and Uher, Tomas and Hagemeier, Jesper and Vaneckova, Manuela and Ramasamy, Deepa P and Tyblova, Michaela and Bergsland, Niels and Seidl, Zdenek and Dwyer, Michael G and Krasensky, Jan and Havrdova, Eva and Horakova, Dana},
journal = {Multiple Sclerosis Journal},
number = {13},
pages = {1709--1718},
title = {{A serial 10-year follow-up study of brain atrophy and disability progression in RRMS patients}},
volume = {22},
year = {2016}
}

@article{Carass2020,
author = {Carass, Aaron and Roy, Snehashis and Gherman, Adrian and Reinhold, Jacob C. and Jesson, Andrew and Arbel, Tal and Maier, Oskar and Handels, Heinz and Ghafoorian, Mohsen and Platel, Bram and Birenbaum, Ariel and Greenspan, Hayit and Pham, Dzung L. and Crainiceanu, Ciprian M. and Calabresi, Peter A. and Prince, Jerry L. and Roncal, William R.Gray and Shinohara, Russell T. and Oguz, Ipek},
journal = {Scientific Reports},
title = {{Evaluating White Matter Lesion Segmentations with Refined S{\o}rensen-Dice Analysis}},
year = {2020},
volume = {10},
pages = {1--19}
}

@book{Huber1981,
 author = {Huber, Peter Jost},
 year = {1981},
 title = {{Robust Statistics}},
 publisher = {John Wiley and Sons},
 address = {New York}
}

@InProceedings{cerri2020,
  title={{A Longitudinal Method for Simultaneous Whole-Brain and Lesion Segmentation in Multiple Sclerosis}},
  author={Cerri, Stefano and Hoopes, Andrew and Greve, Douglas N and M{\"u}hlau, Mark and Van Leemput, Koen},
  booktitle={3rd International Workshop in Machine Learning in Clinical Neuroimaging (accepted)},
  year={2020}
}

\end{document}


\title{Supplementary Material for\\ ``A Contrast-Adaptive Method for Simultaneous Whole-Brain and Lesion Segmentation in Multiple Sclerosis
''}

\maketitle

\begin{figure}
    \centering
    \includegraphics[width=.99\linewidth]{./Images/NewFinal/MICCAIData.png} \\
    \includegraphics[width=.99\linewidth]{./Images/NewFinal/MICCAISegNew.png} \\
    \caption{Contrast-adaptiveness of the proposed method to different combinations of input modalities. Segmentations are shown for one subject of the MSSeg dataset.
    The top row shows slices of the data and the manual lesion annotation;
    the middle row shows the lesion probability map and Dice score computed by the proposed method for specific input combinations; and the bottom row shows the corresponding complete segmentations produced by the method.}
    \label{fig:supMICCAI}
\end{figure}

\begin{figure}
    \centering
    \includegraphics[width=.99\linewidth]{./Images/NewFinal/TRIOData.png} \\
    \includegraphics[width=.99\linewidth]{./Images/NewFinal/TRIOSegNew.png} \\
    \caption{Contrast-adaptiveness of the proposed method to different combinations of input modalities. Segmentations are shown for one subject of the Trio dataset.
    The top row shows slices of the data and the manual lesion annotation;
    the middle row shows the lesion probability map and Dice score computed by the proposed method for specific input combinations; and the bottom row shows the corresponding complete segmentations produced by the method.}
    \label{fig:supTRIO}
\end{figure}

\begin{figure}
    \centering
    \includegraphics[width=.99\linewidth]{./Images/NewFinal/FATData.png} \\
    \includegraphics[width=.99\linewidth]{./Images/NewFinal/FATSegNew.png} \\
    \caption{Contrast-adaptiveness of the proposed method to different combinations of input modalities. Segmentations are shown for one subject of the Achieva dataset.
    The top row shows slices of the data and the manual lesion annotation;
    the middle row shows the lesion probability map and Dice score computed by the proposed method for specific input combinations; and the bottom row shows the corresponding complete segmentations produced by the method.}
    \label{fig:supFAT}
\end{figure}

\begin{figure}
    \centering
    \includegraphics[width=.8\linewidth]{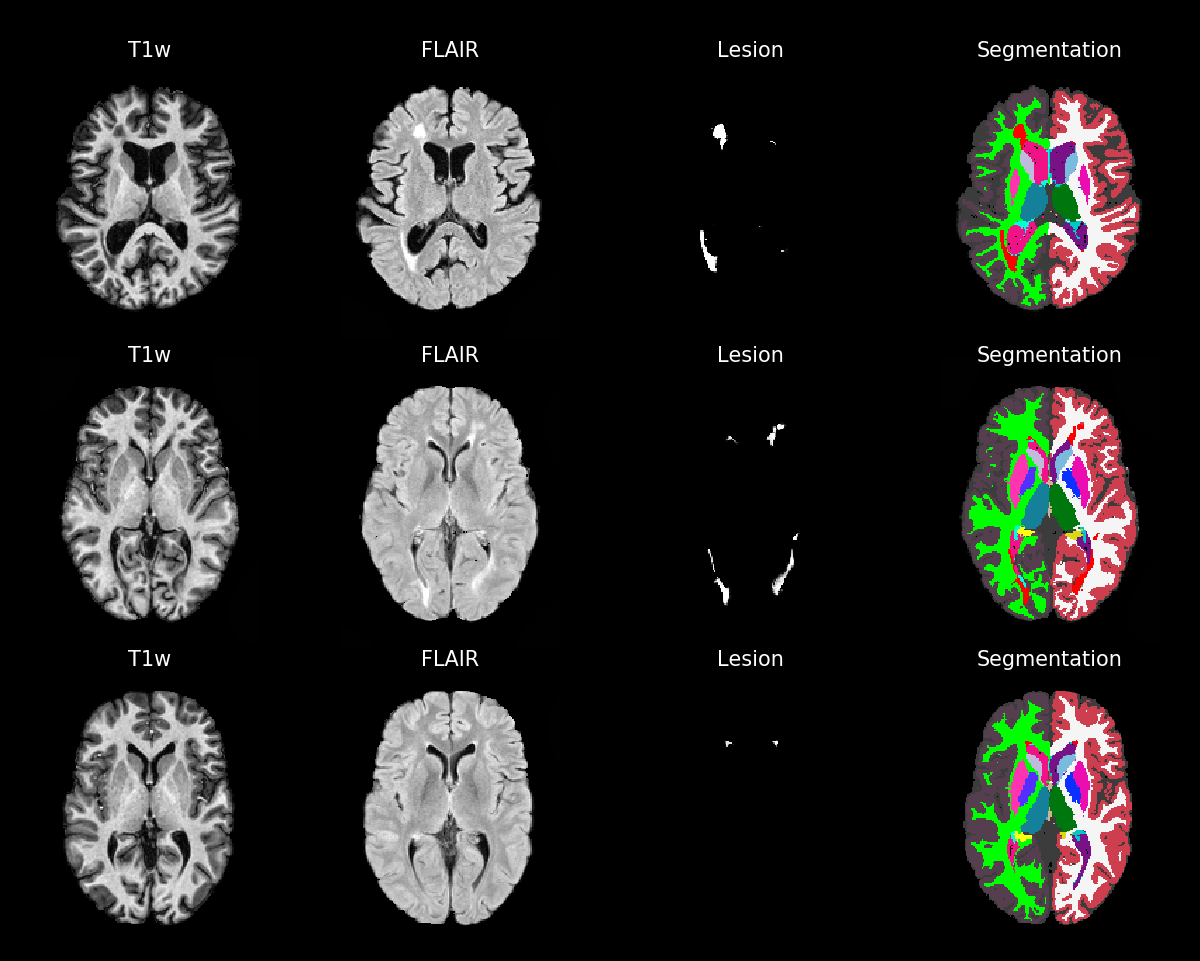} \\
    \caption{Segmentations of three subjects of the ISBI dataset from the proposed method on T1w-FLAIR input.
    From top to bottom: high, median and low score lesion performance obtained on the website evaluation platform of the ISBI challenge.
    From left to right: T1w, FLAIR, lesion probability map, whole-brain segmentation.}
    \label{fig:supISBI}
\end{figure}